\documentclass[reprint,twocolumn,amsmath,amssymb,amsthm, dsfont,prl,nobibnotes]{revtex4-2}
\usepackage{times}
\usepackage[colorlinks=true,citecolor=blue,linkcolor=blue]{hyperref}
\usepackage{graphicx}
\usepackage{mathrsfs}
\usepackage{dcolumn}
\usepackage{bm}
\usepackage{tikz}
\usepackage{tikz-cd}
\usepackage{hhline}
\usepackage{float}
\usepackage{scalerel}
\usepackage{amsfonts}
\usepackage{bbm}
\usepackage{color}
\usepackage{tabularx}
\usepackage{multirow}
\usepackage{enumitem}

\usepackage{subfig}
\usepackage{import}

\usepackage{CJKutf8}

\usetikzlibrary{decorations.markings,calc}

\usepackage{mathtools}

\usetikzlibrary{decorations.markings}
\usetikzlibrary{3d} % for canvas isometric
\usetikzlibrary{calc}

\usepackage{mathabx}
\usepackage{amsthm}

\theoremstyle{definition}
\newtheorem{defn}{Definition}
\newtheorem{example}{Example}

\newtheorem{remark}{Remark}

\newcommand{\ee}{\mathsf{e}}
\newcommand{\mm}{\mathsf{m}}
\newcommand{\Ker}{\text{Ker\,}}
\newcommand{\Imaa}{\text{Im\,}}
\newcommand{\ra}{\rightarrow}

\newcommand{\avi}{a_v^{(i)}}
\newcommand{\bpi}{b_p^{(i)}}
\newcommand{\aq}{a_1^{(q_2)}}
\newcommand{\bq}{b_1^{(q_2)}}

\newcommand{\NT}[1]{ { \color{purple} (NT: {}) }}
\newcommand{\SZ}[1]{ { \color{orange} (SZ: {}) }}

\newcommand{\TW}[1]{ { \color{brown} (TW: {}) }}
\begin{document}

%\preprint{APS/123-QED}

\title{Coupled-Layer Construction of Quantum Product Codes}

\author{Shuyu Zhang
(\begin{CJK*}{UTF8}{gbsn}张舒予\end{CJK*})}
\affiliation{C. N. Yang Institute for Theoretical Physics and Department of Physics and Astronomy,  Stony Brook University, Stony Brook, NY 11794, USA}

\author{Tzu-Chieh Wei (\begin{CJK*}{UTF8}{bsmi}魏子傑\end{CJK*})}
\affiliation{C. N. Yang Institute for Theoretical Physics and Department of Physics and Astronomy,  Stony Brook University, Stony Brook, NY 11794, USA}

\author{Nathanan Tantivasadakarn}
\affiliation{C. N. Yang Institute for Theoretical Physics and Department of Physics and Astronomy,  Stony Brook University, Stony Brook, NY 11794, USA}

\date{\today}

\begin{abstract}
Product codes are a class of quantum error correcting codes built from two or more constituent codes. They have recently gained prominence for a breakthrough yielding quantum low-density parity-check (qLDPC) codes with favorable scaling of both code distance and encoding rate. However, despite its powerful algebraic formulation, the physical mechanism for assembling a general product code from its constituents remains unclear. In this letter, we show that the tensor and balanced product codes admit an intuitive coupled-layer construction by taking a stack of one code and condensing a set of excitations in the pattern given by the checks of the other code. We also make a connection to concatenated codes by showing that the tensor product code can be obtained by gauging large-weight logicals in concatenated codes, making them qLDPC. Our framework accommodates both classical or quantum CSS input codes, unifies known physical mechanisms for constructing higher dimensional topological phases via anyon condensation, and naturally extends to non-topological codes.
\end{abstract}

\maketitle

\textit{Introduction.} 
Quantum computation is impractical without quantum error correction ~\cite{shor1995scheme}. Among many candidates, the most widely used family of codes to date are topological error correcting codes, which share intimate connections with topological phases of matter~\cite{simon2023topological}. Excitingly, experimental advances  can now implement such topological codes in actual hardware~\cite{satzingerrealizing2021,krinner2022realizing,HetenyiWootton24,IqbalTCprep,Iqbal24,bluvstein2024logical,IqbalZ3,QuEraMagicState2025,lo2026universal}, with some even demonstrating an advantage beyond the break-even point~\cite{GoogleWillow2024,brock2025quantum}. However, topological codes are severely restricted in scalability exactly due to its topological property \cite{Bravyi_2009,bravyi2010tradeoffs,Flammia2017}: it can host at most a constant number of logicals and a sub-linear distance on the number of physical qubits.

Recently, there has been a breakthrough in constructing \emph{good} quantum low-density parity-check (qLDPC) codes: those that boast a linear scaling for both number of logicals and code distance. From the condensed matter point of view, qLDPC codes are $k$-local stabilizer Hamiltonians, which consist of terms with finite number of support.
Thus, there has been substantial interest in interpreting qLDPC codes as phases of matter. Even though these codes lack geometric locality, they still share certain properties to conventional phases of matter, such as their gauge-theoretic interpretation~\cite{Kubica18,RakovszkyLDPC1,RakovszkyLDPC2} and stability to perturbations~\cite{DeRoeck24,Placke2024,Placke25,Yin25,yu2025universal}.

A valuable construction of qLDPC codes is quantum product codes~\cite{Tillich:2013esj,Bravyi14,hastings2021fiber,panteleev2021quantum,breuckmann2021balanced, Panteleev_ACM2022}, which produces a new quantum code from two or more classical or quantum codes. For example, the 2D toric code (TC) can be realized as the tensor product of two classical repetition codes. Given that product constructions have been used to demonstrate the existence of good qLDPC codes, 
a natural question to ask is in what sense the product code is \emph{constructed} from its constituents.  Ref.~\cite{RakovszkyLDPC2} demonstrated that the tensor product construction of two classical codes can be achieved by ``repeating" each of the checks in a two-dimensional grid and gauging the resulting classical code. However, the concept of repeating checks does not generalize naturally to product of two quantum codes. Is there a unifying physical framework of various product constructions?

In this letter, we demonstrate that quantum product codes can be naturally described via a coupled-layer construction. Such constructions are ubiquitous in condensed matter physics, where lower-dimensional phases of matter are coupled together to produce higher-dimensional ones. Prominent examples include the construction of fractional quantum Hall phases by coupling gapless wires~\cite{Kane02,TeoKane14}, or the construction of 3D topological phases from its 2D counterparts~\cite{WangSenthil13,JianQi14,Vijay17,MaLakeChenHermele2017,PremHuangSongHermele2019,schmitz2019distilling,Halasz17,designer,aasen2020topological,Williamson21,SullivanPlanarpstring,Williamson21,Wen20,Wang22, song2023topological, aasen2020topological,Liu23}.

Our construction generalizes the anyon condensation construction of topological codes~\cite{WangSenthil13,JianQi14} to non-topological codes such as qLDPC codes with a single intuitive picture: \emph{the second code is used as a recipe to choose commuting condensation terms in a stack of the first code}. At the same time, it also encompasses both the tensor product and balanced product constructions and allows us to reproduce in one coherent framework many important codes from the input of two codes, either of which can be classical or quantum CSS codes.

\textit{Review of the tensor product.} A CSS quantum code $[[n,k,d]]$ can be encoded in a cochain complex of $\mathbb{F}_2$ vector spaces
\begin{align*}
    A \xrightarrow{\delta_X} Q \xrightarrow{\delta_Z} B.
\end{align*}
where $Q \cong \mathbb F_2^n$, $A$, $B$ denote the qubits, $X$- and $Z$-stabilizers, respectively, and $\delta_X^\top$ and $\delta_Z$ are the corresponding parity checks. The condition that the stabilizers commute is equivalent to $\delta_Z\delta_X = 0$. Given $a\in A$, define $q\in a$ to run over all $q$'s checked by $a$ (i.e., all generators in $\delta_X(a)$). Likewise, given $q\in Q$, define $a\ni q$ runs over all $a$'s checking $q$ (generators in $\delta_X^T(q)$), and similarly for $Z$-stabilizers.

\begin{figure}
    \begin{tikzpicture}[scale=0.65, every node/.style={scale=0.8}]
  % column centers
    \def\LL{-6}
  \def\L{-3}
  \def\C{0}
  \def\R{3}
   \def\RR{6}
   \def\shadedwidth{60}
   \def\shadedheight{160}
    \def\H{1}
    \def\HH{2}

  % Background stripes
  \begin{scope}
    \node[rounded corners=9pt, fill=red!20,  minimum width=\shadedwidth, minimum height=\shadedheight] at (\L,-1) {};
    \node[rounded corners=9pt, fill=black!10, minimum width=\shadedwidth, minimum height=\shadedheight] at (\C,-1) {};
    \node[rounded corners=9pt, fill=blue!20, minimum width=\shadedwidth, minimum height=\shadedheight] at (\R,-1) {};
  \end{scope}

  % Big labels and arrows along the top
  \node[font=\bfseries\Huge, text=red!80!black]  (A) at (\L, -3) {$\mathit{A}$};
  \node[font=\bfseries\Huge, text=black!75]      (Q) at (\C,-3) {$\mathit{Q}$};
  \node[font=\bfseries\Huge, text=blue!80!black] (B) at (\R, -3) {$\mathit{B}$};
  \draw[very thick,->] (A) -- (Q);
  \draw[very thick,->] (Q) -- (B);

  % Column subtitles
  \node[font=\Large, text=red!80!black,  anchor=north] at (\L, -3.5) {$X$-checks};
  \node[font=\Large, text=black!80,      anchor=north] at (\C, -3.5) {Qubits};
  \node[font=\Large, text=blue!80!black, anchor=north] at (\R, -3.5) {$Z$-checks};

  % Centered \oplus in each column
  \node[font=\Large, text=red!80!black]  at (\L, 0) {$\oplus$};
  \node[font=\Large, text=black]      at (\C, \H) {$\oplus$};
    \node[font=\Large, text=black]      at (\C, -\H) {$\oplus$};
  \node[font=\Large, text=blue!80!black] at (\R, 0) {$\oplus$};

  % External boundary labels
  \node (AA) at (\LL, 0) {$A_{1}\!\otimes\! A_{2}$};
  \node (BB) at ( \RR, 0) {$B_{1}\!\otimes\! B_{2}$};

  % Nodes inside stripes
  % Left stripe (A)
  \node[text=red!80!black]  (A1Q2) at (\L,  \H) {$A_{1}\otimes Q_{2}$};
  \node[text=red!80!black]  (Q1A2) at (\L, -\H) {$Q_{1}\otimes A_{2}$};

  % Middle stripe (Q)
  \node[text=black!80]      (A1B2) at (\C,  \HH) {$A_{1}\otimes B_{2}$};
  \node[text=black!80]      (Q1Q2) at (\C,       0.0) {$Q_{1} \otimes Q_{2}$};
  \node[text=black!80]      (B1A2) at (\C, -\HH) {$B_{1}\otimes A_{2}$};

  % Right stripe (B)
  \node[text=blue!80!black] (Q1B2) at (\R,  \H) {$Q_{1}\otimes B_{2}$};
  \node[text=blue!80!black] (B1Q2) at (\R, -\H) {$B_{1}\otimes Q_{2}$};

  % Arrows from the far left into A stripe
  \draw[->] (AA) -- (A1Q2);
  \draw[ ->] (AA) -- (Q1A2);

  % Arrows A stripe -> Q stripe
  \draw[->] (A1Q2) -- (A1B2);
  \draw[->] (A1Q2) -- (Q1Q2);
  \draw[->] (Q1A2) -- (Q1Q2);
  \draw[->] (Q1A2) -- (B1A2);
    \draw[->] (Q1Q2) -- (Q1B2);

  % Arrows Q stripe -> B stripe
  \draw[black, ->] (A1B2) -- (Q1B2);
  \draw[black, ->] (Q1Q2) -- (B1Q2);
  \draw[black, ->] (B1A2) -- (B1Q2);

  % Arrows from B stripe to far right
  \draw[black, ->] (Q1B2) -- (BB);
  \draw[black, ->] (B1Q2) -- (BB);
\end{tikzpicture}
    \caption{The tensor product complex}
    \label{fig:productcomplex}
\end{figure}
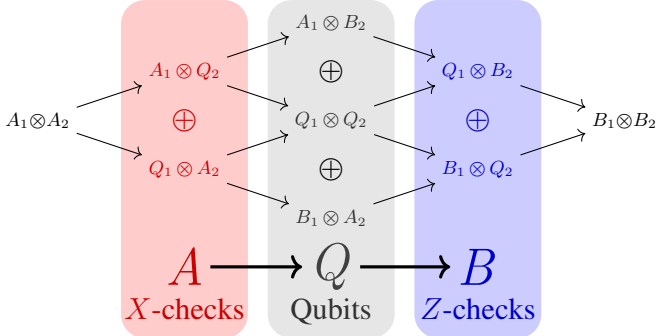

Given two cochain complexes $\mathcal{C} = (C^\bullet,\delta_C^\bullet)$ and $\mathcal{D} = (D^\bullet,\delta_D^\bullet)$ (where ${}^\bullet$ denotes the degree),
one can form a new complex by the tensor product, where $(\mathcal{C}\otimes \mathcal{D})^k = \oplus_{i+j=k} C^i\otimes D^j$, and $\delta^k_{\mathcal{C}\otimes\mathcal{D}} = \sum_{i+j=k}\delta_\mathcal{C}^i\otimes id +  id\otimes \delta_\mathcal{D}^j$. \ \ Applying the tensor product to two CSS codes CSS$_i$: $A_i\xrightarrow{\delta_{X,i}} Q_i \xrightarrow{\delta_{Z,i}}B_i$, yields the chain complex in Figure~\ref{fig:productcomplex}.

Placing qubits at degree $0$, the tensor product yields the product quantum code with
\begin{align}
\label{eq:ABQproduct}
   &A = (A_1\otimes Q_2) \oplus (Q_1\oplus A_2), \ \ B =(B_1\otimes Q_2) \oplus (Q_1\oplus B_2), \nonumber\\
   & Q= (A_1 \otimes B_2) \oplus (Q_1 \otimes Q_2) \oplus (B_1 \otimes A_2).
\end{align}
Note that the usual homological/hypergraph product of two classical codes also fits into this construction~\cite{Tillich:2013esj,Bravyi14}, but we need to gauge/Hadamard one of the classical codes in our convention~\cite{supp}. A familiar construction is the 2D TC, which can be realized as a product of two classical repetition codes: one with $ZZ$ checks, and one with $XX$ checks.

\textit{Motivating example.} Our construction starts with $n_2$ copies of CSS$_1$. Then, the layers are coupled together using the pattern of checks in CSS$_2$. Before introducing the formalities, let us demonstrate with a simple example. Let CSS$_1$ be the 2D TC $[[2L^2,2,L]]$ (with $A,Q,B$ generated by vertices $v$, edges $e$ and plaquettes $p$ of the $L\times L$ square lattice), and CSS$_2$ be the [[4,2,2]] code with checks $X_1X_2X_3X_4$ and $Z_1Z_2Z_3Z_4$. We therefore start with four copies of the TC. Denote $X_{e,i}$ and $Z_{e,i}$ the Pauli matrices acting on the qubit on edge $e$ in the $i^\text{th}$ copy for $i=1,\ldots,4$. The stabilizers  are $\avi = \prod_{e\in v} X_{e,i}$ and $\bpi=\prod_{e\in p} Z_{e,i}$. The violations of $\avi$ and $\bpi$ will be called $\ee_i$ and $\mm_i$ anyons, respectively.

Next, we introduce ancilla qubits: one for each vertex with stabilizer $X_v$, and one for each plaquette with stabilizer $Z_p$. The total stabilizer group is therefore $\mathcal{S}_0 =\langle \avi, \bpi, X_v, Z_p \rangle$.

\begin{figure}[t!]
    \centering
    \begin{tikzpicture}
    [baseline=0ex]
    \node at (0,1.2) {$\underline{\xi:X_v \avi}$};
        \node at (0,0) {$X_v$};
        \node at (-0.45, 0) {$X_i$};
        \node at (0.45,0) {$X_i$};
        \node at (0, 0.45) {$X_i$};
        \node at (0,-0.45) {$X_i$};
        \draw[   step = 1.5] (-0.8,-0.8) grid (0.8,0.8);
    \end{tikzpicture}
    \hspace{5pt}
    \begin{tikzpicture}
    [baseline=5ex]
    \node at (0.75,1.2+0.75) {$\underline{\zeta :Z_p  \bpi}$};
        \node at (0.75,0.75) {$Z_p$};
        \node at (0, 0.75) {$Z_i$};
        \node at (1.5,0.75) {$Z_i$};
        \node at (0.75, 0) {$Z_i$};
        \node at (0.75,1.5) {$Z_i$};
        \draw[   step = 1.5] (0,0) grid (1.5,1.5);
    \end{tikzpicture}
    \hspace{5pt}
    \begin{tikzpicture}
    [baseline=0ex]
     \node at (0.7,1.2) {$\underline{\alpha_e}$};
        \node at (0.7,0.5) {$X$};
        \node at (0.7, -0.5) {$X$};
        \node at (0.7,0) {$\prod_i X_{e,i}$};
        \draw[   step = 1.5] (-0.2,-0.5) grid (1.7,0.5);
    \end{tikzpicture}
    \hspace{5pt}
    \begin{tikzpicture}
    [baseline=0ex]
    \node at (0,1.2) {$\underline{\beta_e}$};
        \node at (-0.75,0) {$Z$};
        \node at (0.75, 0) {$Z$};
        \node at (0,0) {$\prod_i Z_{e,i}$};
        \draw 
        (-1,0) -- (1,0)
        (-0.75,-0.5) -- (-0.75,0.5)
        (0.75,-0.5) -- (0.75,0.5);
    \end{tikzpicture}
    \caption{\label{fig:1} The stabilizers of the 2D toric code $\otimes$ [[4,2,2]]. 
    }
\end{figure}
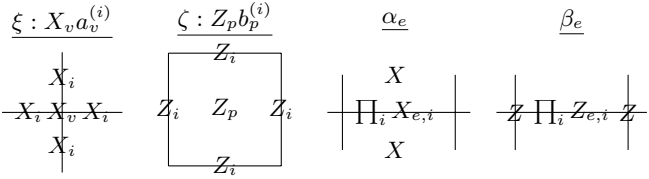

Finally, we perform a code switching by adding the following two stabilizers for each edge $e$ of the lattice:
\begin{align}
\label{eq:422TCcondensation}
\alpha_e &= \prod_{p \ni e} X_p \prod_{i=1}^{4} X_{e,i}, & \beta_e=\prod_{v \ni e} Z_v \prod_{i=1}^{4} Z_{e,i},
\end{align}
where the product over $p\ni e$ and $v\ni e$ ranges over, respectively, $p$ and $v$ containing $e$. Keeping terms that commute, the final stabilizer group can be shown to be $\langle X_v  \avi, Z_p  \bpi,   \alpha_e, \beta_e \rangle $, visualized in Fig.~\ref{fig:1}. One can check that this stabilizer group is exactly the tensor product code CSS$_1 \otimes$ CSS$_2$.

We now interpret the above code switching in terms of anyon condensation. We view the generators of the original stabilizer group as terms in a commuting Hamiltonian $H = -\sum_{s \in \mathcal{S}_0}s$. The code switching corresponds to adding $-\Lambda \sum_e (\alpha_e + \beta_e)$ to the Hamiltonian and taking $\Lambda \rightarrow \infty$. Performing degenerate perturbation theory in $1/\Lambda$, the effective Hamiltonian at second order contains exactly $X_v \avi$ and $Z_p \bpi$.

To physically interpret this perturbation, first consider $\beta_e$ in Eq.~\eqref{eq:422TCcondensation}. Each $Z_{e,i}$ creates a pair of $\mathsf{e}_{i}$ anyons in the $i^\text{th}$ TC on the boundary vertices $v_1, v_2$ of $e$, while $Z_{v_1}Z_{v_2}$ is an Ising term which creates two $\mathbb{Z}_2$ charges in the paramagnet $X_v$ defined on the vertices. Thus, this term can be thought of as condensing the anyon $\ee_1\ee_2\ee_3\ee_4$ (along with the $\mathbb{Z}_2$ charge). More precisely, we are gauging diagonal $Z$-logical of the four TCs~\cite{TantivasadakarnMeasurement,williamson2024low} (i.e., the 1-form symmetry generated by all the electric lines, where the ancilla plays the role of the 2-form gauge field). Similarly, the term $\alpha_e$ condenses $\mm_1\mm_2\mm_3\mm_4$ (gauges the diagonal $X$-logical). Since we condensed away two copies worth of TCs, we are effectively left with two remaining copies worth of TCs as a topological phase. Indeed, the resulting product code has $4$ logical qubits. The logicals of the product code can be thought of as taking the logicals of the $[[4,2,2]]$ code and naively replacing $Z$ with $\ee$ and $X$ with $\mm$. Namely, they are strings of the anyons $\ee_1\ee_2$, $\ee_1\ee_3$, $\mm_1\mm_2$, $\mm_1\mm_3$.

\smallskip
\textit{Coupled-layer construction.} We now provide the general construction given two codes CSS$_1$ and CSS$_2$. For each qubit $q_2 \in Q_2$ in CSS$_2$, introduce a copy of CSS$_1$ labeled by CSS$_1^{(q_2)}$. The stabilizers in that layer are labeled $\aq = \prod_{q_1 \in a_1} X_{q_1,q_2}$ and $\bq = \prod_{q_1 \in b_2} Z_{q_1,q_2}$. Additionally, we introduce the following ancillas: For each $a_1 \in A_1$ and $b_2 \in B_2$, introduce an $X$-check $X_{a_1,b_2}$. Similarly, for $b_1\in B_1$ and $a_2 \in A_2$, introduce $Z_{b_1,a_2}$. 
The starting stabilizer group is therefore equivalent to $n_2$ copies of CSS$_1$ and ancillas given by
$\mathcal{S}_0 = \langle \aq, \bq, X_{a_1,b_2}, Z_{b_1,a_2} \rangle$. Moreover, $\mathcal{S}_0$ lives in the space $Q$ in Eq.~\eqref{eq:ABQproduct}, exactly the qubits of the product code.

Now we add the following to the stabilizer group: 
\vspace{-0.5em}
\begin{itemize}
    \item For each $q_1 \in Q_1$ and $a_2 \in A_2$, introduce $X$-stabilizer $\alpha(q_1,a_2) := \prod_{b_1\ni q_1} X_{b_1,a_2}\prod_{q_2\in a_2}X_{q_1,q_2}$.    \vspace{-0.5em}
    \item For each $q_1 \in Q_1$ and $b_2 \in B_2$, introduce $Z$-stabilizers $\beta(q_1,b_2) := \prod_{a_1\ni q_1} Z_{a_1,b_2} \prod_{q_2\in b_2} Z_{q_1,q_2}$.
\end{itemize}
\vspace{-0.5em}
One can check that these terms commute, so they can be added simultaneously.

\begin{figure}
    \begin{tikzpicture}[scale=0.8]
  % column centers
    \def\L{-4}
    \def\C{0}
    \def\R{4}
    \def\shadedwidth{60}
    \def\shadedheight{160}
    \def\H{1}
    \def\HH{2}
    \def\offs{2}

  % Background stripes
  \begin{scope}
    \node[rounded corners=9pt, fill=red!20,  minimum width=\shadedwidth, minimum height=30] at (\L,-1) {};
    \node[rounded corners=9pt, fill=blue!20, minimum width=\shadedwidth, minimum height=30] at (\R,1) {};
  \end{scope}

  % Centered \oplus in each column
  \node[font=\Large, text=black]  at (\L, 0) {$\oplus$};
  \node[font=\Large, text=black]      at (\C, \H) {$\oplus$};\node[font=\Large, text=black]      at (\C, -\H) {$\oplus$};
  \node[font=\Large, text=black] at (\R, 0) {$\oplus$};

  % Left stripe
  \node  (A1Q2) at (\L,  \H) {};
  \node at ($(\L,  \H) + (0, 0)$) {$A_{1}\otimes Q_{2}$};
  \node at ($(\L,  \H) + (\offs, 0)$) {$\xi(a_1, q_2)$};

  \node (Q1A2) at (-2, -\H) {};
  \node[text=red!80!black] at ($(\L, -\H) +(0, 0)$) {$Q_{1}\otimes A_{2}$};
  \node[text=red!80!black] at ($(\L, -\H) + ( \offs, 0)$) {$\alpha(q_1, a_2)$};
  
  \node[text=red!80!black] at (\L, -2*\H) {Measure $X$};

  % Middle stripe (Q)
  \node[text=black!80]      (A1B2) at (\C,  \HH) {$A_{1}\otimes B_{2}$};
  \node[text=red]      (A1B2caption) at (\C, +1.3*\HH) {$X$-ancilla};
  \node[text=black!80]      (Q1Q2) at (\C,       0.0) {$Q_{1} \otimes Q_{2}$};
  \node[text=black!80]      (B1A2) at (\C, -\HH) {$B_{1}\otimes A_{2}$};
  \node[text=blue]      (B1A2caption) at (\C, -1.3*\HH) {$Z$-ancilla};

  % Right stripe (B)
  % \node (Q1B2) at (2,  \H) {};
  
  \node[text=blue!80!black] at ($(\R,  \H) + (0, 0)$) {$Q_{1}\otimes B_{2}$};
  \node[text=blue!80!black] at ($(\R,  \H) - (\offs,0)$) {$\beta(q_1, b_2)$};
  
  \node[text=black] (B1Q2) at (\R, -\H) {};
  \node[text=black] at ($(\R, -\H) + (0, 0)$) {$B_{1}\otimes Q_{2}$};
  \node[text=black] at ($(\R, -\H) - (\offs, 0)$) {$\zeta(b_1, q_2)$};
  
  \node[text=blue!80!black] at (\R, 2*\H) {Measure $Z$};

  % Arrows A stripe -> Q stripe
  \draw[->,dashed,red] (-3, 1.2) -- (A1B2);
  \draw[->, ultra thick] (-3, 0.8) -- (Q1Q2);
  \draw[->,red!80!black, thick] (-3, -0.8) -- (Q1Q2);
  \draw[->,red!80!black, thick] (-3, -1.2) -- (B1A2);

  % Arrows Q stripe -> B stripe
  \draw[ ->,blue!80!black, thick] (A1B2) -- (3, 1.2);
  \draw[->,blue!80!black, thick] (Q1Q2) -- (3, 0.8);
  \draw[ ->,ultra thick] (Q1Q2) -- (3, -0.8);
  \draw[ dashed,->,blue] (B1A2) -- (3, -1.2);

  \def\pad{0.35}   % how much to extend *past* A1⊗Q2 and B1⊗Q2 along the line
  
  \def\th{0.21}    % half-thickness; try 0.45–0.70
  
  \fill[black!40, fill opacity=0.22, draw opacity=0, rounded corners=8pt]
  % upper edge from left to right
  ($ (A1Q2) + ({-3*\pad},{\pad}) + ({\th},{3*\th}) $) --
  ($ (B1Q2) + ({ 3*\pad},{-\pad}) + ({\th},{3*\th}) $) --
  % lower edge back from right to left
  ($ (B1Q2) + ({ 3*\pad},{-\pad}) + ({-\th},{-3*\th}) $) --
  ($ (A1Q2) + ({-3*\pad},{\pad}) + ({-\th},{-3*\th}) $) -- cycle;

\end{tikzpicture}
    \caption{The coupled layer construction from the chain complex perspective. Starting with stacks of CSS$_1$ given by
    the complex in the grey band along with $X/Z$ ancillas, Measuring $X/Z$-stabilizers given
    by the red/blue arrow ($\alpha(q_1,a_2)/ \beta(q_1,b_2)$) induces the blue/red dashed arrow, completing $\zeta(b_1,q_2)/\xi(a_1,q_2)$.}
    \label{fig:coupledlayerchaincomplex}
\end{figure}
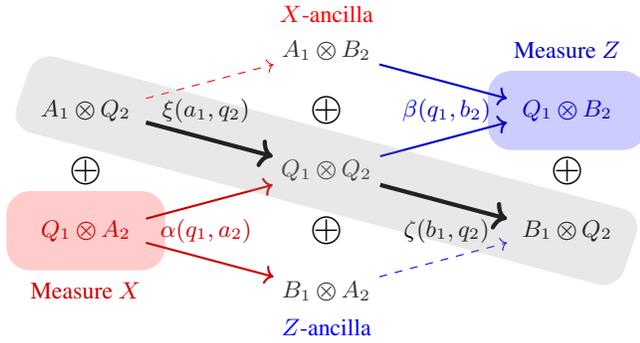

To compute the resulting stabilizer group, we find terms in $\mathcal{S}_0$ that commute with the above two types of stabilizers. One can check that the following terms in $\mathcal{S}_0$ remain
\vspace{-0.5em}
\begin{itemize}
    \item $X$ terms from $A_1\otimes Q_2$: Each $a_1\otimes q_2$ gives $\xi(a_1,q_2) : = \aq\left(\prod_{b_2\ni q_2}X_{a_1,b_2}\right)$;
    \vspace{-0.5em}
    \item $Z$ terms from $B_1\otimes Q_2$: Each $b_1\otimes q_2$ gives $\zeta(b_1, q_2) := \bq\left(\prod_{a_2\ni q_2}Z_{b_1,a_2}\right)$.
\end{itemize}
\vspace{-0.5em}
It is clear they are indeed products of generators of $\mathcal{S}_0$, and the four terms form exactly the tensor product code CSS$_1\otimes$CSS$_2$. The corresponding chain complex perspective is shown in Fig.~\ref{fig:coupledlayerchaincomplex}.

Looking at $\alpha(q_1,a_2)$ from a condensation point of view, each $X_{q_1,q_2}$ anti-commutes with the set of $Z$-stabilizers $\{\bq\,|\, b_1\ni q_1\}$ in CSS$_1^{(q_2)}$, hence creating a number of excitations. On the other hand, each $X_{b_1,a_2}$ creates a single $\mathbb{Z}_2$ charge in the $Z$-paramagnet labeled by $a_2$. Hence, this term can be interpreted as condensing excitations in layers CSS$_1(q_2)$ where $q_2\in a_2$. Similarly, $\beta(q_1,b_2)$ condenses excitations created by $Z_{q_1,q_2}$ in layers CSS$_1(q_2)$ where $q_2\in b_2$.

\begin{figure}[h!]
    \centering
    % \hspace*{-1cm}
    \subfloat{\resizebox{0.45\textwidth}{!}{
    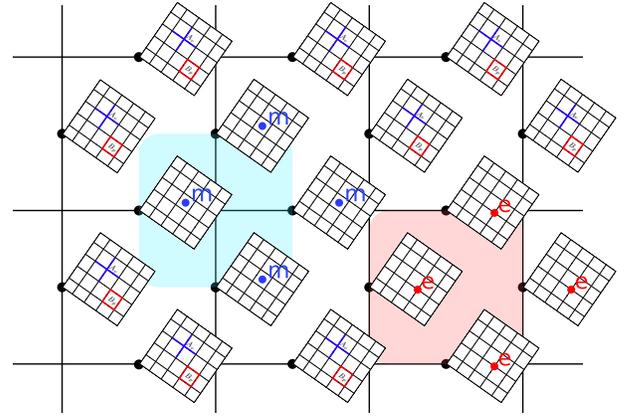}}
    \caption{4D toric code from condensing anyons in a grid of 2D toric codes placed on each edge of the square lattice.}
    \label{fig:4DTCcoupledlayer}
\end{figure}

Our construction reproduces a well-known construction of the 3D TC by taking a stack of 2D TC and condensing $\ee$ pairs in adjacent layers~\cite{JianQi14}. In this framework, this is just the 2D TC tensored with the repetition code, where the $\ee$ pairs condensed correspond exactly to the weight-2 checks of the repetition code. Moreover, we may exchange the roles of the two codes in the coupled layer construction, which gives an alternative construction of the 3D toric codes by coupling repetition codes laid in a 2D grid~\cite{supp}.  As another more interesting application, we provide a new coupled-layer construction of the (self-correcting) 4D toric code by tensoring the 2D toric code with itself. Consider a 2D toric code living at each edge of the square lattice. We may perform two types of condensations: for each vertex, condense four $\mm$ anyons from the four toric codes surrounding each vertex. Similarly, condense four $\ee$ anyons surrounding each plaquette as in Fig.~\ref{fig:4DTCcoupledlayer}. This results in the 4D toric code~\cite{supp}.

\begin{figure}[b!]
    \centering
    % \hspace*{-1cm}
    \includegraphics[scale=0.7]{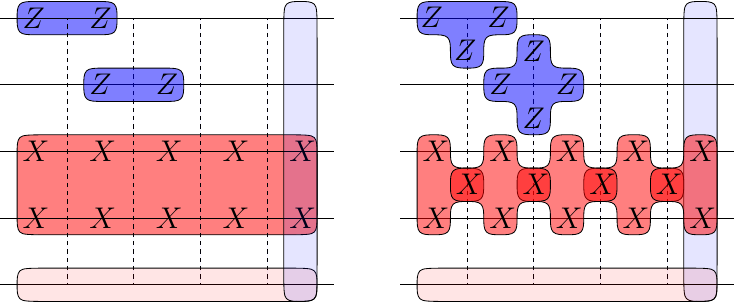}
    \caption{(Left) Stabilizers of Shor's code, obtained by stacking bit-flip repetition codes and enforcing two consecutive  $X$-logicals. (Right) Gauging these $X$-stabilizers breaks them down into local plaquette terms, while the $Z$-stabilizers are minimally-coupled into vertex terms of the surface code. The logicals of Shor's code, a horizontal $X$-string (red) and a vertical $Z$-string (blue) is inherited into the surface code.}
    \label{fig:shor}
\end{figure}

\textit{Relation to concatenated codes}. It is insightful to compare the tensor product to concatenated codes~\cite{knill1996concatenated} (here we assume they are CSS for simplicity). Let CSS$_1$ and CSS$_2$ be the outer and inner codes, respectively. The concatenated code replaces physical qubits of CSS$_2$ by logical qubits of CSS$_1$. Thus, to construct the concatenated code, we take $n_2$ copies of CSS$_1$ and enforce the combination of logicals given by checks in CSS$_2$ by adding them to the stabilizer group. A problem of concatenated codes is that if the logicals of CSS$_1$ have large distance, the concatenated code will have large-weight stabilizers. For example, take the two codes to be the bit-flip and phase-flip repetition codes on $L$ qubits. The concatenated code is a generalization of Shor's code with weight $2L$ stabilizers (see Fig.~\ref{fig:shor}). Moreover, the stabilizer group of concatenated codes is not symmetric under swapping CSS$_1$ and CSS$_2$.

The tensor product solves the problem of large weight stabilizers by \emph{gauging} them. With the help of ancillas, the global enforcement of the stabilizers can instead be broken down into local enforcements given by $\alpha$/$\beta$ for the $X$/$Z$-logicals, respectively. Importantly, the product of these local terms reproduces the original large stabilizers in the concatenated code, which is why they are still enforced. Moreover, the resulting product code is symmetric, and the logicals still take the same form. For Shor's code, the resulting code after gauging the large-weight stabilizers is a surface code, which is the tensor product of the repetition codes~\cite{supp}.

\textit{Subtleties with metachecks.} There are instances where our coupled layered construction (code switching) can give more terms in the stabilizer group than those from the naive tensor product. For example, taking the product of the 2D classical Ising model in the $Z$-basis, and the repetition code in the $X$-basis, produces the 3D TC, but with one orientation of plaquette stabilizers missing. In contrast, the coupled-layer construction will automatically include such terms. This subtlety is related to the existence of local metachecks of the 2D Ising model, which is not present in previous examples.  By including local metachecks, which are plaquettes of the square lattice, into the chain complex, one can recover the 3D TC. In general, by extending the chain complex to include all low weight metachecks (and relations among them) before taking the tensor product, one can recover the correct coupled-layer construction~\cite{supp}.

\smallskip
\textit{Balanced product}. 
Given the recent breakthrough of balanced product of quantum codes~\cite{tian2020haah,hastings2021fiber,panteleev2021quantum, breuckmann2021balanced, Panteleev_ACM2022}, here, we rederive its construction using the coupled-layer approach. For simplicity of the presentation, we assume the group action is free (i.e., there are no non-trivial fixed points)~\cite{supp2}.

Recall, given two vector spaces $V$ and $W$ acted on by a group $G$ from right and left respectively, the balanced product $V\otimes_G W$ is defined by $V\otimes W$ quotiented by the relation $v \cdot g \otimes w \sim v\otimes g \cdot w $ for $v\in V$, $w\in W$ and $g\in G$. Take two CSS codes, and let $G$ act on CSS$_1$ and CSS$_2$ from right and left respectively, the balanced product CSS code is given by Eq.~(\ref{eq:ABQproduct}), but with $\otimes$ replaced by $\otimes_G$.

To realize the coupled-layer construction, the only difference from tensor product is the quotient by the $G$ action, hence we need to fix representatives of orbits. We leave the stacks of CSS$_1$ untouched and perform the quotients on CSS$_2$. To this end, we first choose representatives $\{\tilde a_2\}$, $\{\tilde b_2\}$ and $\{\tilde q_2\}$ for each orbit in $G\backslash A_2$, $G\backslash B_2$ and $G\backslash Q_2$ respectively. The remaining steps mirrors closely the tensor product. For each $\tilde q_2$, introduce a copy of CSS$_1$ denoted CSS$_1^{(\tilde q_2)}$. For each $\Tilde{a}_2\in \Tilde{A}_2$ and $b_1 \in B_1$, introduce a $Z$-ancilla $Z_{b_1,  \tilde a_2}$. For each $\Tilde{b}_2\in \Tilde{B}_2$ and $a_1 \in A_1$, introduce an $X$-ancilla $X_{a_1,  \tilde b_2}$. The stabilizer group is (noticing the identical structure as in the above coupled-layer construction)
\begin{align*}
    \mathcal{S}_0 = \langle a_1^{(\tilde q_2)}, b_1^{(\tilde q_2)}, X_{a_1,\tilde b_2}, Z_{b_1,\tilde a_2} \rangle.
\end{align*}

Next, we perform code switching as before, the following terms are added to the stabilizer group: 
\vspace{-0.5em}
\begin{itemize}
    \item For each pair $(q_1, \tilde a_2)$, introduce the $X$-stabilizer 
    
    $\alpha(q_1, \tilde a_2) := \prod_{b_1\ni q_1} X_{b_1, \tilde a_2}
    \prod_{q_2\in \tilde a_2}X_{q_1 \cdot g_{q_2}, \tilde q_2}$.    \vspace{-0.5em}
    \item For each pair $(q_1, \tilde b_2)$, introduce the $Z$-stabilizer
    
    $\beta(q_1, \tilde b_2) := \prod_{a_1\ni q_1} Z_{a_1, \tilde b_2} \prod_{q_2\in \tilde b_2} Z_{q_1 \cdot g_{q_2}, \tilde q_2}$.
\end{itemize}
\vspace{-0.5em}
where $g_{q_2}$ is the unique group element which takes the representative $\tilde q_2$ to $q_2$, i.e., $q_2 = g_{q_2} \cdot \tilde q_2$. This construction has a similar interpretation from the condensation point of view as before. Looking at $\alpha(q_1,\tilde a_2)$, each $X_{b_1, \tilde a_2}$ creates a $\mathbb{Z}_2$ charge, while each $X_{q_1\cdot g_{q_2}, \tilde q_2}$ creates a number of excitations in CSS$_1^{(\tilde q_2)}$. However, now there can be more than one set of excitations created in the same layer, which are twisted by the $G$-action. For instance, if $q_2, q_2' \in \tilde a_2$, such that $[q_2] = [q_2'] = [\tilde q_2]$, then the excitations created by $X_{q_1\cdot g_{q_2}, \tilde q_2}$ and $X_{q_1\cdot g_{q_2'}, \tilde q_2}$ in CSS$_1^{(\tilde q_2)}$ will be condensed together.

Keeping the terms in $\mathcal{S}_0$ which commute with the condensation, the following combinations remain
\vspace{-0.5em}
\begin{itemize}
    \item For each pair $(a_1, \tilde q_2)$, there is an $X$-stabilizer 
    $\xi(a_1,\tilde q_2) = a_1^{(\tilde q_2)}   \prod_{b_2 \ni \tilde q_2}
    X_{a_1\cdot g_{b_2}, \tilde b_2}$,
    \vspace{-0.5em}
    \item For each pair $(b_1, \tilde q_2)$, there is a $Z$ stabilizer $\zeta(b_1,\tilde q_2) = b_1^{(\tilde{q}_2)}  \prod_{a_2 \ni \tilde q_2} 
    Z_{b_1\cdot g_{a_2}, \tilde a_2}$,
\end{itemize}
\vspace{-0.5em}
Again, $g_{a_2}$ and $g_{b_2}$ are similarly defined by $a_2 = g_{a_2} \cdot \tilde a_2$ and $b_2 = g_{b_2} \cdot \tilde b_2$. These precisely form the stabilizer group of the balanced product CSS$_1 \otimes_G$ CSS$_2$~\cite{supp2}.

\textit{Example}.
Haah's code~\cite{haah2011local} can be understood as a balanced product between two classical fractal codes~\cite{Panteleev_ACM2022,tan2025fracton}. Both codes are defined on the 3D cubic lattice, with qubits on vertices. We take CSS$_1$/ CSS$_2$ to be in $Z$/ $X$-basis respectively. Their stabilizers are given by
\begin{equation*}
    \text{CSS$_1$}:\begin{tikzpicture}[scale = 0.75, baseline = 0.5cm]
    % Define coordinates
    \coordinate (O) at (0,0,0); %back-bottom-left
    \coordinate (A) at (2,0,0);%back-bottom-right
    \coordinate (B) at (2,2,0);%back-top-right
    \coordinate (C) at (0,2,0);%back-top-left
    \coordinate (D) at (0,0,2);
    \coordinate (E) at (2,0,2);
    \coordinate (F) at (2,2,2);
    \coordinate (G) at (0,2,2);
    
    \node  at (0, 0, 0) {$Z$};
    \node  at (2, 0, 0) {$Z$};
    \node at (0, 2, 0) {$Z$};
    \node at (0, 0, 2) {$Z$};

    % Draw edges
    \draw [dotted] (O) -- (A);
    \draw (A) -- (B) -- (C); 
    \draw [dotted] (O) -- (C);% back face
    \draw (D) -- (E) -- (F) -- (G) -- cycle; % front face
    \draw [dotted] (O) -- (D); 
    \draw (A) -- (E);
    \draw (B) -- (F);
    \draw (C) -- (G);% vertical edges
    \end{tikzpicture}
    \hspace{25pt}
    \text{CSS$_2$}:
    \begin{tikzpicture}[scale = 0.75, baseline = 0.5cm]
    % Define coordinates
    \coordinate (O) at (0,0,0); %back-bottom-left
    \coordinate (A) at (2,0,0);%back-bottom-right
    \coordinate (B) at (2,2,0);%back-top-right
    \coordinate (C) at (0,2,0);%back-top-left
    \coordinate (D) at (0,0,2);
    \coordinate (E) at (2,0,2);
    \coordinate (F) at (2,2,2);
    \coordinate (G) at (0,2,2);
    
    \node  at (2, 2, 2) {$X$};
    \node  at (2, 0, 0) {$X$};
    \node at (0, 2, 0) {$X$};
    \node at (0, 0, 2) {$X$};

    % Draw edges
    \draw [dotted] (O) -- (A);
    \draw (A) -- (B) -- (C); 
    \draw [dotted] (O) -- (C);% back face
    \draw (D) -- (E) -- (F) -- (G) -- cycle; % front face
    \draw [dotted] (O) -- (D); 
    \draw (A) -- (E);
    \draw (B) -- (F);
    \draw (C) -- (G);% vertical edges
    \end{tikzpicture}
\end{equation*}
The group $G$ is $\mathbb{Z}^3$ generated by translations $T_x$, $T_y$ and $T_z$, on both codes. We use coordinates $\textbf{n} = (n_1,n_2,n_3)\in \mathbb{Z}^3$ to label both vertices, and the cube $\tilde n_1 \times \tilde n_2 \times \tilde n_3$, where $\tilde n_i$ is the interval $[n_i, n_i+1]$. The quotient $G\backslash$ CSS$_2$ has one orbit for vertices and stabilizers each, and we choose representatives to be the origin $\textbf{0}$ and $-\textbf{1} = (-1,-1,-1)$ respectively.

The stacking is simple, we introduce one copy of CSS$_1$, and an $X$-ancilla for each vertex. The stacked system has two qubits per vertex. We label the qubit of CSS$_1$ to be $1$, and the ancilla to be $2$. The stabilizer group is $\mathcal{S}_0 = \langle b_{\textbf{n}}^{(1)}, Z_{\textbf{n}}^{(2)} \rangle $, where the superscript indicates which qubit the operator acts on. Next we perform code switching by enforcing
\begin{align*}
     \alpha(\textbf{n},-\textbf{1}) = X_{\textbf{n}}^{(1)} X_{\textbf{n}\cdot T_x^{-1}T_y^{-1}}^{(1)}
     X_{\textbf{n}\cdot T_x^{-1}T_z^{-1}}^{(1)}
     X_{\textbf{n}\cdot T_y^{-1}T_z^{-1}}^{(1)}
     \prod_{b_{\textbf{m}}\ni \textbf{n}} X_{\textbf{m}}^{(2)}
\end{align*}
The remaining terms in $\mathcal{S}_0$ are
\begin{align*}
    \zeta(\textbf{n},\textbf{0})
    =
    b_\textbf{n}^{(1)}
    Z_{\textbf{n}}^{(2)} Z_{\textbf{n}\cdot T_xT_y}^{(2)}
    Z_{\textbf{n}\cdot T_xT_z}^{(2)}
    Z_{\textbf{n}\cdot T_yT_z}^{(2)}
\end{align*}
Pictorially, these are
\begin{equation*}
    \begin{tikzpicture}[scale = 0.75, baseline = 0.5cm]
    % Define coordinates
    \coordinate (O) at (0,0,0); %back-bottom-left
    \coordinate (A) at (2,0,0);%back-bottom-right
    \coordinate (B) at (2,2,0);%back-top-right
    \coordinate (C) at (0,2,0);%back-top-left
    \coordinate (D) at (0,0,2);
    \coordinate (E) at (2,0,2);
    \coordinate (F) at (2,2,2);
    \coordinate (G) at (0,2,2);

    \node at (1,2.8,0) {\underline{\,$\alpha(\textbf{n},-\textbf{1})$}\,};
    
    \node[circle, fill = red, inner sep = 1.5pt] at (2,2,2) {};
    
    \node  at (2, 2, 2) {$XX$};
    \node  at (2, 2, 0) {$IX$};
    \node at (0, 2, 2) {$IX$};
    \node at (2, 0, 2) {$IX$};
    \node at (0, 0, 2) {$XI$};
    \node at (0, 2, 0) {$XI$};
    \node at (2, 0, 0) {$XI$};
    \node at (0, 0, 0) {$II$};

    % Draw edges
    \draw [dotted] (O) -- (A);
    \draw (A) -- (B) -- (C); 
    \draw [dotted] (O) -- (C);% back face
    \draw (D) -- (E) -- (F) -- (G) -- cycle; % front face
    \draw [dotted] (O) -- (D); 
    \draw (A) -- (E);
    \draw (B) -- (F);
    \draw (C) -- (G);% vertical edges
    \end{tikzpicture}
    \hspace{15pt}
    \begin{tikzpicture}[scale = 0.75, baseline = 0.5cm]
    % Define coordinates
    \coordinate (O) at (0,0,0); %back-bottom-left
    \coordinate (A) at (2,0,0);%back-bottom-right
    \coordinate (B) at (2,2,0);%back-top-right
    \coordinate (C) at (0,2,0);%back-top-left
    \coordinate (D) at (0,0,2);
    \coordinate (E) at (2,0,2);
    \coordinate (F) at (2,2,2);
    \coordinate (G) at (0,2,2);

    \node at (1,2.8,0) {\underline{\,$\zeta(\textbf{n},\textbf{0})$}\,};
    
    \node[circle, fill = red, inner sep = 1.5pt] at (0,0,0) {};
    
    \node  at (2, 2, 2) {$II$};
    \node  at (2, 2, 0) {$IZ$};
    \node at (0, 2, 2) {$IZ$};
    \node at (2, 0, 2) {$IZ$};
    \node at (0, 0, 2) {$ZI$};
    \node at (0, 2, 0) {$ZI$};
    \node at (2, 0, 0) {$ZI$};
    \node at (0, 0, 0) {$ZZ$};

    % Draw edges
    \draw [dotted] (O) -- (A);
    \draw (A) -- (B) -- (C); 
    \draw [dotted] (O) -- (C);% back face
    \draw (D) -- (E) -- (F) -- (G) -- cycle; % front face
    \draw [dotted] (O) -- (D); 
    \draw (A) -- (E);
    \draw (B) -- (F);
    \draw (C) -- (G);% vertical edges
    \end{tikzpicture}
\end{equation*}
with the vertex $\textbf{n}$ colored in red. These are the stabilizers of Haah's code.

\medskip
%\noindent
\textit{Discussions and Outlook}. We have provided a unifying physical perspective for the tensor and balanced product constructions of qLDPC codes, which apply to both classical and CSS quantum codes. Namely, we showed that the tensor product of two CSS codes can be constructed by gauging a combination of logicals in the stack of one code following the patterns of stabilizers of the other code. In the same flavor, balanced product codes are constructed by gauging a combination of logicals in the stack of one code (appropriately twisted in each layer by the group action) according to the stabilizers of the other code. 

Our focus here was to develop a fundamental understanding of the product construction.  Indeed, we have demonstrated its versatility to reproduce a large class of CSS codes. However, tensor and balanced product codes cannot produce codes that are not CSS, or certain fracton codes such as the X-cube~\cite{vijay2016fracton} or XYZ product code~\cite{2022Quant...6..766L}. To that end, we will explore generalizations of coupled-layer construction beyond product codes in a future work, which can potentially give new families of good qLDPC codes, as well as new topological phases.

We have also pointed out a deep connection between product code and concatenated codes, where the product code gauges the logical instead of enforces the logical of the outer code as large-weight stabilizers. Nevertheless, large-weight stabilizers in itself is not necessarily an obstruction to fault-tolerance~\cite{Concatenatesavequbits}. To that end, it would be fruitful to use insights from one to construct variants of the other, such as using generalized concatenated codes~\cite{grassl2009generalized} to construct new types of product codes, such as the balanced and subsystem concatenated codse~\cite{supp}.

On the condensed matter side, coupling chiral or non-Abelian phases has been performed to construct chiral and non-Abelian fracton phases~\cite{HybridFractonone,HybridFractontwo,designer,Williamson21,SullivanPlanarpstring,Fuji19,Fuji23}.
It would thus be interesting to couple chiral or non-Abelian phases together using the checks of qLDPC codes to construct new chiral and non-Abelian qLDPC phases that go beyond stabilizer codes \cite{christos2026non,mcdonough2026calderbank}.

\smallskip
\textit{Acknowledgments}---
We thank Jin Ming Koh, Mincheol Park, and Zijian Song for useful discussions. This work was supported by the
U.S. National Science Foundation (NSF) under Award No. PHY
2310614.

\onecolumngrid
\clearpage

\onecolumngrid

%=======Supp==========
\setcounter{section}{0}
\setcounter{equation}{0}
\setcounter{figure}{0}
\setcounter{table}{0}
\setcounter{page}{1}

\setcounter{secnumdepth}{3}

\begin{center}
    \textbf{\large Supplementary Material\\for\\
Coupled-Layer Construction of Quantum Product Codes} \\
\end{center}

% \onecolumngrid
\setcounter{tocdepth}{2} % Re-enables ToC entries for the Supplemental sections

\tableofcontents

\section{Review of CSS codes and the tensor product}

In this section we review some aspects of CSS codes, gauging, and the tensor product, as well as establish our convention at the level of chain complexes.

\subsection{Quantum codes as a chain complex}

\begin{defn}
   A CSS quantum code is a length-3 cochain complex $C^\bullet$
\begin{align*}
   \underbrace{C^{-1}}_\text{X-checks} \xrightarrow{\delta^{-1}} \underbrace{C^0}_\text{qubits} \xrightarrow{\delta^0} \underbrace{C^1}_\text {Z-checks}% \xrightarrow{\delta^3} \underbrace{C^4}_\text {Z-metachecks}
\end{align*} 
with qubits placed on degree $0$, Z-checks placed on degree $1$ and X-checks placed on degree $-1$.
\end{defn}

\begin{defn}
The parameters of a CSS quantum code is defined as follows
\begin{alignat*}{2}
    &\text{number of qubits:}
    \hspace{5pt} 
    &&n = \dim C^0  \\
    &\text{number of X-checks:} \hspace{5pt} 
    &&n^X = \dim C^{-1} \\
    &\text{number of Z-checks: }
    \hspace{5pt}
    &&n^Z = \dim C^1\\
    &\text{number of logicals:} \hspace{5pt}
    &&k  = \dim \Ker \delta^{0} -  \dim \Imaa \delta^{-1}= \dim H^{0} \\
    &\text{number of X-metachecks:}
    \hspace{5pt}
    &&k^X = \dim \Ker \delta^{-1} = \dim H^{-1}\\
    &\text{number of Z-metachecks: }
    \hspace{5pt}
    &&k^Z = dim \Ker (\delta^{0})^\top= \dim H^{1}
\end{alignat*}
\end{defn}

\begin{defn}
The Hadamarded code is obtained by Hadamarding all the stabilizers. It is related to the dual chain complex by reversing all the arrows and negating the indices. We will denote such a Hadamarded code $\check C^\bullet= C^{-\bullet}$ with $\check \delta^{m} = (\delta^{-m-1})^\top$
\end{defn}
It is clear that
\begin{align}
   \check n^X &=n^Z,& \check n&= n,&  \check n^Z &= n^X\\
   \check k^X &=k^Z,& \check k&= k,&  \check k^Z &= k^X
\end{align}
\begin{defn}
A $Z$-type classical code is a CSS quantum code for which $n^X=0$. Such a code can be labeled using the map $\delta^0$. Likewise, an $X$-type classical code is a CSS quantum code for which $n^Z=0$. 
\end{defn}

\begin{remark}
The Hadamard of an $X$-type classical code is a $Z$-type classical code labeled by $(\delta^{-1})^\top$.
\end{remark}

\subsection{Gauging}
Given a CSS quantum code, we define the gauging of the $X$-logicals ($X$-symmetries) as follows~\cite{Kubica18}. First, we choose a $\delta^1$ and $C^2$ such that $\delta^1 \circ \delta^0 =0$ to extend the chain complex. Physically, this corresponds to specifying a subset of $Z$-metachecks. Now the chain complex looks like
\begin{align*}
   \underbrace{C^{-1}}_\text{X-checks} \xrightarrow{\delta^{-1}} \underbrace{C^0}_\text{qubits} \xrightarrow{\delta^{0}} \underbrace{C^1}_\text {Z-checks} \xrightarrow{\delta^{1}} \underbrace{C^2}_\text {subset of Z-metachecks}
\end{align*}

The procedure of gauging now corresponds to shifting up the chain complex by one.
\begin{defn}
    The (X-symmetry) gauged code of $C^\bullet$ is the code $C[1]^\bullet = C^{\bullet+1}$
\begin{align*}
   \underbrace{C^{-1}}_\text{subset of X-metachecks} \xrightarrow{\delta^{-1}} \underbrace{C^{0}}_\text{X-checks} \xrightarrow{\delta^{0}} \underbrace{C^{1}}_\text {qubit} \xrightarrow{\delta^{1}} \underbrace{C^{2}}_\text {Z-checks}
\end{align*}
\end{defn}
The parameters of the gauged code are
\begin{align*}
   n[1]^X &= n & n[1] &= n^Z & n[1]^Z &= \dim C^2\\
   k[1]^X &=\Ker\delta^0 =n-(n^z-k^z) & k[1] &= \dim H^1 = k^Z + \dim \Ker \delta^1- n^Z  &  k[1]^Z &=\dim \Ker  (\delta^{1})^\top 
\end{align*}
which depends on the choice of $C^2$. For practical purposes, to get good distance, we will try to include low-weight $Z$-metachecks of the original code, while leaving the non-constant weight metachecks to ensure that the resulting gauged code is also qLDPC.

\begin{defn}
    Similarly, the ($X$-symmetry) ungauged code of $C^\bullet$ is the code $C[-1]^\bullet = C^{\bullet-1}$
    \begin{align*}
   %\underbrace{C^0}_\text{X-metachecks} \xrightarrow{\delta^0} 
        \underbrace{C^{-1}}_\text{subset of X-metachecks} \xrightarrow{\delta^{-1}} \underbrace{C^{0}}_\text{X-checks} \xrightarrow{\delta^{0}} \underbrace{C^{1}}_\text {qubit} \xrightarrow{\delta^{1}} \underbrace{C^{2}}_\text {Z-checks}
    \end{align*}
\end{defn}

\begin{remark}
Ungauging the $X$-symmetry is also sometimes called gauging the $Z$-symmetry. Throughout the text, when we say gauging/ungauging, we will always assume it is of the $X$-symmetry.
\end{remark}
\begin{remark}
Ungauging can be achieved by performing Hadamard, gauging and performing Hadamard again. That is,
\begin{align*}
    \widecheck{C[-1]}^\bullet = \check{C} [1]^\bullet
\end{align*}
\end{remark}

\begin{remark} 
Let $C^\bullet$ be a $Z$-type classical code. Then one often talks about the \emph{transpose code} of $C$, which is a $Z$-type classical code labeled $[n^\top,k^\top,d^\top]$ corresponding to the parity check matrix $(\delta^0)^\top$. In this case, one has $n^\top = n^Z$ and $k^\top = k^Z$. The transpose code corresponds to ungauging followed by Hadamard ($\widecheck{C[-1]}^\bullet$), or equivalently, Hadamard, followed by gauging ($\check{C} [1]$).
\end{remark}

\begin{example}
\label{ex:1Drep}
The classical $Z$-type repetition code (1D Ising model) on an open chain on $L$ bits. 
\begin{align*}
    n^X&=0 & n &= L  & n^Z &= L-1\\
    k^X &= 0 &  k &= 1 &  k^Z &=   0
\end{align*}
The gauged code is an $X$-type classical code
\begin{align*}
        n[1]^X&=L & n[1] &=L-1  & n[1]^Z &= 0\\
    k[1]^X &= 1 &  k[1]&= 0&  k[1]^Z &= 0
\end{align*}
\end{example}

\begin{example} 
% square lattice

$Z$-type Ising model on an $L\times L$ torus, with spins placed on vertices. In this case, the complex coincides with the cell-complex of the torus, so $\delta^{-1}=0$ and $C^{-1}$ is empty. We have 
\begin{align*}
    \dim \Ker \delta^0 & =1 &  \dim \Imaa \delta^0 & =L^2-1
\end{align*}
Therefore
\begin{align*}
    n^X&=0 & n &= L^2  & n^Z &= 2L^2\\
    k^X &= 0 &  k &= 1 &  k^Z &=   L^2+1
\end{align*}
To perform the gauging, we choose $C^2$ to be all plaquettes of the square lattice so $\dim C^2 = L^2$ and
\begin{align*}
\dim \Ker \delta^1 &=L^2+1 & \dim \Imaa \delta^1 &=L^2-1
\end{align*}
The parameters of the gauged code are
\begin{align*}
    n[1]^X&=L^2 & n[1]&= 2L^2  & n[1]^Z &= L^2\\
    k[1]^X &= 1 &  k[1] &= (L^2+1)+(L^2+1)  - 2L^2 =2 &  k[1]^Z &=  L^2 - (L^2-1)=1
\end{align*}
which we identify as the toric code (TC)
\end{example}

%============================================
%============================================

\subsection{Tensor product of chain complexes}

\begin{defn}
   Given two CSS quantum codes $C^\bullet = [[n_1,k_1,d_1]]$ and $D^\bullet = [[n_2,k_2,d_2]]$, the tensor product code corresponds to the tensor product complex $(C\otimes D)^\bullet$
    \begin{align*}
        (C\otimes D)^{-2} \xrightarrow{\delta^{-2}} \underbrace{(C\otimes D)^{-1}}_\text{X-checks} \xrightarrow{\delta^{-1}} \underbrace{(C\otimes D)^{0}}_\text{qubits} \xrightarrow{\delta^{0}} \underbrace{(C\otimes D)^{1}}_\text {Z-checks} \xrightarrow{\delta^{1}} (C\otimes D)^{2}
    \end{align*}
    The qubits in the product code are $(C\otimes D)^{0} = (C^{-1} \otimes D^{1}) \oplus (C^{0} \otimes D^{0}) \oplus (C^{1} \otimes D^{-1})$, and logicals are given by the K\"unneth formula. The parameters of the product code are
    \begin{align*}
        n = n_1n_2 + n_1^Xn_2^Z + n_1^Zn_2^X\\
        k = k_1k_2 + k_1^Xk_2^Z + k_1^Zk_2^X
    \end{align*} 
\end{defn}

\begin{remark} 
Shift is additive under the tensor product,
\begin{align*}
    (C[m_1] \otimes D[m_2])^\bullet = (C \otimes D)[m_1 + m_2]^\bullet
\end{align*}
\end{remark}

\begin{remark} 
Let $C^\bullet$ and $D^\bullet$ be $Z$-type classical codes. The homological product of classical codes\cite{Bravyi14} is a chain complex on degrees $0,1,2$. Therefore, one obtains a quantum code if the qubits are placed on degree one. In contrast, our definition puts qubits on degree zero. Thus, the tensor product of two classical codes is also $Z$-type classical code, and corresponds to the ungauged code of the homological product of classical codes $  (C\otimes D)[1] = C[1]\otimes D = C \otimes D[1]$~\cite{RakovszkyLDPC2}. Thus, to reproduce the usual homological product of classical codes, one should instead perform the tensor product after gauging one of the factors, or gauge the resulting classical code.
\end{remark}

\begin{remark} 
Instead, to reproduce the hypergraph product~\cite{Tillich:2013esj} of $C$ and $D$, one should first Hadamard one of the factors before performing the tensor product  $\check{C}\otimes D$ or $C \otimes \check{D}$ (these two choices differ by Hadamard). 
\end{remark}

\begin{example}
 Let $C$ be the 1D repetition code on an open chain from Example~\ref{ex:1Drep}. Let us consider the various products of $C$ with itself. 
 \begin{itemize}
     \item The tensor product (as defined in our convention) $C \otimes C$ is the 2D Ising model on an open square.
     \item The homological product $(C\otimes C)[1]$ is the surface code on a square with smooth boundaries on all sides, and thus has zero logicals.
     \item The hypergraph product $\check{C}\otimes C$ is the surface code on a square with rough boundary on two sides and smooth boundary on two sides, thus hosting one logical.
 \end{itemize}  
\end{example}

\begin{remark}
    One can take also perform gauging by tensoring  1-term chain complex $0 \ra 0\ra \mathbb{Z}_2$ where the $\mathbb{Z}_2$ lives in degree 1. Similarly, one can ungauge by tensoring with $\mathbb{Z}_2\ra 0 \ra 0$, where the $\mathbb{Z}_2$ is at degree $-1$.
\end{remark}

% =========================================================

%

\section{4D loop-only TC}
In this section, we present two ways of obtaining the 4D loop-only TC. First we show the coupled layer construction obtained from the tensor product of two 2D toric codes. Then, we show the coupled layer construction corresponding to the tensor product of the 3D TC and the $Z$-type  repetition code. In doing so, we will uncover the subtleties on metachecks in the latter construction, which will be studied more systematically in Section~\ref{sec:metacheck}.

\subsection{2D TC $\otimes$ 2D TC}

The chain complex representing the 4D loop-only TC (qubits, $X$-checks and $Z$-checks on 2-cells, 1-cells and 3-cells respectively) is exactly the tensor product of two 2D TC. This is easy to see by noting the chain complex of 2D TC on an infinite lattice is exactly the cellular cochain complex on $\mathbb{R}^2$. The tensor product of two such complexes gives the cellular cochain complex on $\mathbb{R}^4$. The three terms we take to define the product CSS code is centered on 2-cells, giving the loop-only TC in 4D.

\begin{figure}[h!]
    \centering
    % \hspace*{-1cm}
    \subfloat{\resizebox{0.4\textwidth}{!}{
    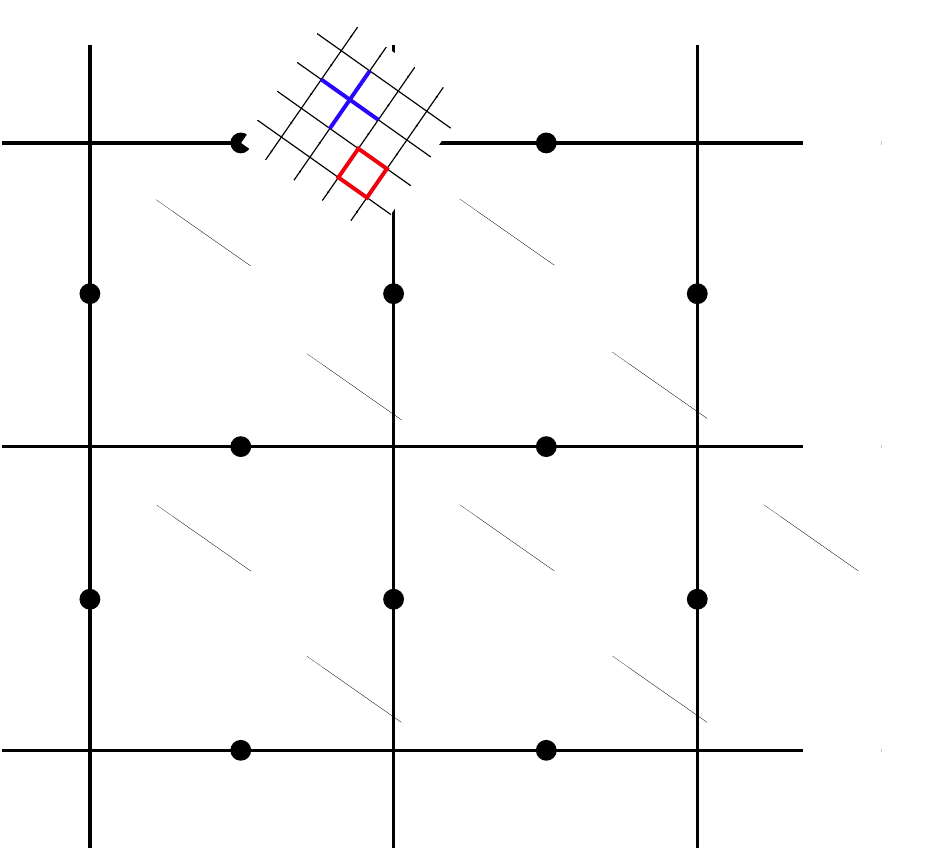}}
    \qquad
    \subfloat{\resizebox{0.34\textwidth}{!}{
    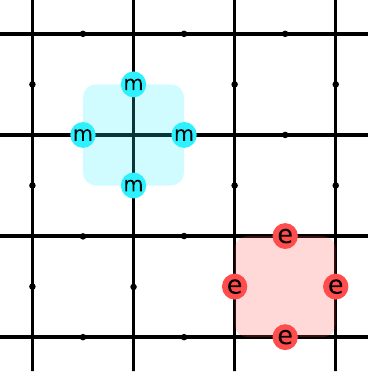}}
    
    \caption{Left: The stacking configuration of 2D TC $\otimes$ 2D TC. The ancillas are not displayed for clarity.
    Right: condensing anyons on a 2D layout. The blue box indicates $\mm$ anyons are condensed around the vertex, and red box shows $\ee$ anyons are condensed around the plaquette. After condensing, both $\mm$ and $\ee$ form loops, giving the 4D loop-only TC. }
    \label{fig:4DTCcoupledlayerapp}
\end{figure}

From our coupled layer construction, it follows that the 4D loop-only TC can be realized by stacking and condensing 2D TC using checks of another 2D TC, which we now illustrate. Both CSS$_1$ and CSS$_2$ are TC, which we denote by TC$_1$ and TC$_2$, and write their complexes as $V_i\rightarrow E_i \rightarrow P_i$ to emphasize that checks and qubits are labeled by vertices, edges and plaquettes. For each qubit in TC$_2$, introduce a copy of TC$_1$. For each pair $p_1\otimes v_2$, introduce a $Z$-ancilla; and for each pair $v_1\otimes p_2$, introduce an $X$-ancilla. The configuration is given by the left of Figure~\ref{fig:4DTCcoupledlayerapp}.

The code switching terms come from $E_1\otimes V_2$ and $E_1 \otimes P_2$ respectively. Given $e_1\otimes v_2$, we have $\alpha(e_1,v_2) = \prod_{e_2\ni v_2}X_{e_1,e_2} \prod_{p_1\ni e_1}X_{p_1,v_2}$. Each $X_{e_1,e_2}$ excites a pair of $\mm$ anyons in layer TC$_1^{(e_2)}$, while each $X_{p_1,v_2}$ excites a $\mathbb{Z}_2$ charges on site $(p_1,v_2)$. We interpret this as condensing a pair of $\mm$ anyons loops in layers of TC around the vertex $v_2$. Similarly, given $e_1\otimes p_2$, we have $\beta(e_1,p_2) = \prod_{e_2\in p_2} Z_{e_1,e_2} \prod_{v_1 \in e_1}Z_{v_1,p_2}$, which condenses $\ee$ anyons in layers of TC around the plaquette $p_2$. The right of Figure~\ref{fig:4DTCcoupledlayerapp} shows this scheme.

It is interesting to observe that the condensation is a coherent moving of loops of particles, which shares a similarity to p-string condensation~\cite{pstring,pstringsagar,pstring1form1,pstring1form2}, that gives rise to fracton order. However, This loop is only allowed to condense in the spatial directions of TC$_1$. Thus the corresponding symmetry that is gauged is not a topological 1-form symmetry in 4D. Rather, it is a (codimension 2) foliated 1-form symmetry.

In comparison to the construction of 3D TC by condensing pairs of anyons in adjacent layers of a stack of 2D TCs, stacking on a 2D layout allows both $\mm$ and $\ee$ anyons to form strings, which commute with each other. Notably, the 2D TC does not have local metachecks since the stabilizers of 2D TC only has global redundancies. This makes the construction straightforward. We will now study a different construction using a 3D TC, and we will see that because of local redundancies of the 3D TC checks, the coupled layer construction becomes more involved.

%============================================
%============================================

\subsection{3D TC $\otimes$ Ising}
\label{sec:3Dx1D}
Take CSS$_1$ to be $3$D Toric code on the cubic lattice with $v$, $e$, $p$ and $c$ to denote the vertices, edges, plaquettes and cubes in $3$D. $X$-stabilizers, qubits, and $Z$-stabilizers are defined on the vertices, edges and plaquettes, respectively. Take CSS$_2$ to be $X$-type $1$D Ising model with $XX$ checks. For convenience, we assume the Ising qubits are on edges, checks are on vertices, and denote vertices by $w\in \mathbb{Z}$, edges by $\Tilde{w} = [w,w+1]$. The resulting product code we will construct is the self-correcting 4D toric code with only loop-like excitations.

The initial stacks is described on the $4$D lattice. The qubits in TC$(\Tilde{w})$ is placed on $e(\Tilde{w}) \equiv e\times [w,w+1]$, which is a $2$-cell in $4$D. The ancillas are placed on $p(w) \equiv p\times \{w\}$, which is also a $2$-cell in $4$D. This covers all the $2$-cells in $4$D. The initial stabilizer of the stacks of 3D toric codes along with ancillas is therefore
\begin{align*}
    \mathcal{S}_0 = \langle a_v^{(\Tilde{w})}, b_p^{(\Tilde{w})}, Z_{p(w)} \rangle
\end{align*}

The code switching in this case is by enforcing $\alpha(e,w) = \prod_{p\ni e}X_{p(w)} \prod_{\Tilde{w}\ni w} X_{e(\Tilde{w})}$. Each $X_{e(\Tilde{w})}$ excites a flux loop around the edge $e$ in layer $\Tilde{w}$, and the product $\prod_{p \ni e}X_{p(w)}$ creates a loop charged under the 1-form $\mathbb{Z}_2$ in layer $w$.  By enforcing this term, flux loops in pairs of nearest layers are condensed. From the condensation perspective, when flux loops are condensed, the electric charges must form loops in order to commute. Hence we expect the $4$D loop-only Toric code to emerge. However there are subtleties to the tensor product chain complex.

Keeping the commuting terms in $\mathcal{S}_0$, we find the stabilizer group in the deformed code is
\begin{align*}
    &\mathcal{S} = \langle
    a_v^{(\Tilde{w})}, \alpha(e,w), \zeta(p,\Tilde{w}),  \nu(c,w)
    \rangle,\\
    &\zeta(p,\Tilde{w}) = b_p^{(\Tilde{w})}\prod_{w\in \Tilde{w}} Z_{p(w)}, 
    \hspace{15pt}
    \nu(c,w) = \prod_{p\in c} Z_{p(w)}.
\end{align*}
The first three terms together form the tensor product code TC $\otimes$ Rep. The interesting term is $\nu(c,w)$, which does not arise from the tensor product, but is nevertheless generated from adding $\alpha(e,w)$ to the stabilizer group.

Let us first demonstrate explicitly that  $\nu(c,w)$ is indeed generated. Each $\nu(c,w)$ obviously belongs to $\mathcal{S}_0$, and it commutes with $a_v^{(\Tilde{w})}$ since their supports never overlap. To check that $\nu(c,w)$ commutes with $\alpha(e,w)$, note that they overlap only on the qubits living on $p(w)$ where $e\in p\in c$. But when fixing $e\in c$, there are exactly two plaquettes satisfying this condition. This shows that they always commute. Finally, let us show that $\mathcal{S}$ is the $4$D TC, the terms $a_v^{(\Tilde{w})}$ and $\alpha(e,w)$ are $X$ stabilizers centered on edges $v(\Tilde{w})$ and $e(w)$ respectively. The terms $\zeta(p,\Tilde{w})$ and $\nu(c,w)$ are $Z$ stabilizers centered on cubes $p(\Tilde{w})$ and $c(w)$ respectively.

Where does the term $\nu(c,w)$ come from? The combination of $Z_{p(w)}$ that forms $\nu(c,w)$ is exactly due to the metachecks, or redundancies, between $Z$ stabilizers in $3$D TC (CSS$_1$), $\prod_{p\in c} b_p = 1$ for each cube $c$. In terms of gauging, this relation generates the dual $1$-form symmetry after gauging the $1$-form symmetry generated by the $X$ vertex terms. In the naive tensor product code, where we only keep the $X$ and $Z$ checks in the chain complex, these additional terms will not be included, and therefore will be treated as low-weight logicals. Thus, a code without the $\nu(c,w)$ terms as $Z$-stabilizers will have a bad distance of $6$. However, we see that they are automatically included in the final stabilizer group, which results in the $4$D loop-only TC, whose code distance scales as $L^2$.

In the general case, let $R_X$ and $M_Z$ be metachecks of CSS$_1$ and $r_x\in R_x$ and $m_Z\in M_Z$, then the following terms also belong to $\mathcal{S}$:
\begin{align*}
    \mu(r_X, b_2) = \prod_{a_1\in r_X} X_{a_1,b_2},
    \hspace{15pt}
    \nu(m_Z,a_2) = \prod_{b_1\in m_Z} Z_{b_1,a_2}.
\end{align*}
To see these are the only extra terms in $\mathcal{S}$, note any other operators in $\mathcal{S}$ has to be a logical of the tensor product code, which is $\bigoplus_{i+j=0}H^i(CSS_1)\otimes H^j(CSS_2)$. The operators $\mu(r_X,b_2)$ and $\nu(m_Z,a_2)$ belong to the classes $H^{-1}($CSS$_1)\otimes H^1($CSS$_2)$ and $H^1($CSS$_1)\otimes H^{-1}($CSS$_2)$ respectively, hence any additional one has to be a logical in the class $H^0($CSS$_1)\otimes H^0($CSS$_2)$. However, no $X$ or $Z$ logicals in $H^0($CSS$_1)$ can be written as a product of stabilizers $a_1$ or $b_1$. We conclude
\begin{align*}
     \mathcal{S} =\Big\langle \alpha(q_1,a_2), \beta(q_1,b_2),
     \xi(a_1,q_2), \zeta(b_1, q_2),
     \mu(r_X,b_2), \nu(r_Z,a_2) \Big\rangle.
\end{align*}
See Section~\ref{sec:metacheck} below on metachecks below for systematic discussions.

\section{Measurement-based resource state}

In this section, we discuss how to obtain a resource state by taking a product of CSS code and a 1D cluster state. A paradigmatic example is the 3D RBH state \cite{raussendorf2005long}. 

First, we recall the definition of the resource state for performing measurement based quantum computation (MBQC) of a given stabilizer code, also called the foliated stabilizer code~\cite{Bolt16}. Given a CSS code $A\rightarrow Q \rightarrow B$, we build a graph $\Gamma = (V,E)$ in the following way. For each $q \in Q$, introduce a 1D chain, whose vertices are labeled by integers. For each $a \in A$, introduce a vertex at every even step, label them by $a_{2i}$, as well as including edges $\{(a_{2i}, q_{2i})\,|\, q\in a,\, i\in \mathbb{Z}\}$. For each $b \in B$, introduce a vertex at every odd step, label them by $b_{2i+1}$, as well as including edges $\{(b_{2i+1}, q_{2i+1})\,|\, q\in b,\, i\in \mathbb{Z}\}$. The foliated stabilizer code is the graph state of $\Gamma$. That is, there is one qubit per vertex initialized in the $X = 1$ state, and apply $CZ$ on all edges. The
stabilizer group is
\begin{align*}
    \mathcal{S} = \left\langle  X_v\prod_{v'\in N(v)} Z_{v' }\,\Bigg|\, v \in V\right\rangle ,
\end{align*}
where $N(v) = \{v'\in V \,|\, (v',v) \in E\}$ is the set of vertices connected to $v$.

Note the graph $\Gamma$ is always bipartite, hence the above stabilizer code can be turned into a CSS code by applying the Hadamard on half of the qubits. We now show this CSS Hamiltonian is in fact a tensor product of the CSS code and the 1D cluster state $H_{\text{CS}} = -\sum_{n} Z_{n-1}X_n Z_{n+1}$. We remark that a related idea was discussed Ref.~\cite{Hillmann2024} which views fault-tolerant complex as a tensor product of a CSS code and a repetition code.

First, we turn the 1D cluster state into a CSS code by applying Hadamard on all even sites. The stabilizer group of the 1D CSS cluster state is
\begin{align*}
   \text{CSS} = \left \langle X_{2i} X_{2i-1}X_{2i+2}, Z_{2i-1} Z_{2i} Z_{2i+1} | i\in \mathbb{Z}\right \rangle
\end{align*}
Denoting the $X$-stabilizers by $A_{\text{odd}}$, $Z$-stabilizers by $B_{\text{even}}$ and qubits by $Q_{\mathbb{Z}}$, we can construct the tensor product CSS $\otimes$ CS. The Hilbert space is the union of $Q\otimes Q_\mathbb{Z}$, $A \otimes B_{\text{even}}$ and $B\otimes A_{\text{odd}}$, which is exactly the Hilbert space of $H_{\text{MBQC}}$. The $X$-stabilizers are
\begin{itemize}
    \item $Q \otimes A_{\text{odd}}$. Given $q\otimes a_{2i+1}$, this gives $\alpha(q,2i+1) = X_{q,2i}X_{q,2i+1} X_{q,2i+2} \prod_{b\ni q} X_{b,2i+1}$.
    \item $A \otimes Q_\mathbb{Z}$. For $a\otimes q_{2i}$, this gives $\alpha(a,2i) = X_{a,2i} \prod_{q\in a} X_{q,2i}$. For $a\otimes q_{2i+1}$, this gives $\alpha(a,2i+1) = X_{a,2i} X_{a,2i+2} \prod_{q \in a} X_{q,2i+1}$.
\end{itemize}
Similarly for $Z$-stabilizers
\begin{itemize}
    \item $Q \otimes B_{\text{even}}$. Given $q\otimes b_{2i}$, this gives $\beta(q,2i) = Z_{q,2i-1}Z_{q,2i} Z_{q,2i+1} \prod_{a \ni q} Z_{a,2i}$. 
    \item $B \otimes Q_\mathbb{Z}$. For $b\otimes q_{2i}$, this gives $\beta(b,2i) = Z_{b,2i-1}Z_{b,2i+1} \prod_{q\in b} Z_{q,2i}$. For $b\otimes q_{2i+1}$, this gives $\beta(b,2i+1) = Z_{b,2i+1} \prod_{q \in b} Z_{q,2i+1}$.
\end{itemize}
However, note that
\begin{align*}
    &\alpha(a,2i+1) = \alpha(a,2i) \alpha(a,2i+2) \prod_{q\in a} \alpha(q,2i+1),\\
    &\beta(b,2i) = \beta(b,2i-1)\beta(b,2i+1)\prod_{q\in b}\beta(q,2i).
\end{align*}
Hence, they are not independent stabilizers. In fact, these form the error detection cells of the resource state. Now, apply Hadamard on the sublattice containing all even sites, that is, qubits $(q, 2i) \in Q \otimes Q_{\mathbb{Z}}$ and $(b, 2i+1) \in B\otimes A_{\text{odd}}$, we have
\begin{alignat*}{2}
    &\alpha(q,2i+1) 
    \rightarrow
    X_{q,2i+1} \prod_{v \in N(q,2i+1)}Z_v, 
    \hspace{35pt}
    &&\alpha(a,2i) 
    \rightarrow 
    X_{a,2i} \prod_{v \in N(a,2i)}Z_v.\\
    &\beta(q,2i) 
    \rightarrow
    X_{q,2i} \prod_{v \in N(q,2i)}Z_v,
    \hspace{35pt}
    &&\beta(b,2i+1) 
    \rightarrow
    X_{b,2i+1} \prod_{v \in N(b,2i+1)}Z_v.
\end{alignat*}
These are precisely the stabilizers of the graph state for $\Gamma$.

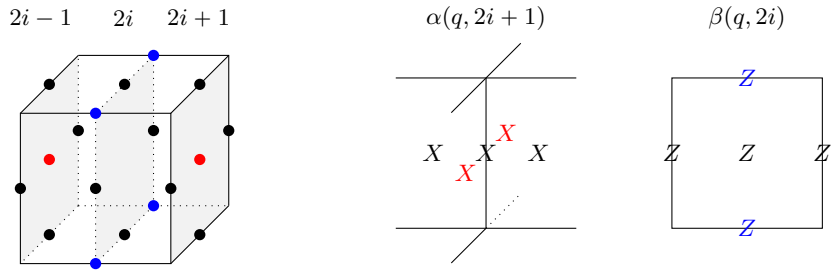
\begin{figure}[h!]
    \begin{tikzpicture}[scale = 1, baseline = 0cm]
    % Define coordinates
    \coordinate (O) at (0,0,0); %back-bottom-left
    \coordinate (A) at (2,0,0);%back-bottom-right
    \coordinate (B) at (2,2,0);%back-top-right
    \coordinate (C) at (0,2,0);%back-top-left
    \coordinate (D) at (0,0,2);
    \coordinate (E) at (2,0,2);
    \coordinate (F) at (2,2,2);
    \coordinate (G) at (0,2,2);

    \node at (-0.5,2.5) {$2i-1$};
    \node at (0.6,2.5) {$2i$};
    \node at (1.6,2.5) {$2i+1$};

    % color the faces
    \fill [gray!10] (O) -- (D) -- (G) -- (C) -- cycle;
    \fill [gray!10] (A) -- (B) -- (F) -- (E) -- cycle;
    \fill [gray!10] ($(O)!0.5!(A)$) -- ($(B)!0.5!(C)$) -- ($(F)!0.5!(G)$) -- ($(D)!0.5!(E)$) -- cycle;

    % Draw edges
    \draw [dotted] (O) -- (A);
    \draw (A) -- (B) -- (C); 
    \draw [dotted] (O) -- (C);% back face
    \draw (D) -- (E) -- (F) -- (G) -- cycle; % front face
    \draw [dotted] (O) -- (D); 
    \draw (A) -- (E);
    \draw (B) -- (F);
    \draw (C) -- (G);% vertical edges
    \draw[dotted] ($(O)!0.5!(A)$) -- ($(B)!0.5!(C)$) -- ($(F)!0.5!(G)$) -- ($(D)!0.5!(E)$) -- cycle;

    \foreach \X/\Y in {A/B, C/O, E/F, G/D, O/D, A/E, B/F, C/G} 
    \node[fill=black, circle, inner sep=1.5pt] at ($( \X )!0.5!( \Y )$) {}; %black dot edges

    \foreach \X/\Y in {O/E, O/B, D/F, G/B}
    \node[fill=black, circle, inner sep=1.5pt] at ($( \X )!0.5!( \Y )$) {}; %black dot faces

    \foreach \X/\Y in {C/D, A/F}
    \node[fill=red, circle, inner sep=1.5pt] at ($( \X )!0.5!( \Y )$) {}; %red dot faces

    \foreach \X/\Y in {O/A, B/C, D/E, F/G}
    \node[fill=blue, circle, inner sep=1.5pt] at ($( \X )!0.5!( \Y )$) {}; %blue dot edges 
    \end{tikzpicture}
    \hspace{55pt}
    \begin{tikzpicture}[scale = 1, baseline = 0.3cm]
    % Define coordinates
    \coordinate (O) at (0,0,0); 
    \coordinate (A) at (1.2,0,0);
    \coordinate (B) at (-1.2,0,0);
    \coordinate (C) at (0,0,1.2);
    \coordinate (D) at (0,0,-1.2);
    \coordinate (E) at (0,2,0);
    \coordinate (F) at (1.2,2,0);
    \coordinate (G) at (-1.2,2,0);
    \coordinate (H) at (0,2,1.2);
    \coordinate (I) at (0,2,-1.2);

    \node at (0.7, 1, 0) {$X$};
    \node at (-0.7, 1, 0) {$X$};
    \node [red] at (0, 1, 0.7) {$X$};
    \node [red] at (0, 1, -0.7) {$X$};
    \node at (0, 1, 0) {$X$};

    \node at (0,2.8,0) {$\alpha(q, 2i+1)$};
    
    % Draw edges
    \draw (A) -- (B);
    \draw (O) -- (C);
    \draw [dotted] (O) -- (D);
    \draw (O) -- (E);
    \draw (F) -- (G);
    \draw (H) -- (E) -- (I);

    \end{tikzpicture}    
    \hspace{25pt}
    \begin{tikzpicture}[scale = 1, baseline = 0.3cm]
    % Define coordinates
    \coordinate (O) at (0,0,0); 
    \coordinate (A) at (2,0,0);
    \coordinate (B) at (2,2,0);
    \coordinate (C) at (0,2,0);
    
    % Draw edges
    \draw (O)-- (A) -- (B) -- (C) -- cycle;

    \foreach \X/\Y in {O/B, O/C, A/B} 
    \node at ($( \X )!0.5!( \Y )$) {$Z$}; 

     \node at (1,2.8,0) {$\beta(q, 2i)$};

    \foreach \X/\Y in {O/A, C/B}
    \node[blue] at ($( \X )!0.5!( \Y )$) {$Z$}; 
    \end{tikzpicture}
    \begin{minipage}{0.95\textwidth}
        \centering
        \caption{Left: a unit cell of the RBH cluster state. The gray layers are three stacks of 2D TC labeled by $2i-1$, $2i$ and $2i+1$. The black qubits are the TC qubits. The blue qubits are $X$-ancillas, and red qubits are $Z$-ancillas. Right: the first term condenses $\mm$ anyons in three TC centered on an odd layer. The second term condenses three $\ee$ anyons in three TC centered on an even layer. The qubit $q$ is chosen to be on a vertical edge in each term.}
        \label{fig:RBH1}
    \end{minipage}
\end{figure}

We now describe the resource state from the perspective of foliation. The tensor product is symmetric in the two codes, thus we may take either to be CSS$_1$. First take the CSS code to be CSS$_1$, and cluster state to be CSS$_2$. In this case, for each integer $\mathbb{Z}$ we foliate a layer of CSS. For each pair of $a\in A$ and even interger $2i$, we introduce an $X$-ancilla. For each pair of $b\in B$ and an odd integer $2i+1$, we introduce a $Z$-ancilla. The condensation follows the patterns of stabilizers in the cluster state. The $X$-stabilizers indicate excitations created by a single $X$ in CSS are being condensed in three consecutive layers, centered on an odd layer. This is given by the space $Q\otimes A_{\text{odd}}$ with terms $\alpha(q,2i+1) = X_{q,2i}X_{q,2i+1} X_{q,2i+2} \prod_{b\ni q} X_{b,2i+1}$. Similarly, $Z$-stabilizers indicate excitations created by a single $Z$ are being condensed in three consecutive layers, centered on an even layer. This is given by the space $Q\otimes B_{\text{even}}$ with terms $\beta(q,2i) = Z_{q,2i-1}Z_{q,2i} Z_{q,2i+1} \prod_{a \ni q} Z_{a,2i}$. In Figure~\ref{fig:RBH1}, we take CSS to be 2D TC for an illustration, whose resource state is the RBH cluster state. In this example, $\ee$ anyons are condensed on three consecutive layers centered on even layers, while $\mm$ anyons are condensed on three consecutive layers centered on odd layers. This condensation pattern in fact matches the coupled layer construction for bosonic topological insulators presented in \cite{WangSenthil13}.

\begin{figure}[h!]
    \centering
    \begin{minipage}{0.35\textwidth}
        \begin{tikzpicture}[scale = 1.2]
        % Define coordinates
        \coordinate (O) at (0,0,0); %back-bottom-left
        \coordinate (A) at (2,0,0);%back-bottom-right
        \coordinate (B) at (2,2,0);%back-top-right
        \coordinate (C) at (0,2,0);%back-top-left
        \coordinate (D) at (0,0,2);
        \coordinate (E) at (2,0,2);
        \coordinate (F) at (2,2,2);
        \coordinate (G) at (0,2,2);
        
        \node at (-0.5,2.5) {$2i-1$};
        \node at (0.6,2.5) {$2i$};
        \node at (1.6,2.5) {$2i+1$};

        % Draw edges
        \draw [dotted] (O) -- (A);
        \draw (A) -- (B) -- (C); 
        \draw [dotted] (O) -- (C);% back face
        \draw (D) -- (E) -- (F) -- (G) -- cycle; % front face
        \draw [dotted] (O) -- (D); 
        \draw (A) -- (E);
        \draw (B) -- (F);
        \draw (C) -- (G);% vertical edges
        % \draw[dotted] ($(O)!0.5!(A)$) -- ($(B)!0.5!(C)$) -- ($(F)!0.5!(G)$) -- ($(D)!0.5!(E)$) -- cycle;
    
        % Draw stacks of cluster state
        \draw [very thick] (-0.5,2,1) -- (2.5,2,1);
        \draw [very thick] (-0.5,1,0) -- (2.5,1,0);
        \draw [very thick] (-0.5,1,2) -- (2.5,1,2);
        \draw [very thick] (-0.5,0,1) -- (2.5,0,1);

        \foreach \X/\Y in {A/B, C/O, E/F, G/D, O/D, A/E, B/F, C/G} 
        \node[fill=black, circle, inner sep=1.5pt] at ($( \X )!0.5!( \Y )$) {}; %black dot edges
    
        \foreach \X/\Y in {O/E, O/B, D/F, G/B}
        \node[fill=black, circle, inner sep=1.5pt] at ($( \X )!0.5!( \Y )$) {}; %black dot faces
    
        \foreach \X/\Y in {C/D, A/F}
        \node[fill=blue, circle, inner sep=1.5pt] at ($( \X )!0.5!( \Y )$) {}; %red dot faces
    
        \foreach \X/\Y in {O/A, B/C, D/E, F/G}
        \node[fill=red, circle, inner sep=1.5pt] at ($( \X )!0.5!( \Y )$) {}; %blue dot edges 
        \end{tikzpicture}
    \end{minipage}
    \begin{minipage}{0.55\textwidth}
        \hspace{-5pt}
        $\alpha(a, 2i)$:
        \hspace{5pt}
        \begin{tikzpicture}[scale = 0.8, baseline = 0cm]
            % Define coordinates
            \coordinate (O) at (0,0,0);
            \coordinate (A) at (2,0,0);
            \coordinate (B) at (0,-1.5,0);
            \coordinate (C) at (0,1.5,0);
            \coordinate (D) at (2,-1.5,0);
            \coordinate (E) at (2,1.5,0);
            \coordinate (F) at (0,0,1.5);
            \coordinate (G) at (0,0,-1.5);
            \coordinate (H) at (2,0,1.5);
            \coordinate (I) at (2,0,-1.5);
        
            % Draw edges
            \draw (O) -- (A);
            \draw (B) -- (C); 
            \draw (D) -- (E);
            \draw (F) -- (G);
            \draw (H) -- (I);
            
            \foreach \X/\Y in {O/D, O/E, O/H, O/I}
            \node at ($( \X )!0.5!( \Y )$) {$X$}; 
            
            \foreach \X/\Y in {O/A}
            \node[red] at ($( \X )!0.5!( \Y )$) {$X$}; 
        \end{tikzpicture}
        \hspace{10pt}
        $\alpha(a, 2i+1)$:
        \hspace{5pt}
        \begin{tikzpicture}[scale = 0.8, baseline = 0cm]
            % Define coordinates
            \coordinate (O) at (0,0,0);
            \coordinate (A) at (-1.5,0,0);
            \coordinate (B) at (1.5,0,0);
            \coordinate (C) at (0,-1.5,0);
            \coordinate (D) at (0,1.5,0);
            \coordinate (E) at (0,0,1.5);
            \coordinate (F) at (0,0,-1.5);
            
            % Draw edges
            \draw (A) -- (B);
            \draw (C) -- (D); 
            \draw (E) -- (F);
            
            \foreach \X/\Y in {O/C, O/D, O/E, O/F}
            \node at ($( \X )!0.5!( \Y )$) {$X$}; 
            
            \foreach \X/\Y in {O/A, O/B}
            \node[red] at ($( \X )!0.5!( \Y )$) {$X$}; 
        \end{tikzpicture}\\
        \vspace{0.5cm}
        $\beta(b, 2i+1)$:
        \hspace{5pt}
        \begin{tikzpicture}[scale = 0.8, baseline = 0.2cm]
            % Define coordinates
            \coordinate (O) at (0,0,0);
            \coordinate (A) at (0,2,0);
            \coordinate (B) at (0,2,2);
            \coordinate (C) at (0,0,2);
            % Draw edges
            \draw (O) -- (A) --(B) -- (C) -- cycle;
            
            \foreach \X/\Y in {O/A, A/B, B/C, C/O}
            \node at ($( \X )!0.5!( \Y )$) {$Z$}; 
            
            \foreach \X/\Y in {O/B}
            \node[blue] at ($( \X )!0.5!( \Y )$) {$Z$}; 
        \end{tikzpicture}
        \hspace{35pt}
        $\beta(b, 2i)$:        \hspace{5pt}
        \begin{tikzpicture}[scale = 0.8, baseline = 0.2cm]
            % Define coordinates
            \coordinate (O) at (0,0,0); %back-bottom-left
            \coordinate (A) at (2,0,0);%back-bottom-right
            \coordinate (B) at (2,2,0);%back-top-right
            \coordinate (C) at (0,2,0);%back-top-left
            \coordinate (D) at (0,0,2);
            \coordinate (E) at (2,0,2);
            \coordinate (F) at (2,2,2);
            \coordinate (G) at (0,2,2);
        
            % Draw edges
            \draw [dotted] (O) -- (A);
            \draw (A) -- (B) -- (C); 
            \draw [dotted] (O) -- (C);% back face
            \draw (D) -- (E) -- (F) -- (G) -- cycle; % front face
            \draw [dotted] (O) -- (D); 
            \draw (A) -- (E);
            \draw (B) -- (F);
            \draw (C) -- (G);% vertical edges
            % \draw[dotted] ($(O)!0.5!(A)$) -- ($(B)!0.5!(C)$) -- ($(F)!0.5!(G)$) -- ($(D)!0.5!(E)$) -- cycle;
        
            \foreach \X/\Y in {O/E, O/B, D/F, G/B}
            \node[] at ($( \X )!0.5!( \Y )$) {$Z$}; %black dot faces
        
            \foreach \X/\Y in {C/D, A/F}
            \node[blue] at ($( \X )!0.5!( \Y )$) {$Z$}; %red dot faces
        \end{tikzpicture}
    \end{minipage}
    \begin{minipage}{0.95\textwidth}
    \centering
    \caption{Left: a unit cell of the RBH cluster state. The thick lines are copies of the 1D cluster state. The blue and red qubits are 1D $X$-ancillas and $Z$-ancillas respectively. Right: the top two terms condense $\mathbb{Z}_2^{\text{even}}$ charges and $\mathbb{Z}_2^{\text{odd}}$ domain walls. Bottom two terms condense $\mathbb{Z}_2^{\text{even}}$ domain walls and $\mathbb{Z}_2^{\text{odd}}$ charges. }
    \label{fig:RBH2}
    \end{minipage}
\end{figure}
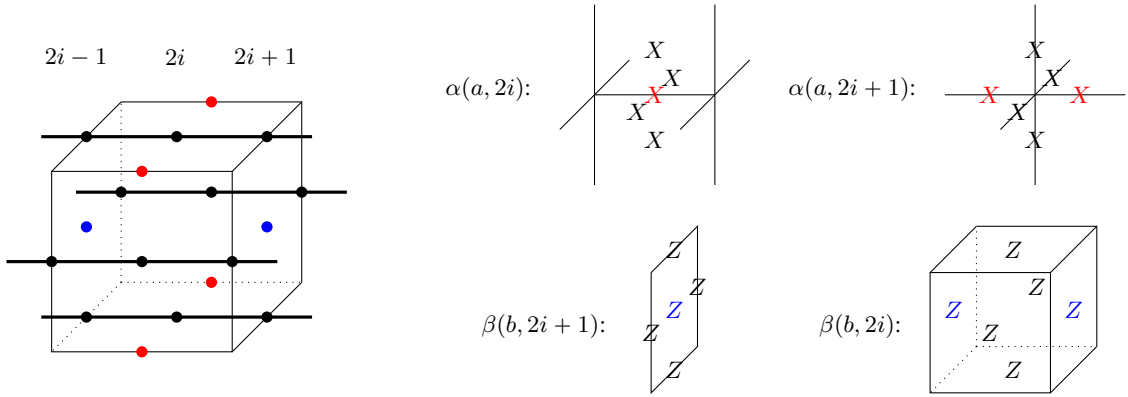

Alternatively, we can take the cluster state to be CSS$_1$, and the CSS code to be CSS$_2$. In this case, for each qubit in CSS, we foliate a layer of cluster state. For each $X$-stabilizer in CSS, we foliate a line of $Z$-ancilla, with qubits on even sites; while for each $Z$-stabilizer in CSS, we foliate a line of $X$-ancilla, with qubits on odd sites. The code switching is by enforing $X$-stabilizers in space $A\otimes Q_\mathbb{Z}$ with terms $\alpha(a,2i) = X_{a,2i} \prod_{q\in a} X_{q,2i}$ and $\alpha(a,2i+1) = X_{a,2i} X_{a,2i+2} \prod_{q \in a} X_{q,2i+1}$; as well as $Z$-stabilizers in space $B\otimes Q_\mathbb{Z}$ with terms $\beta(b,2i) = Z_{b,2i-1}Z_{b,2i+1} \prod_{q\in b} Z_{q,2i}$ and $\beta(b,2i+1) = Z_{b,2i+1} \prod_{q \in b} Z_{q,2i+1}$. In Figure~\ref{fig:RBH2}, we again use the RBH cluster state as an illustrating example.

In order to see what are the excitations being condensed, recall the cluster state has two $\mathbb{Z}_2$ symmetries, $\mathbb{Z}_2^{\text{even}}$ and $\mathbb{Z}_2^{\text{odd}}$, which act on even and odd sublattices respectively,
\begin{align*}
    \mathbb{Z}_2^{\text{even}} = \prod_{i}Z_{2i},
    \hspace{35pt}
    \mathbb{Z}_2^{\text{odd}} = \prod_{i}X_{2i+1}.
\end{align*}
First consider the $X$-type condensation. Each $X_{q,2i}$ creates a $\mathbb{Z}_2^{\text{even}}$ charge (in $\alpha(a,2i)$), while each $X_{q,2i+1}$ creates a pair of $\mathbb{Z}_2^{\text{odd}}$ domain walls on sites $2i$ and $2i+2$ (in $\alpha(a,2i+1)$). Both are condensed via the red ancillas. Hence the $\mathbb{Z}_2^{\text{even}}$ charges and $\mathbb{Z}_2^{\text{odd}}$ domain walls in layers $q\in a$ are condensed together. Similarly for the $Z$-type condensation, where the $\mathbb{Z}_2^{\text{even}}$ domain walls and $\mathbb{Z}_2^{\text{odd}}$ charges in layers $q\in b$ are condensed together.

In particular, if the 1D cluster state has boundaries, then it hosts one logical qubit per boundary~\cite{BrownRoberts20}, hence it has nontrivial cohomology $H^0(\text{CS})$. The tensor product structure makes it clear that the foliated CSS code encodes the CSS code on its boundary because $H^0(\text{CSS} \otimes \text{CS}) = H^0(\text{CSS}) \otimes H^0(\text{CS})$. Indeed, in measurement based quantum computation, the CSS code on the boundary is teleported layer by layer via measurements.

\section{Balanced product}
In this section, we will explain the coupled layer construction corresponding to the balanced product~\cite{breuckmann2021balanced,Panteleev_ACM2022}. This is the general case, and does not assume the group action is free as in the main text.

We begin with some preliminaries. Let $V$ and $W$ be two vector spaces with a group $G$ action from the right and left respectively. Assume there are bases $\{v_i\}$ and $\{w_j\}$ where $G$ acts by permutation. The balanced product is the vector space $V\otimes_G W$ spanned by $v_i\otimes w_j$, quotiented by the equivalence relation $v\cdot g\otimes w \sim v\otimes g\cdot w$. The basis for $V\otimes_G W$ is formed by equivalence classes $[v_i\otimes w_j]$. Given an element $v_i\otimes w_j$, we can represent its equivalence class in the following way. Let $K_{w_j}$ be the subgroup of $G$ which fixes $w_j$, denote the equivalence classes of $V/K_{w_j}$ by $[v_i]^{(w_j)}$. Then $[v_i\otimes w_j]$ can be represented by $[v_i]^{(w_j)} \otimes w_j$. 
To be more precise, for each equivalence class in $G\backslash W$, we fix a representative in $\{w_j\}$. Given a basis element $w_j$, we denote by $\tilde w_j$ the representative of its equivalence class $[w_j]$. Then $\{[v_i]^{(\tilde w_j)} \otimes  \tilde w_j\}$ form a basis for $V\otimes_G W$. A generic element $v_i\otimes w_j$ can be taken to this basis as follows. Find $g_{w_j}$ which relates $w_j$ and its representative $\tilde w_j$ in $G\backslash W$, that is $w_j = g_{w_j} \cdot \tilde w_j$.  (Note that $g_{w_j}$ is not unique if the action is not free.) Then $v_i\otimes w_j \sim [v_i \cdot g_{w_j}]^{(\tilde w_j)} \otimes \tilde w_j$ uniquely.

Now, take two CSS codes, and let $G$ act on CSS$_1$ and CSS$_2$ from right and left respectively. Choose a set of representatives $\Tilde{A}_2$, $\Tilde{B}_2$ and $\Tilde{Q}_2$ for $G\backslash A_2$, $G\backslash B_2$ and $G\backslash Q_2$ respectively. For each $\Tilde{q}_2\in \Tilde{Q}_2$, introduce a copy of CSS$_1/K_{\Tilde{q}_2}$ denoted CSS$_1^{(\tilde q_2)}$. We write $[a_1]^{(\bullet)} \ni [q_1]^{(\bullet)}$ and $[b_1]^{(\bullet)}\ni [q_1]^{(\bullet)}$  to denote respectively the $X$-stabilizers and $Z$-stabilizers containing qubit $[q_1]^{(\bullet)}$ in the quotient code CSS$_1/K_{\bullet}$. Similarly, we use $[q_1]^{(\bullet)}\in [a_1]^{(\bullet)}$ and  $[q_1]^{(\bullet)}\in  [b_1]^{(\bullet)}$ to indicate $[q_1]^{(\bullet)}$ is checked by the $X$-stabilizer $[a_1]^{(\bullet)}$ and $Z$-stabilizer $[b_1]^{(\bullet)}$ respectively. For each $\Tilde{a}_2\in \Tilde{A}_2$ and $[b_1]^{(\tilde a_2)} \in B_1/K_{\tilde a_2}$, we introduce a $Z$-ancila $Z_{[b_1]^{(\tilde a_2)}, \tilde a_2}$; for each $\Tilde{b}_2\in \Tilde{B}_2$ and $[a_1]^{(\tilde b_2)} \in A_1/K_{\tilde b_2}$, we introduce an $X$-ancila $X_{[a_1]^{(\tilde b_2)}, \tilde b_2}$. The stabilizer group is (noticing the identical structure as in the coupled layer construction)
\begin{align*}
    \mathcal{S}_0 = \left\langle 
    [a_1]^{({\Tilde{q}_2})}, [b_1]^{({\Tilde{q}_2})}, X_{[a_1]^{(\Tilde{b}_2)},\Tilde{b}_2}, Z_{[b_1]^{(\Tilde{a}_2)},\Tilde{a}_2}
    \right\rangle.
\end{align*}

Next, we perform code switching as before. The $X$ terms come from $Q_1\otimes_G A_2$. Formally, this is
\begin{align*}
    \alpha([q_1]^{(\tilde a_2)},\Tilde{a}_2) = 
    \hspace{5pt}
    \prod_{\mathclap{\scriptscriptstyle [b_1]^{(\Tilde{a}_2)}\ni [q_1]^{(\Tilde{a}_2)}}} 
    \hspace{10pt}X_{[b_1]^{(\Tilde{a}_2)},\Tilde{a}_2} 
    \hspace{5pt}\prod_{\mathclap{\scriptscriptstyle[q_2]^{(q_1)}\in [\Tilde{a}_2]^{(q_1)}}}
    \hspace{10pt}
    X_{[q_1\cdot g_{q_2}]^{(\Tilde{q}_2)},\Tilde{q}_2}.
\end{align*}
where the second product represents the following procedure: first pick any element $q_1\in [q_1]^{(\tilde a_2)}$, then $[q_1]^{(\tilde a_2)}\otimes \Tilde{a}_2 \sim q_1 \otimes [\Tilde{a}_2]^{( q_1)}$. Then the product ranges over qubits checked by $[\Tilde{a}_2]^{( q_1)}$ in the code $K_{q_1}\backslash$CSS$_2$. For each such qubit $[q_2]^{(q_1)}$, pick an element $q_2$, then $q_1\otimes [q_2]^{(q_1)} \sim [q_1\cdot g_{q_2}]^{(\Tilde{q}_2)}\otimes \Tilde{q}_2$, which brings it back to the basis we chose.

The $Z$ terms come from $Q_1\otimes_G B_2 $. Formally,
\begin{align*}
    \beta([q_1]^{(\tilde b_2)},\Tilde{b}_2) = 
    \hspace{5pt}
    \prod_{\mathclap{\scriptscriptstyle [a_1]^{(\Tilde{b}_2)}\ni [q_1]^{(\Tilde{b}_2)}}} 
    \hspace{10pt}Z_{[a_1]^{(\Tilde{b}_2)},\Tilde{b}_2} 
    \hspace{5pt}
    \prod_{\mathclap{\scriptscriptstyle[q_2]^{(q_1)}\in [\Tilde{b}_2]^{(q_1)}}}
    \hspace{10pt}
    Z_{[q_1\cdot g_{q_2}]^{(\Tilde{q}_2)},\Tilde{q}_2}.
\end{align*}

This has a similar interpretation as before from the condensation point of view. Looking at $\alpha([q_1]^{(\tilde a_2)},\Tilde{a}_2)$, each $X_{[q_1\cdot g_{q_2}]^{(\Tilde{q}_2)},\Tilde{q}_2}$ creates a number of excitations in CSS$_1^{(\tilde q_2)}$. These excitations are condensed through $X$-checks in CSS$_2$, but after permuted by the action of $G$.

Keeping the terms in $\mathcal{S}_0$ which commute with code switching, the following terms remain
\vspace{-0.5em}
\begin{itemize}
    \item $X$ terms from $A_1\otimes_G Q_2$: Each $[a_1]^{(\tilde q_2)}\otimes \tilde q_2$ gives $\xi([a_1]^{(\Tilde{q}_2)},\Tilde{q}_2) = [a_1]^{(\Tilde{q}_2)}  \hspace{10pt} \prod\limits_{\mathclap{[b_2]^{(a_1)}\ni [\Tilde{q}_2]^{(a_1)}}} 
    \hspace{10pt}
    X_{[a_1\cdot g_{b_2}]^{(\tilde b_2)}, \Tilde{b}_2}$.
    \vspace{-0.5em}
    \item $Z$ terms from $B_1\otimes_G Q_2$: Each $[b_1]^{(\tilde q_2)}\otimes \tilde q_2$ gives $\zeta([b_1]^{(\Tilde{q}_2)},\Tilde{q}_2) = [b_1]^{(\Tilde{q}_2)}  \hspace{10pt} \prod\limits_{\mathclap{[a_2]^{(b_1)}\ni [\Tilde{q}_2]^{(b_1)}}} 
    \hspace{10pt}
    Z_{[b_1\cdot g_{a_2}]^{(\tilde a_2)}, \Tilde{a}_2}$.
\end{itemize}
\vspace{-0.5em}
This is precisely the stabilizer group of the balanced product CSS$_1 \otimes_G$ CSS$_2$.

Below we provide several examples of the balanced product via the coupled-layer construction. The first example gives a physical interpretation. The second example uses a non-abelian group, and is a simple analogy of the construction used to produce good qLDPC codes. We then  produc two well known examples, which are the color code and Haah's code. The last example is a balanced product between two quantum codes, rather than two classical codes.

\subsection{3D TC with twisted boundary}

In this example, we take CSS$_1$ to be 2D TC, and CSS$_2$ to be decoupled Ising chains extending in the diagonal direction on a 2D square lattice, with qubits on vertices. We label the qubits of CSS$_2$ by 2D coordinates $(q,w) \in \mathbb{Z} \times \mathbb{Z}$. The stabilizers of CSS$_2$ are given by the figure below.
\begin{figure}[h!]
    \begin{tikzpicture}[scale = 1, baseline = 0ex]
        \node at (0,0) {$Z$};
        \node at (2,2) {$Z$};
        
        \draw[step = 2] (-0.5,-0.5) grid (2.5, 2.5);
    \end{tikzpicture}
    \caption{A stabilizer of CSS$_2$. The rest are given by horizontal and vertical translations.}
\end{figure}
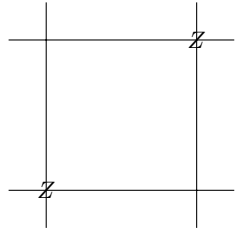
Let the group $\mathbb{Z} = \langle T \rangle $ act on both CSS$_1$ and CSS$_2$ by translating to the right, with $T$ being translation by one site. We consider the balanced product CSS$_1 \otimes_\mathbb{Z} $ CSS$_2$.

We describe the stacking procedure. First we find the left quotient $\mathbb{Z}\backslash$CSS$_2$. The action collapses CSS$_2$ into a 1D chain. The equivalence classes of qubits are labeled by integers $[w]$, $w\in \mathbb{Z}$, and we choose representative to be $(0, w)$. Equivalence class of checks are labeled by segments $[\tilde w]$, $\tilde w = [w, w+1]$, and we choose the representative to be $Z_{0,w}Z_{1,w+1}$. Therefore, we introduce a copy of CSS$_1$ for each $w\in \mathbb{Z}$, and an $X$-ancilla for each pair $(v, [\tilde w])$. These form the stacked system.

\begin{figure}[h!]
    \begin{tikzpicture}[scale = 1, baseline = -3ex]
        % First plaquette
        % Define coordinates
        \coordinate (O) at (0,0,0); 
        \coordinate (A) at (2,0,0);
        \coordinate (B) at (2,2,0);
        \coordinate (C) at (0,2,0);
        \coordinate (D) at (4,2,0);

        \node at (-0.6, 0) {$w$};
        \node at (-0.6, 2) {$w+1$};
    
        % Draw edges
        \draw (C) -- (O)-- (A) -- (B)
        (B) -- (D);
        \draw[red] (O) -- (A)
        (B) -- (D);
        \draw [densely dotted] (B) -- (C);

        % Draw arrow
        \draw[->,  yscale=0.7] (1,3) arc[start angle=180, end angle=5, radius=1cm];
        \node at (2,3) {$T$};

        \foreach \X/\Y in { O/C, A/B} 
        \node[black] at ($( \X )!0.5!( \Y )$) {$Z$}; 

        \foreach \X/\Y in {O/A, D/B}
        \node[black] at ($( \X )!0.5!( \Y )$) {$Z$}; 

        % Second plaquette
        % Define coordinates
        \coordinate (O) at (7.5,0.5,0);
        \coordinate (A) at (7.5,2.5,0);
        \coordinate (B) at (7.5,2.5,2);
        \coordinate (C) at (7.5,0.5,2);
        \coordinate (D) at (9.5,2.5,0);
        \coordinate (E) at (9.5,2.5,2);
        % Draw edges
        \draw (B) -- (C) -- (O) -- (A)
        (D)--(E);
        \draw[densely dotted] (A) -- (B)
        (A) -- (D)
        (B) -- (E);

        \draw[red] (O) -- (C)
        (D) -- (E);

        \node at (6.5, 0.5, 1) {$w$};
        \node at (6.5, 2.5, 1) {$w+1$};

        % Draw arrow
        \draw[->,  yscale=0.7] (7.5,3.6,1) arc[start angle=180, end angle=5, radius=1cm];
        \node at (8.15, 3.1) {$T$};
        
        \foreach \X/\Y in {D/E, C/O}
        \node at ($( \X )!0.5!( \Y )$) {$Z$}; 
        \foreach \X/\Y in {O/A,  B/C}
        \node[black] at ($( \X )!0.5!( \Y )$) {$Z$};
    \end{tikzpicture}
    \begin{minipage}{0.95\textwidth}
        \centering
        \caption{The term $\beta(e,\tilde w)$ form a horizontal edge (left) and vertical edge (right) respectively. The edge $e$ as well as its shift $e\cdot T$ are colored in red.}
        \label{fig:twistedTCcond}
    \end{minipage}
\end{figure}
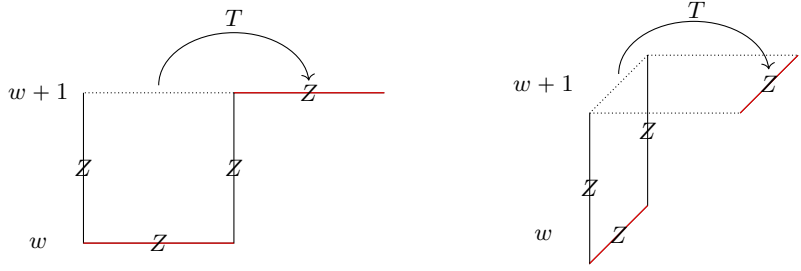

Next we perform code switching. There are only $Z$-type terms given by pairs $(e, [\tilde w])$. The qubits checked by $[\tilde w]$ in $\mathbb{Z} \backslash$CSS$_2$ are $(0,w)$ and $(1,w+1)$. The first qubit is the representative of $[w]$, while the second is not the representative of $[w+1]$. It is related via the group action, $T\cdot (0,w+1) = (1,w+1)$. Thus the code switching is given by
\begin{align*}
    \beta(e, \tilde w) = Z_{e, w}Z_{e\cdot T, w+1} \prod_{v \ni e} Z_{v,\tilde w}.
\end{align*}
Pictorially, this term is given by Figure~\ref{fig:twistedTCcond}. Since each $Z$ in a CSS$_1$ layer creates a pair of $\ee$ anyons, $\beta(e,\tilde w)$ condenses $\ee$ in neighboring layers but shifted by one site. 

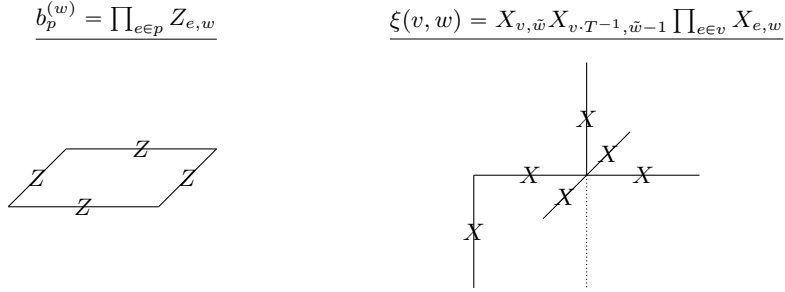
\begin{figure}[h!]
    \begin{tikzpicture}[scale = 1, baseline = 1.65cm]
        \node at (0.8,3.7,0) {\underline{$b_p^{(w)} = \prod_{e\in p}Z_{e,w}$}};
        
        % Define coordinates
        \coordinate (B) at (2,2,0);%back-top-right
        \coordinate (C) at (0,2,0);%back-top-left
        \coordinate (F) at (2,2,2);
        \coordinate (G) at (0,2,2);

        % Draw edges    
        \draw  (B) -- (C); 
        \draw  (F) -- (G) ;
        \draw (B) -- (F);
        \draw (C) -- (G);

        \foreach \X/\Y in { B/F, C/G}
        \node[circle, inner sep=1.5pt] at ($( \X )!0.5!( \Y )$) {$Z$};
    
        \foreach \X/\Y in { B/C,  F/G}
        \node[circle, inner sep=1.5pt] at ($( \X )!0.5!( \Y )$) {$Z$}; 
    \end{tikzpicture}
    \hspace{55pt}
    \begin{tikzpicture}[scale = 1, baseline = 0cm]

        \node at (0,2,0) {\underline{$\xi(v,w) = X_{v,\tilde w} X_{v \cdot T^{-1}, \tilde w -1} 
    \prod_{e \in v} X_{e, w}$}};
        % Define coordinates
        \coordinate (O) at (0,0,0);
        \coordinate (A) at (-1.5,0,0);
        \coordinate (B) at (1.5,0,0);
        \coordinate (C) at (0,-1.5,0);
        \coordinate (D) at (0,1.5,0);
        \coordinate (E) at (0,0,1.5);
        \coordinate (F) at (0,0,-1.5);
        \coordinate (G) at (-1.5,-1.5,0);

        % Draw edges
        \draw (A) -- (B);
        \draw [densely dotted] (O) -- (C);
        \draw (O) -- (D); 
        \draw (E) -- (F);
        \draw (A) -- (G);
            
        \foreach \X/\Y in { O/E, O/F}
        \node at ($( \X )!0.5!( \Y )$) {$X$};
        \foreach \X/\Y in { A/G, O/D}
        \node[black] at ($( \X )!0.5!( \Y )$) {$X$};
            
        \foreach \X/\Y in {O/A, O/B}
        \node[black] at ($( \X )!0.5!( \Y )$) {$X$}; 
    \end{tikzpicture}
    \caption{Stabilizers after the code switching.}
    \label{fig:twistedTCdeform}
\end{figure}

After code switching, the plaquette terms in TC survives, and a new $X$-stabilizer is generated. These terms are given by Figure~\ref{fig:twistedTCdeform}
The stabilizer group of the deformed code is thus $\mathcal{S} = \langle  b_p^{(w)}, \beta(e, \tilde w), \xi(v,w) \rangle$. This resembles the 3D TC, but as we go in $z$-direction, the horizontal layers are shifted. If we put the system on an $L\times L \times L$ torus, then we can shift all the layers back, such that in the bulk we recover the 3D TC. However, we shift by $L$ sites when the boundaries in $z$-direction are glued. This code can be understood as a fiber-bundle code \cite{hastings2021fiber}.

\begin{figure}[h!]
    \centering
    \resizebox{0.5\textwidth}{!}{\includegraphics[width=0.5\linewidth]{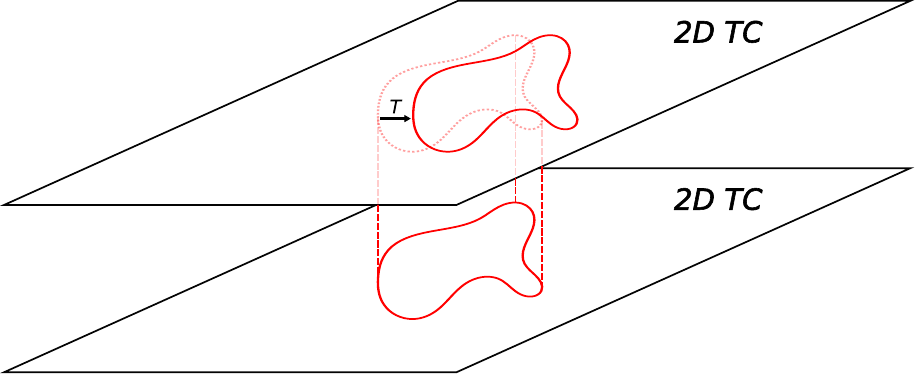}}
    
    \begin{minipage}{0.95\textwidth}
        \caption{The $\mathbb{Z}_2$ 1-form symmetry we gauge to obtain the balanced product CSS$_1\otimes_{\mathbb Z}$ CSS$_2$. In going up in $z$-direction, we need to conjugate the symmetry operators by $T$. }
        \label{fig:twisted3DTCGauge}
    \end{minipage}
\end{figure}

Similar to tensor product, we can interpret the code switching as gauging logicals. In the case of 2D TC $\otimes$ 1D $Z$-Ising, we need to gauge the diagonal subgroup $\mathbb{Z}_2\subset \mathbb{Z}_2^{w} \times \mathbb{Z}_2^{w+1}$ of the 1-form symmetry generated by $\ee$ loops in neighboring layers, where $\mathbb{Z}_2^{w}$ is the symmetry in layer $w$. In the present case, we must gauge the diagonal $\mathbb{Z}_2$ subgroup in $ \mathbb{Z}_2^{[w]} \times T\mathbb{Z}_2^{[w+1]}T^{-1}$ for each pair of neighboring layers. The schematics is shown in Figure~\ref{fig:twisted3DTCGauge}. However, because the 2D TC is a topological code, the action of translation is trivial, and the resulting code is still a topological code, which is the 3D TC.

This interpretation can be generalized to balanced product in general. Assume the action is free, in which case we start with stacks of CSS$_1$. We treat the set of $X$-stabilizers and $X$-logicals as $X$-type symmetries of CSS$_1$, denoted $S_X$, and similarly the $Z$-type symmetries $S_Z$. Then a representative $\tilde a_2 \in G\backslash A_2$ indicate that we gauge the logical $\prod_{q_2\in \tilde a_2} g_{q_2} S_X g_{q_2}^{-1} $, where $gS_Xg^{-1}$ shifts the supports of operators in $S_X$ by $g$. Similarly, $\tilde b_2 \in G\backslash B_2$ indicate that we gauge the logical $\prod_{q_2\in \tilde b_2} g_{q_2} S_Z g_{q_2}^{-1}$. These logical are diagonal in layers of CSS$_1$ up to modulation by the group actions according to the patterns of stabilizers in CSS$_2$. Gauging these symmetries gives the balanced product between CSS$_1$ and CSS$_2$.

%============================================
%============================================

\subsection{$D_6$ balanced product}

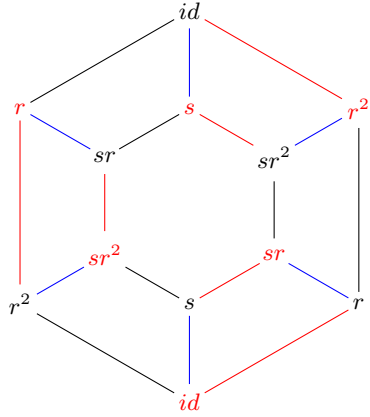
\begin{figure}[h!]
    \begin{tikzpicture}[scale=1.3, every node/.style={circle,inner sep=1pt,font=\small}]

    % Outer hexagon
    \node (b_id)  at (90:2) {$id$};
    \node (r_r2) [red]  at (30:2) {$r^2$};
    \node (b_r)  at (-30:2) {$r$};
    \node (r_id) [red] at (-90:2) {$id$};
    \node (b_r2)  at (-150:2) {$r^2$};
    \node (r_r) [red] at (150:2) {$r$};

    % Inner hexagon 
    \node (r_s) [red]  at (90:1) {$s$};
    \node (b_sr2) at (30:1) {$sr^2$};
    \node (r_sr) [red] at (-30:1) {$sr$};
    \node (b_s) at (-90:1) {$s$};
    \node (r_sr2) [red] at  (-150:1) {$sr^2$};
    \node (b_sr) at (150:1) {$sr$};

    % Red edges
    \draw[red] (b_id) -- (r_r2)
                (r_s) -- (b_sr2)
                (b_sr)-- (r_sr2)
                (r_r) -- (b_r2)
                (b_s) -- (r_sr)
                (b_r) -- (r_id);

    % Black edges
    \draw[black] (b_id) -- (r_r)
                (b_sr) -- (r_s)
                (b_r2) -- (r_id)
                (b_s) -- (r_sr2)
                (b_r) -- (r_r2)
                (b_sr2) -- (r_sr);

    % Blue edges
    \draw[blue] (b_id) -- (r_s)
                (b_sr) -- (r_r)
                (b_r2) -- (r_sr2)
                (b_s) -- (r_id)
                (b_r) -- (r_sr)
                (b_sr2) -- (r_r2);
    \end{tikzpicture}
    \begin{minipage}{0.95\textwidth}
        \centering
        \caption{The Double cover $\Gamma$ of the left Cayley graph of $D_6$. $D_6$ acts freely on vertices by multiplication on the right, and the actions on edges follow from the action on vertices. The two $D_6$ orbits of vertices are colored by black and red, and the three orbits of edges are colored black, red and blue.}
        \label{fig:cay_d6}
    \end{minipage}
\end{figure}

Asymptotically good qLDPC codes are constructed using balanced product with Tanner codes \cite{tanner1981recursive} defined on the Cayley graph of PSL$(2,q)$ \cite{Panteleev_ACM2022}.
In this section, we give a simple example using the dihedral group of order 6, $D_6 \cong S_3$. Choose generators $S = \{s, r, r^{-1}\}$, satisfying $r^3=s^2 =(sr)^2=1$. consider the double cover $\Gamma = (V, E)$ of the Cayley graph as in~\cite{Panteleev_ACM2022}. There is a free right $D_6$ action on $\Gamma$ explained in Figure \ref{fig:cay_d6}.

For simplicity, consider the classical code defined by the cellular chain complex of $\Gamma$ with bits on edges, and checks on vertices, $\mathcal{C}: E \rightarrow V$. Alternatively, this can be thought of as a Tanner code on $\Gamma$ with subcode parity check $H_v = \begin{pmatrix}1 & 1 & 1\end{pmatrix}$ for each vertex $v$. We treat this as a CSS code with only $Z$-stabilizers $b_v = \prod_{e\in v} Z_e$. The code $\mathcal{C}$ inherits the free right action of $D_6$. It follows that the transpose $\mathcal{C}^*: V \rightarrow E$ has an induced left $D_6$ action, for instance $g \cdot V  := V \cdot g^{-1}$, which is also free. We treat the transpose code $\mathcal{C}^*$ as a CSS code with only $X$-stabilizers, and we now describe the coupled layer construction of the balanced product $\mathcal{C} \otimes_{D_6} \mathcal{C}^*$.

The left quotient graph $D_6\backslash \mathcal{C}^*$ three edges as shown by the coloring in Figure~\ref{fig:cay_d6}, therefore we introduce three copies of $\mathcal{C}$ labeled by $1$, $2$ and $3$ correspond to black, red and blue edge-orbits respectively. $D_6\backslash \mathcal{C}^*$ has two vertices, hence we introduce two $Z$-ancillas for eacu vertex of $\Gamma$, labeled by $1$ and $2$ correspond to black and red vertex-orbits respectively. 
The representatives of edges and vertices in $D_6\backslash \mathcal{C}^*$ is shown in Figure~\ref{fig:quo_cay_d6}. In summary, the Hilbert space of the balanced product has three qubits per edge and two qubits per vertex. The stabilizer group is
\begin{align*}
    \mathcal{S}_0 = \langle b_v^{(i)}, Z_{v,n}\rangle,
    \hspace{15pt}
    i = 1,2,3
    ,\hspace{5pt}
    n = 1,2.
\end{align*}

\begin{figure}[h!]
    \begin{tikzpicture}[>=Stealth, node distance=3cm, auto, 
    every node/.style={inner sep=1pt}, scale = 0.5]

    % nodes
    \node (L) {$[id]$};
    \node (R) [right of=L] {$[\textcolor{red}{id}]$};

    % arrows
    \draw (L.north east) .. controls +(1.5,1) and +(-1.5,1) .. node[midway,above] {$[id-\textcolor{red}{r}]$} (R.north west);
    \draw[blue] (L) -- node[midway, above, black] {$[id-\textcolor{red}{s}]$} (R);
    \draw[ red] (L.south east) .. controls +(1.5,-1) and +(-1.5,-1) .. node[midway, below, black] {$[id-\textcolor{red}{r^{2}}]$} (R.south west);

    \end{tikzpicture}
    \caption{The left quotient graph defining $D_6\backslash C^*$. Representatives of each orbit is printed.}
    \label{fig:quo_cay_d6}
\end{figure}

Since CSS$_2$ has only $X$-stabilizers, the code switching introduces only $X$-type stabilizers $\alpha(e,1)$ and $\alpha(e,2)$, for each edge $e\in E$. The first term is simple
\begin{align*}
    \alpha(e,1) = \prod_{i=1}^3 X_{e,i} \prod_{v\ni e}X_{v,1}. 
\end{align*}
Due the representatives that we choose, this term looks exactly like the term in the usual tensor product. $X_1$, $X_2$, $X_3$ on each edge each creates a pair of excitations, which is paired up with $X_1$ on the ancillas. An example of this term is shown in Figure~\ref{fig:cay_d6_cond} below. Next consider the second term. The edges connecting to the representative $[\textcolor{red}{id}]$ are $r^2 - \textcolor{red}{id}$, $r - \textcolor{red}{id}$ and $s - \textcolor{red}{id}$ in order of copy $1$, $2$ and $3$. These are not the representatives we chose in Figure~\ref{fig:quo_cay_d6}, and we need to find the group elements which take the chosen representatives to these edges, which are $r$, $r^2$ and $s$ respectively. The actions by these group elements are moved onto copies of $\mathcal{C}$ via the balanced product. Therefore we have
\begin{align*}
    \alpha(e,2) = X_{e\cdot r,1} X_{e\cdot r^2,2}X_{e\cdot s,3}  \prod_{v\ni e}X_{v,2}.
\end{align*}
Examples of this term is shown in Figure~\ref{fig:cay_d6_cond}. In particular, we condense excitations in different layers up to a $D_6$ action.
\begin{figure}[h!]
    \begin{tikzpicture}[scale=1.3, dot/.style={circle, fill= black,inner sep=1pt,font=\small}, baseline = 0cm]

    \node at (90:2.6)  {\underline{\,$\alpha(id - \textcolor{red}{r}, 1)$\,}};
    
    % Outer hexagon
    \node [dot] (b_id)  at (90:2) {};
    \node[dot] (r_r2)   at (30:2) {};
    \node[dot] (b_r)  at (-30:2) {};
    \node[dot] (r_id)  at (-90:2) {};
    \node[dot] (b_r2)  at (-150:2) {};
    \node [dot ](r_r) at (150:2) {};

    % Inner hexagon 
    \node[dot] (r_s)   at (90:1) {};
    \node[dot] (b_sr2) at (30:1) {};
    \node[dot] (r_sr)  at (-30:1) {};
    \node[dot] (b_s) at (-90:1) {};
    \node[dot] (r_sr2)  at  (-150:1) {};
    \node[dot] (b_sr) at (150:1) {};

    \node at ($(b_id)!0.5!(r_r)$) {$X_1 X_2 X_3$};
        
    % Red edges
    \draw (b_id) -- (r_r2)
                (r_s) -- (b_sr2)
                (b_sr)-- (r_sr2)
                (r_r) -- (b_r2)
                (b_s) -- (r_sr)
                (b_r) -- (r_id);

    \draw[red](b_id) -- (r_r);

    % Black edges
    \draw       (b_sr) -- (r_s)
                (b_r2) -- (r_id)
                (b_s) -- (r_sr2)
                (b_r) -- (r_r2)
                (b_sr2) -- (r_sr);

    % Blue edges
    \draw (b_id) -- (r_s)
                (b_sr) -- (r_r)
                (b_r2) -- (r_sr2)
                (b_s) -- (r_id)
                (b_r) -- (r_sr)
                (b_sr2) -- (r_r2);

    \node  at (90:2.2) {$X_1$};
    \node at (150:2.3) {$X_1$};

    \end{tikzpicture}
    \hspace{35pt}
    \begin{tikzpicture}[scale=1.3, dot/.style={circle, fill= black,inner sep=1pt,font=\small}, baseline = 0cm]

    \node at (90:2.6)  {\underline{\,$\alpha(r - \textcolor{red}{id}, 2)$\,}};
    
    % Outer hexagon
    \node [dot] (b_id)  at (90:2) {};
    \node[dot] (r_r2)   at (30:2) {};
    \node [dot] (b_r)  at (-30:2) {};
    \node [dot] (r_id) at (-90:2) {};
    \node[dot] (b_r2)  at (-150:2) {};
    \node [dot] (r_r)  at (150:2) {};

    % Inner hexagon 
    \node[dot] (r_s)   at (90:1) {};
    \node[dot] (b_sr2) at (30:1) {};
    \node[dot] (r_sr)  at (-30:1) {};
    \node[dot] (b_s) at (-90:1) {};
    \node[dot] (r_sr2)  at  (-150:1) {};
    \node[dot] (b_sr) at (150:1) {};

    \node at ($(b_r2)!0.5!(r_r)$) {$X_1$};
    \node at ($(b_id)!0.5!(r_r2)$) {$X_2$};
    \node at ($(r_s)!0.5!(b_sr2)$) {$X_3$};
        
    % Red edges
    \draw (b_id) -- (r_r2)
                (r_s) -- (b_sr2)
                (b_sr)-- (r_sr2)
                (r_r) -- (b_r2)
                (b_s) -- (r_sr);
                
    \draw[red] (b_r) -- (r_id);

    % Black edges
    \draw (b_id) -- (r_r)
                (b_sr) -- (r_s)
                (b_r2) -- (r_id)
                (b_s) -- (r_sr2)
                (b_r) -- (r_r2)
                (b_sr2) -- (r_sr);

    % Blue edges
    \draw (b_id) -- (r_s)
                (b_sr) -- (r_r)
                (b_r2) -- (r_sr2)
                (b_s) -- (r_id)
                (b_r) -- (r_sr)
                (b_sr2) -- (r_r2);

    \node  at (-30:2.25) {$X_2$};
    \node  at (-90:2.2) {$X_2$};
    
    \end{tikzpicture}
    \begin{minipage}{0.95\textwidth}
        \centering
        \caption{Examples of the code switching terms $\alpha(e,n)$. The edges $e$ are highlighted in red.}
        \label{fig:cay_d6_cond}
    \end{minipage}
\end{figure}
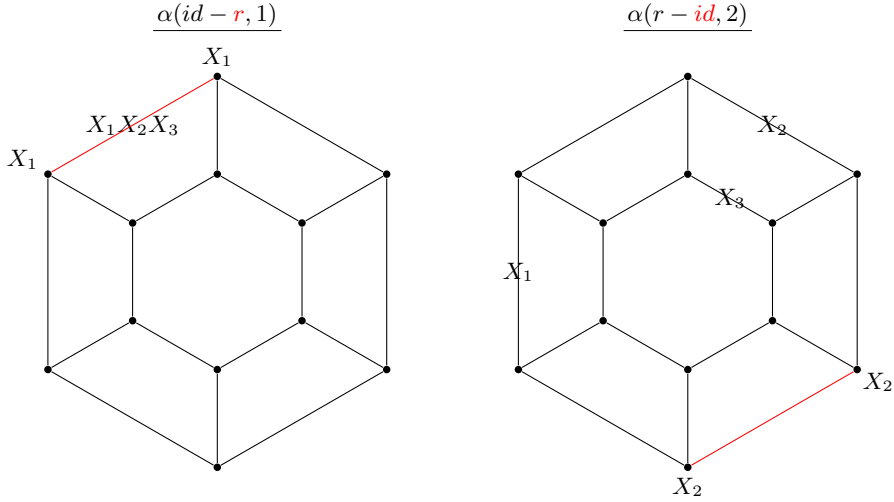

\begin{figure}[h!]
    \begin{tikzpicture}[scale=1.3, dot/.style={circle, fill= black,inner sep=1pt,font=\small}, baseline = 0cm]

    \node at (90:2.6)  {\underline{\,$\zeta(id,1)$\,}};
     
    % Outer hexagon
    \node [dot, red](b_id)  at (90:2) {};
    \node [dot] (r_r2)   at (30:2) {};
    \node [dot] (b_r)  at (-30:2) {};
    \node [dot] (r_id) at (-90:2) {};
    \node [dot] (b_r2)  at (-150:2) {};
    \node [dot] (r_r)  at (150:2) {};

    % Inner hexagon 
    \node[dot] (r_s)   at (90:1) {};
    \node[dot] (b_sr2) at (30:1) {};
    \node[dot] (r_sr)  at (-30:1) {};
    \node[dot] (b_s) at (-90:1) {};
    \node[dot] (r_sr2)  at  (-150:1) {};
    \node[dot] (b_sr) at (150:1) {};

    \node at ($(b_id)!0.5!(r_r2)$) {$Z_1$};
    \node at ($(b_id)!0.5!(r_s)$) {$Z_1$};
    \node at ($(b_id)!0.5!(r_r)$) {$Z_1$};
        
    % Red edges
    \draw (b_id) -- (r_r2)
                (r_s) -- (b_sr2)
                (b_sr)-- (r_sr2)
                (r_r) -- (b_r2)
                (b_s) -- (r_sr)
                (b_r) -- (r_id);

    % Black edges
    \draw (b_id) -- (r_r)
                (b_sr) -- (r_s)
                (b_r2) -- (r_id)
                (b_s) -- (r_sr2)
                (b_r) -- (r_r2)
                (b_sr2) -- (r_sr);

    % Blue edges
    \draw (b_id) -- (r_s)
                (b_sr) -- (r_r)
                (b_r2) -- (r_sr2)
                (b_s) -- (r_id)
                (b_r) -- (r_sr)
                (b_sr2) -- (r_r2);

    \node  at (90:2.2) {$Z_1$};
    \node  at (-150:2.2) {$Z_2$};

    \end{tikzpicture}
    \hspace{25pt}
    \begin{tikzpicture}[scale=1.3, dot/.style={circle, fill= black,inner sep=1pt,font=\small}, baseline = 0cm]

    \node at (90:2.6)  {\underline{\,$\zeta(id,2)$\,}};

    % Outer hexagon
    \node [dot,red](b_id)  at (90:2) {};
    \node[dot] (r_r2)   at (30:2) {};
    \node [dot](b_r)  at (-30:2) {};
    \node [dot] (r_id) at (-90:2) {};
    \node[dot] (b_r2)  at (-150:2) {};
    \node [dot] (r_r)  at (150:2) {};

    % Inner hexagon 
    \node[dot] (r_s)   at (90:1) {};
    \node[dot] (b_sr2) at (30:1) {};
    \node[dot] (r_sr)  at (-30:1) {};
    \node[dot] (b_s) at (-90:1) {};
    \node[dot] (r_sr2)  at  (-150:1) {};
    \node[dot] (b_sr) at (150:1) {};

    \node at ($(b_id)!0.5!(r_r2)$) {$Z_2$};
    \node at ($(b_id)!0.5!(r_s)$) {$Z_2$};
    \node at ($(b_id)!0.5!(r_r)$) {$Z_2$};
        
    % Red edges
    \draw (b_id) -- (r_r2)
                (r_s) -- (b_sr2)
                (b_sr)-- (r_sr2)
                (r_r) -- (b_r2)
                (b_s) -- (r_sr)
                (b_r) -- (r_id);

    % Black edges
    \draw (b_id) -- (r_r)
                (b_sr) -- (r_s)
                (b_r2) -- (r_id)
                (b_s) -- (r_sr2)
                (b_r) -- (r_r2)
                (b_sr2) -- (r_sr);

    % Blue edges
    \draw (b_id) -- (r_s)
                (b_sr) -- (r_r)
                (b_r2) -- (r_sr2)
                (b_s) -- (r_id)
                (b_r) -- (r_sr)
                (b_sr2) -- (r_r2);

    \node at (90:2.2) {$Z_1$};
    \node at (-30:2.25) {$Z_2$};

    \end{tikzpicture}
    \hspace{25pt}
    \begin{tikzpicture}[scale=1.3, dot/.style={circle, fill= black,inner sep=1pt,font=\small}, baseline = 0cm]

    \node at (90:2.6)  {\underline{\,$\zeta(id,3)$\,}};

    % Outer hexagon
    \node [dot, red] (b_id)  at (90:2) {};
    \node[dot] (r_r2)   at (30:2) {};
    \node [dot] (b_r)  at (-30:2) {};
    \node [dot] (r_id) at (-90:2) {};
    \node[dot] (b_r2)  at (-150:2) {};
    \node [dot] (r_r)  at (150:2) {};

    % Inner hexagon 
    \node[dot] (r_s)   at (90:1) {};
    \node[dot] (b_sr2) at (30:1) {};
    \node[dot] (r_sr)  at (-30:1) {};
    \node [dot] (b_s) at (-90:1) {};
    \node[dot] (r_sr2)  at  (-150:1) {};
    \node[dot] (b_sr) at (150:1) {};

    \node at ($(b_id)!0.5!(r_r2)$) {$Z_3$};
    \node at ($(b_id)!0.5!(r_s)$) {$Z_3$};
    \node at ($(b_id)!0.5!(r_r)$) {$Z_3$};
        
    % Red edges
    \draw (b_id) -- (r_r2)
                (r_s) -- (b_sr2)
                (b_sr)-- (r_sr2)
                (r_r) -- (b_r2)
                (b_s) -- (r_sr)
                (b_r) -- (r_id);

    % Black edges
    \draw (b_id) -- (r_r)
                (b_sr) -- (r_s)
                (b_r2) -- (r_id)
                (b_s) -- (r_sr2)
                (b_r) -- (r_r2)
                (b_sr2) -- (r_sr);

    % Blue edges
    \draw (b_id) -- (r_s)
                (b_sr) -- (r_r)
                (b_r2) -- (r_sr2)
                (b_s) -- (r_id)
                (b_r) -- (r_sr)
                (b_sr2) -- (r_r2);

    \node  at (90:2.2) {$Z_1$};
    \node  at (-90:0.8) {$Z_2$};

    \end{tikzpicture}
    \begin{minipage}{0.95\textwidth}
        \centering
        \caption{Examples of $\zeta(v,i)$. The vertex $v$ are taken to be $id$ in all three cases, which is the top vertex colored red.}
        \label{fig:cay_d6_deform}
    \end{minipage}
\end{figure}
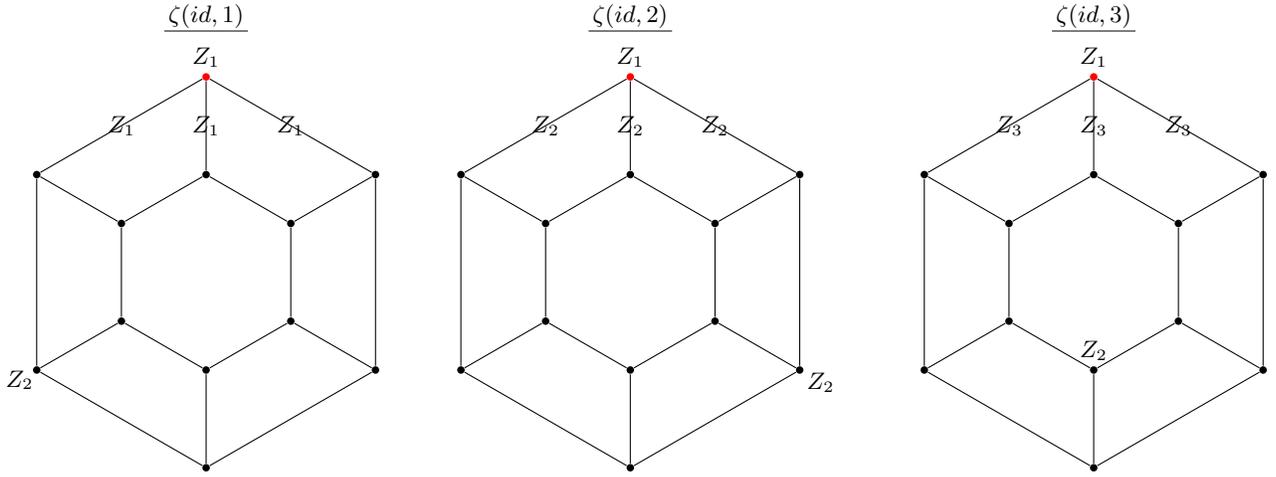

After code switching, the remaining terms in $\mathcal{S}_0$ are
\begin{align*}
    \zeta(v,1) = b_v^{(1)} Z_{v,1} Z_{v\cdot r^2,2},
    \hspace{15pt}
    \zeta(v,2) = b_v^{(2)} Z_{v,1} Z_{v\cdot r,2},
    \hspace{15pt}
    \zeta(v,3) = b_v^{(3)} Z_{v,1} Z_{v\cdot s,2}.
\end{align*}
It is clear that they are indeed products of generators of $\mathcal{S}_0$, and it is not hard to verify that each of them commute with all of $\alpha(e,n)$. Examples of these terms are shown in Figure~\ref{fig:cay_d6_deform}. The stabilizer group $\mathcal{S} = \langle \alpha(e,n), \, \zeta(v, i) \rangle$ form the balanced product $\mathcal{C} \otimes_{D_6} \mathcal{C}^*$.

%============================================
%============================================

\subsection{Color code}
\label{sec:cc}

The color code is expressed as a balanced product of two Newman-Moore models~\cite{tan2025fracton}. Here, we show how our formalism shows that we can obtain the color code by gauging a particular fractal symmetry in the Newman-Moore (NM) model~\cite{NewmanMoore1999}. The NM is a classical code defined on a triangular lattice, with bits on vertices, and parity check for each upward triangle. In $Z$-basis, the stabilizers are
\begin{equation*}
    b_{\triangle} =  \begin{tikzpicture}[baseline = 0.2cm]
        \coordinate (A) at (90:1);
        \coordinate (B) at (-30:1);
        \coordinate (C) at (210:1);

        \foreach \X in {A,B,C}
        \node at (\X) {$Z$};
        
        \draw (A) -- (B) -- (C) -- cycle;
    \end{tikzpicture}
\end{equation*}
In order to construct the color code, we take CSS$_1$ to be the NM model in $Z$-basis, and CSS$_2$ also the NM model but in $X$-basis. Let the group be $G = \mathbb{Z} \otimes \mathbb{Z}$, acting by translation in the directions given by vectors $(-\cos{\pi/3}, -\sin{\pi/3})$  and $(\cos{\pi/3}, -\sin{\pi/3})$.

Next we describe code switching. The quotient code $G\backslash$CSS$_2$ has one orbit for each of qubits and stabilizers. We arbitrarily fix a vertex $\tilde v$ as the qubit representative, and choose the stabilizer representative to be the triangle $\tilde \triangle$, which contains $\tilde v$ as its top vertex, shown visually as
\begin{equation*}
    \begin{tikzpicture}[baseline = 0.2cm]
        \coordinate (A) at (90:1);
        \coordinate (B) at (-30:1);
        \coordinate (C) at (210:1);

        \node at (0:0) {$\tilde \triangle$};

        \node at (90:1.2) {$\tilde v$}; 
        
        \draw (A) -- (B) -- (C) -- cycle;
    \end{tikzpicture}
\end{equation*}
As a result, we only need one copy of CSS$_1$ (NM model in $Z$-basis), and one $Z$-ancilla for each upward triangle. The stacked stabilizer group is thus $\mathcal{S}_0 = \left\langle Z_{\triangle}, b_{\triangle} \right\rangle$, where
\begin{equation*}
    Z_{\triangle} = \begin{tikzpicture}[baseline = 0.2cm]
        \coordinate (A) at (90:1);
        \coordinate (B) at (-30:1);
        \coordinate (C) at (210:1);

        \node at (0:0) {$Z$};

        \draw (A) -- (B) -- (C) -- cycle;
    \end{tikzpicture}
    \hspace{35pt}
    b_{\triangle} =  \begin{tikzpicture}[baseline = 0.2cm]
        \coordinate (A) at (90:1);
        \coordinate (B) at (-30:1);
        \coordinate (C) at (210:1);

        \foreach \X in {A,B,C}
        \node at (\X) {$Z$};
        
        \draw (A) -- (B) -- (C) -- cycle;
    \end{tikzpicture}
\end{equation*}
The code switching enforces $X$-stabilizers $\alpha$, and the remaining commuting terms in $\mathcal{S}_0$ are $\zeta$, which are shown below
\begin{align*}
    \alpha(v, \tilde \triangle) &= \prod_{\triangle \ni v}X_{\triangle} \prod_{v'\in \tilde \triangle}X_{v \cdot g_{v'}}
&
\zeta(\triangle, \tilde v) &= b_\triangle \prod_{\triangle' \ni \tilde v} Z_{\triangle \cdot g_{\triangle'}} \\
    &=
    \begin{tikzpicture}[scale = 1.5, baseline = 0cm]
        \coordinate (A) at (0:1);
        \coordinate (B) at (60:1);
        \coordinate (C) at (120:1);
        \coordinate (D) at (180:1);
        \coordinate (E) at (240:1);
        \coordinate (F) at (300:1);
        \node[circle, fill = red, inner sep = 1.5pt] at (0:0) {};
        \node at (30:0.55) {$X$};
        \node at (150:0.55) {$X$};
        \node at (270:0.55) {$X$};
        \node at (0:0) {$X$};
        \node at (-60:1) {$X$};
        \node at (-120:1) {$X$};
        % Draw boundary
        \draw (A) -- (B) -- (C) -- (D) -- (E) -- (F) -- cycle;
        % Draw diagonal lines
        \foreach \X/\Y in {A/D,B/E,C/F}
        \draw (\X) -- (\Y);
    \end{tikzpicture}
&&
    =    
    \begin{tikzpicture}[scale = 1.5, baseline = 0cm]
        \coordinate (A) at (0:1);
        \coordinate (B) at (60:1);
        \coordinate (C) at (120:1);
        \coordinate (D) at (180:1);
        \coordinate (E) at (240:1);
        \coordinate (F) at (300:1);
        % Draw boundary
        \draw (A) -- (B) -- (C) -- (D) -- (E) -- (F) -- cycle;
        % Draw diagonal lines
        \foreach \X/\Y in {A/D,B/E,C/F}
        \draw (\X) -- (\Y);
        \draw[red] (0:0) -- (E) -- (F) -- cycle;
        \node at (30:0.55) {$Z$};
        \node at (150:0.55) {$Z$};
        \node at (270:0.55) {$Z$};
        \node at (0:0) {$Z$};
        \node at (-60:1) {$Z$};
        \node at (-120:1) {$Z$};
    \end{tikzpicture}
\end{align*}
where in the pictures, the vertex $v$ in $\alpha(v, \tilde \triangle)$ is colored red, as well as the triangle $\triangle$ in $\zeta(\triangle ,\tilde v)$. By shifting each qubit on the vertices upward so that it lies inside a downward triangle, we can see these are exactly have stabilizers of color code.

\begin{figure}[h!]
\begin{minipage}{0.45\textwidth}
    \centering
    \includegraphics[width=0.9\linewidth]{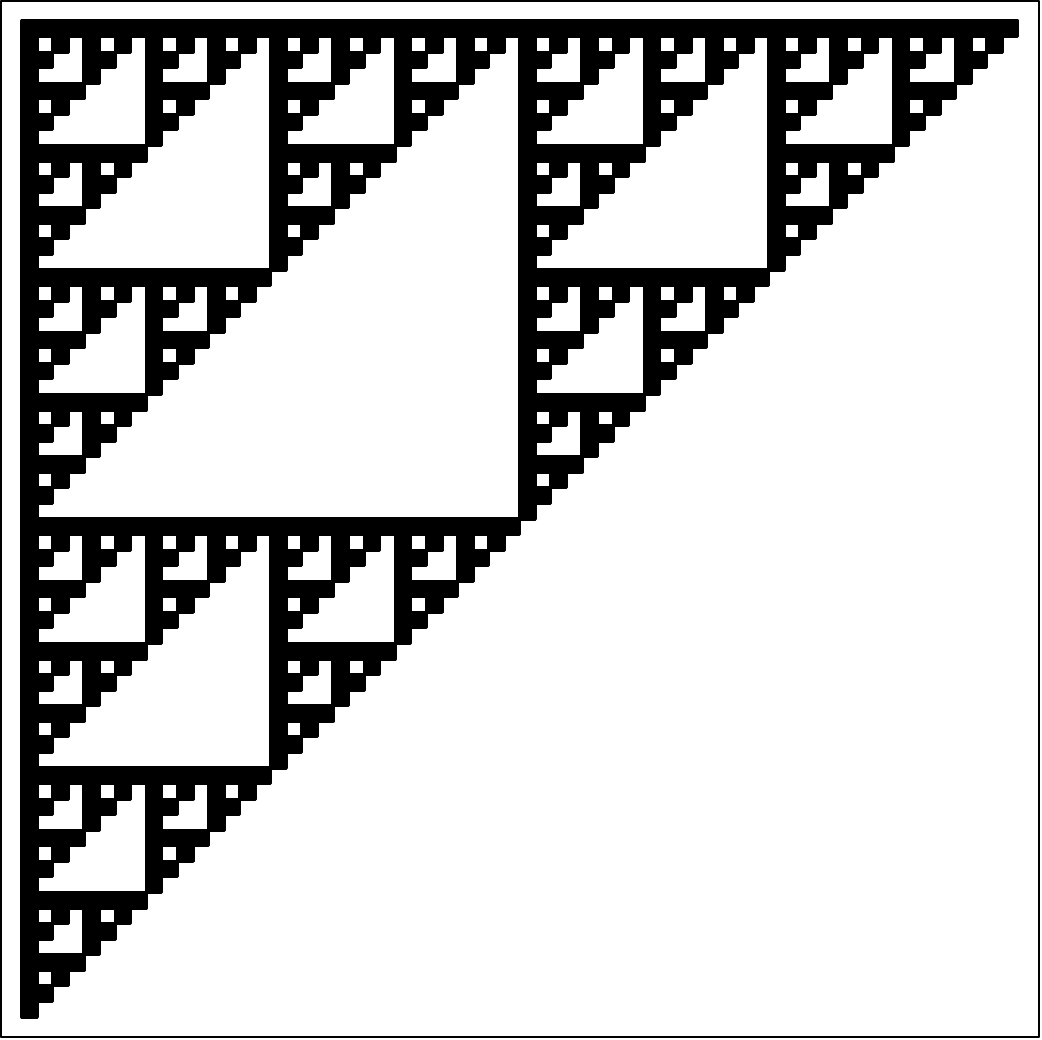}
\end{minipage}
\begin{minipage}{0.45\textwidth}
    \centering
    \includegraphics[width=0.9\linewidth]{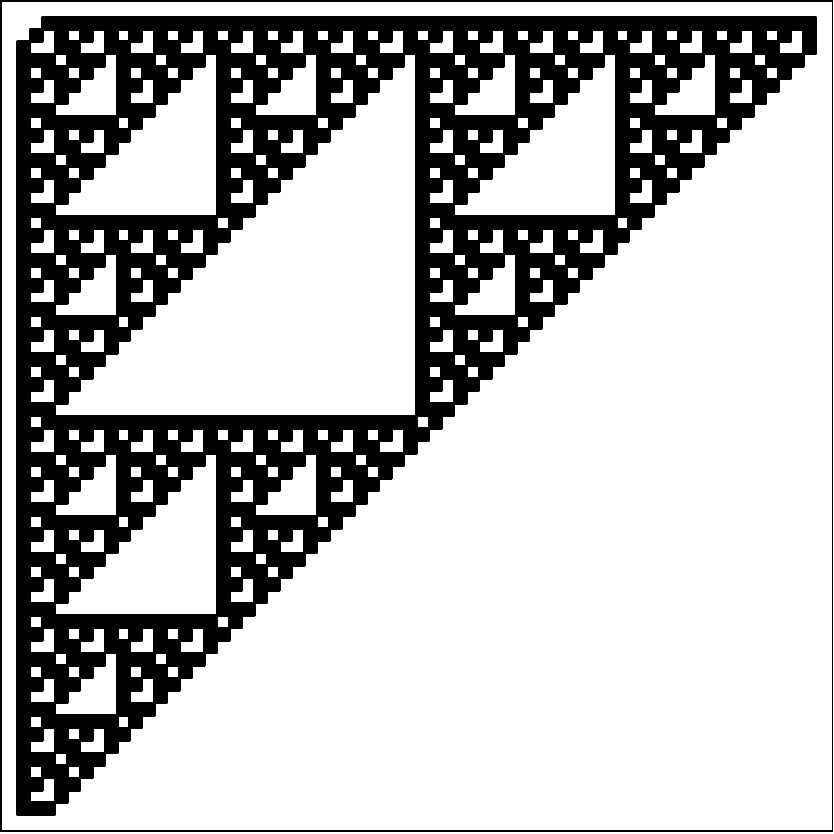}
\end{minipage}
\caption{Left: a logical operator on the NM model. Right: fractal logical in NM model which is gauged to obtain the color code.}
    \label{fig:sierpinski}
\end{figure}

The usual logical generators in the NM model take the shape of the Sierpinski gasket, an example is shown by the left of Figure~\ref{fig:sierpinski}, where each pixel is a qubit, and black qubits indicate the onces the logical acts on. When drawing the figures, we have deformed the triangular lattice to a square lattice, so that the stabilizers are 
\[
\begin{tikzpicture}[scale = 1.732]
        \node at (0,0) {$Z$};
        \node at (1,0) {$Z$};
        \node at (1,1) {$Z$};
        \draw (0,0) -- (1,0) -- (1,1) -- cycle;
\end{tikzpicture}
\]
Gauging all these logicals will take us to a dual NM model in the X basis on all the down triangles. Thus, one can ask  what subset of logicals in the NM model that we are gauging to arrive at the color code in this case. The answer is given by the coupled layer construction. It turns out to be a subgroup generated by the product of three fractal logicals, one at the origin, and the other two are shifted by the inverse translation vectors. An exmaple of the resulting fractal operators is shown in the right of Figure~\ref{fig:sierpinski}.

%============================================
%============================================

\subsection{Haah's code}

We show how Haah's code is expressed as a balanced product between two classical codes. Let CSS$_1$ and CSS$_2$ to be the following codes in $X$-basis and $Z$-basis respectively.
\begin{equation*}
    \text{CSS$_1$}:
    \hspace{5pt} 
    \begin{tikzpicture}[scale = 1, baseline = 0.5cm]
    % Define coordinates
    \coordinate (O) at (0,0,0); %back-bottom-left
    \coordinate (A) at (2,0,0);%back-bottom-right
    \coordinate (B) at (2,2,0);%back-top-right
    \coordinate (C) at (0,2,0);%back-top-left
    \coordinate (D) at (0,0,2);
    \coordinate (E) at (2,0,2);
    \coordinate (F) at (2,2,2);
    \coordinate (G) at (0,2,2);
    
    \node  at (0, 0, 0) {$X$};
    \node  at (2, 0, 0) {$X$};
    \node at (0, 2, 0) {$X$};
    \node at (0, 0, 2) {$X$};

    % Draw edges
    \draw [dotted] (O) -- (A);
    \draw (A) -- (B) -- (C); 
    \draw [dotted] (O) -- (C);% back face
    \draw (D) -- (E) -- (F) -- (G) -- cycle; % front face
    \draw [dotted] (O) -- (D); 
    \draw (A) -- (E);
    \draw (B) -- (F);
    \draw (C) -- (G);% vertical edges
    \end{tikzpicture}
    \hspace{35pt}
    \text{CSS$_2$}:
    \hspace{5pt} 
    \begin{tikzpicture}[scale = 1, baseline = 0.5cm]
    % Define coordinates
    \coordinate (O) at (0,0,0); %back-bottom-left
    \coordinate (A) at (2,0,0);%back-bottom-right
    \coordinate (B) at (2,2,0);%back-top-right
    \coordinate (C) at (0,2,0);%back-top-left
    \coordinate (D) at (0,0,2);
    \coordinate (E) at (2,0,2);
    \coordinate (F) at (2,2,2);
    \coordinate (G) at (0,2,2);
    
    \node  at (2, 2, 2) {$Z$};
    \node  at (2, 0, 0) {$Z$};
    \node at (0, 2, 0) {$Z$};
    \node at (0, 0, 2) {$Z$};

    % Draw edges
    \draw [dotted] (O) -- (A);
    \draw (A) -- (B) -- (C); 
    \draw [dotted] (O) -- (C);% back face
    \draw (D) -- (E) -- (F) -- (G) -- cycle; % front face
    \draw [dotted] (O) -- (D); 
    \draw (A) -- (E);
    \draw (B) -- (F);
    \draw (C) -- (G);% vertical edges
    \end{tikzpicture}
    \hspace{35pt}
    \begin{tikzpicture}[baseline = 0cm]
        \node at (1.2,0,0) {$y$};
        \node at (0,1.2,0) {$z$};
        \node at (0,0,1.3) {$x$};
        \draw[->] (0,0,0) -- (1,0,0);
        \draw[->] (0,0,0) -- (0,1,0);
        \draw[->] (0,0,0) -- (0,0,1);
    \end{tikzpicture}
\end{equation*}
In both cases, each qubit is labeled by a triple of integers $\textbf{n} \in \mathbb{Z}^3$, and each stabilizer (cube) is labeled by the vertex where the dotted lines meet. The group is taken to be $\mathbb{Z}^3$, acting as the full translation symmetry in three dimensions. 

Next we describe code switching. The quotient $G\backslash$CSS$_2$ has one orbit for each of qubits and stabilizers. We choose the representative of qubits to be at origin $\textbf{0} = (0,0,0)$, and the representative of stabilizers to be at $-\textbf{1} = (-1,-1,-1)$. Since there is one orbit of qubits, we introduce one copy of CSS$_1$. Similarly, since there is one orbit of stabilizers, we introduce a $X$-ancilla for each vertex $v$. This gives two qubits per vertex, and we label the qubits in CSS$_1$ as qubit $1$, and the ancillas as qubit $2$. The stabilizer group is
\begin{align*}
    \mathcal{S}_0 = \langle a_v\otimes I, I \otimes X_v \rangle
\end{align*}
where $a_v\otimes I$ means the stabilizer $a_v$ only acts on qubit $1$ at each vertex. The code switching is by enforing the following terms
\begin{equation*}
    \beta(v, -\textbf{1}) 
    =
    \prod_{a_{v'}\ni v} I\otimes Z_{v'} \prod_{v''\in -\textbf{1}} Z_{v\cdot g_{v''}} \otimes I
    \hspace{5pt} 
    =
    \hspace{5pt}
    \begin{tikzpicture}[scale = 1, baseline = 0.5cm]
    % Define coordinates
    \coordinate (O) at (0,0,0); %back-bottom-left
    \coordinate (A) at (2,0,0);%back-bottom-right
    \coordinate (B) at (2,2,0);%back-top-right
    \coordinate (C) at (0,2,0);%back-top-left
    \coordinate (D) at (0,0,2);
    \coordinate (E) at (2,0,2);
    \coordinate (F) at (2,2,2);
    \coordinate (G) at (0,2,2);
    
    \node[circle, fill = red, inner sep = 1.5pt] at (2,2,2) {};
    
    \node  at (2, 2, 2) {$ZZ$};
    \node  at (2, 2, 0) {$IZ$};
    \node at (0, 2, 2) {$IZ$};
    \node at (2, 0, 2) {$IZ$};
    \node at (0, 0, 2) {$ZI$};
    \node at (0, 2, 0) {$ZI$};
    \node at (2, 0, 0) {$ZI$};
    \node at (0, 0, 0) {$II$};

    % Draw edges
    \draw [dotted] (O) -- (A);
    \draw (A) -- (B) -- (C); 
    \draw [dotted] (O) -- (C);% back face
    \draw (D) -- (E) -- (F) -- (G) -- cycle; % front face
    \draw [dotted] (O) -- (D); 
    \draw (A) -- (E);
    \draw (B) -- (F);
    \draw (C) -- (G);% vertical edges
    \end{tikzpicture}
\end{equation*}
where the vertex $v$ is colored red. The remaining term in $\mathcal{S}_0$ is
\begin{equation*}
    \zeta(v, \textbf{0}) = a_v\otimes I \prod_{b_{v'} \ni \textbf{0}}I\otimes  X_{v \cdot g_{b_{v'}}}
    \hspace{5pt} 
    =
    \hspace{5pt}
    \begin{tikzpicture}[scale = 1, baseline = 0.5cm]
    % Define coordinates
    \coordinate (O) at (0,0,0); %back-bottom-left
    \coordinate (A) at (2,0,0);%back-bottom-right
    \coordinate (B) at (2,2,0);%back-top-right
    \coordinate (C) at (0,2,0);%back-top-left
    \coordinate (D) at (0,0,2);
    \coordinate (E) at (2,0,2);
    \coordinate (F) at (2,2,2);
    \coordinate (G) at (0,2,2);
    
    \node[circle, fill = red, inner sep = 1.5pt] at (0,0,0) {};
    
    \node  at (2, 2, 2) {$II$};
    \node  at (2, 2, 0) {$IX$};
    \node at (0, 2, 2) {$IX$};
    \node at (2, 0, 2) {$IX$};
    \node at (0, 0, 2) {$XI$};
    \node at (0, 2, 0) {$XI$};
    \node at (2, 0, 0) {$XI$};
    \node at (0, 0, 0) {$XX$};

    % Draw edges
    \draw [dotted] (O) -- (A);
    \draw (A) -- (B) -- (C); 
    \draw [dotted] (O) -- (C);% back face
    \draw (D) -- (E) -- (F) -- (G) -- cycle; % front face
    \draw [dotted] (O) -- (D); 
    \draw (A) -- (E);
    \draw (B) -- (F);
    \draw (C) -- (G);% vertical edges
    \end{tikzpicture}
\end{equation*}
where the vertex $v$ is colored red. These form the stabilizer group of Haah's code \cite{haah2011local}. 

Similarly to the color code, the logicals of the classical codes are Sierpinski pyramids. The logicals we are gauging in the classical code are products of four logicals of CSS$_1$, where the other three are shifted by the (inverted) position of the checks in CSS$_2$.

%============================================
%============================================

\subsection{Quantum balanced product}

We give an example of the balanced product between two quantum codes. Let CSS$_1$ be 2D TC, and CSS$_2$ be the following $[[2L,2,2]]$ code defined on an $L\times 2$ lattice with periodic boundary conditions in the $L$ direction
\begin{equation*}
    \text{CSS}_2 = 
    \langle a_n ,\, b_n \,|\, n\in \mathbb{Z}
    \rangle
    \hspace{35pt}
    a_n := \hspace{5pt}
    \begin{tikzpicture}
    [baseline=0.8cm]
        \node at (0,-0.5) {$n$};
        \node at (0,0) {$X$};
        \node at (0, 2) {$X$};
        \node at (2,0) {$X$};
        \node at (2, 2) {$X$};
        
        \draw[step = 2] (-0.5,0) grid (2.5,2);
    \end{tikzpicture}
    \hspace{25pt}
    b_n := \hspace{5pt}
    \begin{tikzpicture}
    [baseline=0.8cm]
        \node at (0,-0.5) {$n$};
        \node at (0,0) {$Z$};
        \node at (0, 2) {$Z$};
        \node at (2,0) {$Z$};
        \node at (2, 2) {$Z$};
        
        \draw[step = 2] (-0.5,0) grid (2.5, 2);
    \end{tikzpicture}
\end{equation*}
Let $\mathbb{Z}$ act on both systems, with the generator $T$ acting by translating to the right by one site. First we need to choose representatives of $\mathbb{Z}\backslash$CSS$_2$. The qubits have two orbits. We choose the representatives to be the ones with $2$D coordinates $(0,0)$ and $(0,1)$, and denote them orbit $[1]$ and $[2]$ respectively. Each set of $X$ and $Z$-stabilizers has one orbit, we choose the representatives to be $a_0$ and $b_0$.

\begin{figure}[h!]
    \underline{\,$\alpha(e, [a_0])$\,}
    \hspace{205pt}
    \underline{\,$\beta(e, [b_0])$\,}
    \vspace{0.3cm}
    \\
    \begin{tikzpicture}[scale = 0.9, baseline = 0cm]
        \node at (1,1) {$X$};
        \node at (3,1) {$X$};
        \node at (2,1) {$X_1 X_2$};
        \node at (4,1) {$X_1 X_2$};
        
        \draw[step = 2] (0,0) grid (4,2);
        \draw [red] (2,0) -- (2,2);
    \end{tikzpicture}
    \hspace{15pt}
    \begin{tikzpicture}[baseline = 0cm, scale = 0.9]
        \node at (1,-1) {$X$};
        \node at (1,1) {$X$};
        \node at (1,0) {$X_1 X_2$};
        \node at (3,0) {$X_1 X_2$};
        
        \draw (0, 0) -- (4, 0);
        \draw[step = 2] (0,-2) grid (2,2);
        \draw [red] (0,0) -- (2,0);
    \end{tikzpicture}
    \hspace{15pt}
    \begin{tikzpicture}[baseline = 0.5cm, scale = 0.9]
        \node at (0,0) {$Z$};
        \node at (0, 2) {$Z$};
        \node at (0,1) {$Z_1 Z_2$};
        \node at (2, 1) {$Z_1 Z_2$};
        
        \draw[step = 2] (-0.5,-0.5) grid (2.5,2.5);
        \draw[red] (0,0) -- (0,2);
    \end{tikzpicture}
    \hspace{15pt}
    \begin{tikzpicture}[baseline = -1cm, scale = 0.9]
        \node at (0,0) {$Z$};
        \node at (2, 0) {$Z$};
        \node at (1,0) {$Z_1 Z_2$};
        \node at (3, 0) {$Z_1 Z_2$};
        
        \draw[step = 2] (-0.5,-0.5) grid (4.5,0.5);
        \draw[red] (0,0) -- (2,0);
    \end{tikzpicture}
    \begin{minipage}{0.95\textwidth}
        \caption{Left two figures are $\alpha(e, [a_0])$. Right two figures are $\beta(e, [b_0])$. The edge $e$ is colored red in both cases.}
        \label{fig:balan_prod_quant_cond}
    \end{minipage}
\end{figure}
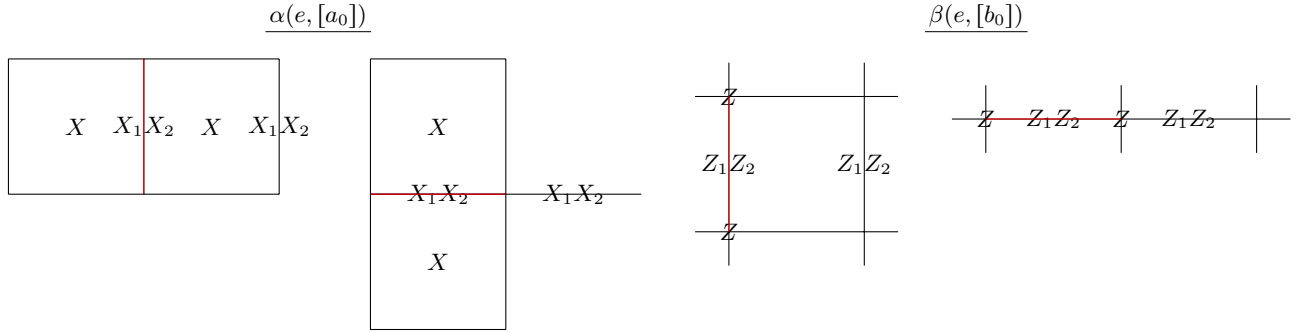

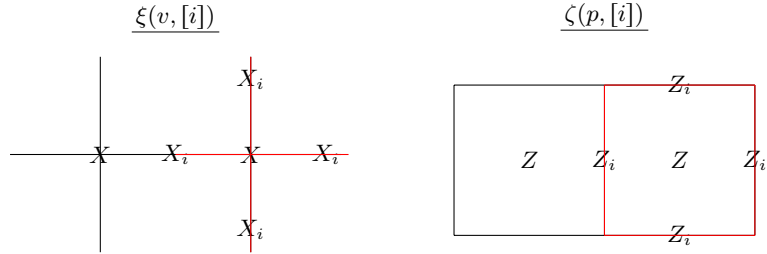
\begin{figure}[h!]
    \begin{tikzpicture}
        \node at (-2,0) {$X$};
        \node at (0,0) {$X$};
        \node at (-1,0) {$X_i$};
        \node at (1,0) {$X_i$};
        \node at (0,-1) {$X_i$};
        \node at (0,1) {$X_i$};

        \node at (-1, 1.8) {\underline{\,$\xi(v,[i])$\,}};
        
        \draw [step = 2] (-3.2,-1.3) grid (1.2,1.3);
        \draw [step = 2, red] (-1,-1.3) grid (1.3,1.3);
    \end{tikzpicture}
    \hspace{35pt}
    \begin{tikzpicture}
        \node at (1,1) {$Z$};
        \node at (3,1) {$Z$};
        \node at (3,0) {$Z_i$};
        \node at (3,2) {$Z_i$};
        \node at (2,1) {$Z_i$};
        \node at (4,1) {$Z_i$};

        \node at (2, 2.9) {\underline{\,$\zeta(p,[i])$\,}};

        \draw [step = 2] (0,0) grid (4,2);
        \draw [step = 2, red] (2,0) grid (4,2);
    \end{tikzpicture}
    \begin{minipage}{0.95\textwidth}
        \caption{Left: $X$-stabilizer $\xi(v,[i])$ with the vertex $v$ highlighted in red. Right: $Z$-stabilizer $\zeta(p,[i])$ with the plaquette $p$ highlighted in red.}
        \label{fig:balan_prod_quant_post_cond}
    \end{minipage}
\end{figure}

Now we describe the coupled layer construction. For each orbit $[1]$, $[2]$ of qubits in CSS$_2$, we stack a copy of TC, so that there are two qubits per edge. Since there is one orbit $[a_0]$ for $X$-stabilizers, we introduce a $Z$-ancilla living on the plaquettes. Similarly, the one orbit $[b_0]$ associates an $X$-ancilla living on vertices. The stacked system is
\begin{align*}
    \mathcal{S}_0 = \{a_v^{(i)}, b_p^{(i)}, X_v, Z_p\}
    \hspace{15pt}
    i = 1,2
\end{align*}
Next we perform code switching. The $X$-terms are
\begin{align*}
    \alpha(e,[a_0]) = X_{e,1} X_{e,2} X_{e \cdot T, 1} X_{e \cdot T, 2} \prod_{p \ni e} X_{p},
    \hspace{35pt}
    \beta(e,[b_0]) = Z_{e,1} Z_{e,2} Z_{T\cdot e, 1} Z_{T\cdot e, 2} \prod_{v \ni e} Z_{v}.
\end{align*}
Diagrammatically, these are given in Figure~\ref{fig:balan_prod_quant_cond}. After code switching, the following terms in $\mathcal{S}_0$ remain
\begin{align*}
    \xi(v,[i]) = a_{v}^{(i)} X_v X_{v\cdot T^{-1}}
    \hspace{35pt}
    \zeta(p,[i]) = b_{p}^{(i)} Z_p Z_{p \cdot T^{-1}}
    \hspace{20pt}
    i = 1,2
\end{align*}
Diagrammatically these are given in Figure~\ref{fig:balan_prod_quant_post_cond}. One can check directly these terms indeed commute. 

Using the K\"unneth formula \cite{breuckmann2021balanced}, one concludes there are $6$ qubits in the ground state on a torus. Note to apply Lemma $19$ in \cite{breuckmann2021balanced}, we need to take $L$ to be odd, and quotient by the group $\mathbb{Z}_L$. We can write down the $Z$-logicals in the following way. Choose a vertical non-contactible loop $l_v$ and a horizontal non-contractible loop $l_h$ on the direct lattice, for each $i = 1,2$, we have
\begin{equation*}
    \mathcal{Z}_{l_v,i}
    =\,
    \begin{tikzpicture}[baseline = 0cm]
        \node at (0,1) {$Z_i$};
        \node at (0,-1) {$Z_i$};
        \node at (-1,1) {$Z$};
        \node at (-1,-1) {$Z$};
        \node at (-0.8, 2.8) {$\vdots$};
        \node at (-0.8, -2.8) {$\vdots$};

        \node at (0, 2.8) {$l_v$};
        
        \draw (0, -2.5) -- (0, 2.5)
        (-1.5,-2) -- (0,-2)
        (-1.5,0) -- (0,0)
        (-1.5,2) -- (0,2);
    \end{tikzpicture}
    \hspace{45pt}
    \mathcal{Z}_{l_h,i}
    =\,
    \begin{tikzpicture}[baseline = 0cm]
        \node at (-1,0) {$Z_i$};
        \node at (1,0) {$Z_i$};
        \node at (-2.9, 0) {$\cdots$};
        \node at (2.9, 0) {$\cdots$};

        \node at (3.5, 0) {$l_h$};
        
        \draw (-2.5, 0) -- (2.5, 0)
        (-2,-0.5) -- (-2,0.5)
        (0,-0.5) -- (0,0.5)
        (2,-0.5) -- (2,0.5);
    \end{tikzpicture}
\end{equation*}
These operators are topological. They can be deformed using the stabilizers $\xi(p, [i])$. More generally, given a loop $l$ on the direct lattice. Denote $H(l)$ by the horizontal edges in $l$, and $V(l)$ by the vertical edges in $l$. For a vertical edge $e$, let $p_e$ denote the plaquette directly to the left of $e$. Then the following is a logical operator
\begin{align*}
    \mathcal{Z}_{l,i} = \prod_{e\in H(l)} Z_{e,i} \prod_{e\in V(l)} (Z_{e,i} Z_{p_e})
\end{align*}
In particular, the stabilizers $\zeta(p,[i])$ are contractible loops. Furthermore, given a vertex $v$, we have the following logicals
\begin{equation*}
    \mathcal{Z} = \prod_{p} Z_p
    \hspace{35pt}
    \mathcal{Z}_v = 
    \begin{tikzpicture}[baseline = 0cm]
        \node at (0,0) {$Z$};
        \node at (1,0) {$Z_1Z_2$};
        
        \draw (-0.8, 0) -- (2,0)
        (0, -0.8) -- (0,0.8);
    \end{tikzpicture}
\end{equation*}
The choice of $v$ in $\mathcal{Z}_v$ is irrelevent. $v$ can be moved horizontally using the stabilizers $\beta(e,[b_0])$, and vertically using a combination of $\beta(e, [b_0])$ and $\prod_{i=1,2}\zeta(p, [i])$. Together, we have $6$ $Z$-logicals
\begin{align*}
    \mathcal{L}_Z = \{\mathcal{Z}_{l_v, \,1}, \mathcal{Z}_{l_h, \,1},
    \mathcal{Z}_{l_v, \,2}, \mathcal{Z}_{l_h, \,2},
    \mathcal{Z}, \mathcal{Z}_v\}
\end{align*}

The conjugate $X$-logicals can be written down similarly. Choose a vertical non-contactible loop $l_v^*$, a horizontal non-contractible loop $l_h^*$ on the dual lattice, for each $i = 1,2$ we have
\begin{equation*}
    \mathcal{X}_{l_v^*,i}
    =\,
    \begin{tikzpicture}[baseline = 0cm]
        \node at (1,2) {$X_i$};
        \node at (1,0) {$X_i$};
        \node at (1,-2) {$X_i$};
        \node at (0,2) {$X$};
        \node at (0,0) {$X$};
        \node at (0,-2) {$X$};
        
        \node at (1, 3.2) {$\vdots$};
        \node at (1, -3.0) {$\vdots$};

        \node at (0.7, 2.9) {$l_v^*$};
        
        \draw (0, -2.5) -- (0, 2.5)
        (2, -2.5) -- (2, 2.5)
        (0,-2) -- (2,-2)
        (0,0) -- (2,0)
        (0,2) -- (2,2);
        \draw [dotted] (1,-2.8) -- (1,2.8);
    \end{tikzpicture}
    \hspace{45pt}
    \mathcal{X}_{l_h^*,i}
    =\,
    \begin{tikzpicture}[baseline = 1cm]
        \node at (-2,1) {$X_i$};
        \node at (0,1) {$X_i$};
        \node at (2,1) {$X_i$};
        
        \node at (-3.2, 1) {$\cdots$};
        \node at (3.2, 1) {$\cdots$};

        \node at (2.7, 0.7) {$l_h^*$};
        
        \draw (-2.5, 0) -- (2.5, 0)
        (-2.5, 2) -- (2.5, 2)
        (-2,0) -- (-2,2)
        (0,-0) -- (0,2)
        (2,0) -- (2,2);
        \draw [dotted] (-2.8,1) -- (2.8,1);
    \end{tikzpicture}
\end{equation*}
Choose a plaquette $p$, we have
\begin{equation*}
    \mathcal{X} = \prod_{v} X_v
    \hspace{35pt}
    \mathcal{X}_p = 
    \,
    \begin{tikzpicture}[baseline = 1cm]
        \node at (1,1) {$X$};
        \node at (2,1) {$X_1X_2$};
        
        \draw [step = 2] (0,0) grid (2,2);
    \end{tikzpicture}
\end{equation*}
Similar to the $Z$-logicals, we can define an $X$-logical for each loop $l^*$ on the dual lattice, which is topological
\begin{align*}
    \mathcal{X}_{l^*,i} = \prod_{e\in V(l^*)} X_{e,i}
    \prod_{e\in H(l^*)} (X_{e,i} X_{v_e})
\end{align*}
where $v_e$ is the vertex to the immediate left of the horizontal edge $e$. The $X$-logicals and $Z$-logicals form conjugate pairs
\begin{align*}
    \{\mathcal{Z}_{l_h,i} ,\,  \mathcal{X}_{l^*_v,i}\}
    \hspace{15pt}
    \{\mathcal{Z}_{l_v,i} ,\,  \mathcal{X}_{l^*_h,i}\}
    \hspace{15pt}
    \{\mathcal{Z}, \mathcal{X}_p\}
    \hspace{15pt}
    \{\mathcal{Z}_v, \mathcal{X}\}
\end{align*}
One can check that each logical commutes with all other logicals except the one which it conjugates to.

From the form of logical operators, it is easy to apply finite-depth local unitary operators to disentangle the stabilizer group into a direct sum of two copies of TC, and two copies of 2D Ising. Let $\mathcal{H}$ and $\mathcal{V}$ denote the set of all horizontal edges, and vertical edges respectively. As defined above, for a horizontal edge (resp. vertical edge) $e$, we use the notation $v_e$ (resp. $p_e$) to denote the vertex (resp. plaquette) to the immediate left of $e$. We use $e(1)$ and $e(2)$ to denote the first and second qubit on edge $e$. Consider the unitary
\begin{align*}
    U = \prod_{e\in \mathcal{V}} \text{CNOT}_{p_e,e(1)}
    \text{CNOT}_{p_e,e(2)}
    \prod_{e\in \mathcal{H}}
    \text{CNOT}_{e(1),v_e}
    \text{CNOT}_{e(2),v_e}
\end{align*}
Conjugating by this unitary, we see for a horizontal edge $e\in \mathcal{H}$,
\begin{equation*}
    U \alpha(e, [a_0]) U^\dagger = \,
    \begin{tikzpicture}[baseline = 0cm, scale = 1]
        \node at (1,-1) {$X$};
        \node at (1,1) {$X$};
        \node at (1,0) {$X_1 X_2$};
        \node at (3,0) {$X_1 X_2$};
        \node at (2,1) {$X_1 X_2$};
        \node at (2,-1) {$X_1 X_2$};
        
        \draw (0, 0) -- (4, 0);
        \draw[step = 2] (0,-2) grid (2,2);
        \draw [red] (0,0) -- (2,0);
    \end{tikzpicture}
    \hspace{35pt}
    U \beta(e, [b_0]) U^\dagger = \,
    \begin{tikzpicture}[baseline = -0.1cm, scale = 1]
        \node at (0,0) {$Z$};
        \node at (2, 0) {$Z$};
        
        \draw[step = 2] (-0.5,-0.5) grid (2.5,0.5);
        \draw[red] (0,0) -- (2,0);
    \end{tikzpicture}
\end{equation*}
For a vertical edge $e\in \mathcal{V}$,
\begin{equation*}
    U \alpha(e, [a_0]) U^\dagger = \,
    \begin{tikzpicture}[scale = 1, baseline = 0.9cm]
        \node at (1,1) {$X$};
        \node at (3,1) {$X$};
        
        \draw[step = 2] (0,0) grid (4,2);
        \draw [red] (2,0) -- (2,2);
    \end{tikzpicture}
    \hspace{35pt}
    U \beta(e, [b_0]) U^\dagger = \,
    \begin{tikzpicture}[baseline = 1cm, scale = 0.9]
        \node at (0,0) {$Z$};
        \node at (0, 2) {$Z$};
        \node at (0,1) {$Z_1 Z_2$};
        \node at (2, 1) {$Z_1 Z_2$};
        \node at (1,0) {$Z_1 Z_2$};
        \node at (1,2) {$Z_1 Z_2$};        
        \draw[step = 2] (-0.5,-0.5) grid (2.5,2.5);
        \draw[red] (0,0) -- (0,2);
    \end{tikzpicture}
\end{equation*}
Finally,
\begin{equation*}
    U \xi(v, [i]) U^\dagger = \,
    \begin{tikzpicture}[baseline = 0cm]
        \node at (-0.9,0) {$X_i$};
        \node at (0.9,0) {$X_i$};
        \node at (0,-0.9) {$X_i$};
        \node at (0,0.9) {$X_i$};
        
        \draw [step = 2] (-1.3,-1.3) grid (1.3,1.3);
    \end{tikzpicture}
    \hspace{35pt}
    U \beta(p, [i]) U^\dagger = \,
    \begin{tikzpicture}[baseline = 1cm]
        \node at (3,0) {$Z_i$};
        \node at (3,2) {$Z_i$};
        \node at (2,1) {$Z_i$};
        \node at (4,1) {$Z_i$};

        \draw [step = 2] (2,0) grid (4,2);
    \end{tikzpicture}
\end{equation*}
These clearly form the stabilizer group of two decoupled copies of TC and two 2D Ising.

\section{Concatenated codes, subsystem tensor product code, and their balancing}
\label{sec:concat}

In this section, we provide a coupled layer construction for concatenated codes, and show how the tensor product code solves the problem of large weight logicals in concatenated codes by gauging them. We also make a connection to the subsystem tensor (homological) product code~\cite{ZengPryadko20}. We note that similar relations have also been pointed out in~\cite{Tillich:2013esj,LiYoder20}. Finally, by balancing, we introduce a new family of codes, which we call the balanced concatenated code, and the subsystem balanced product code.

\subsection{Relation to concatenated codes and the subsystem tensor product code}

 We will restrict our interest to CSS codes. A concatenated code can be constructed from two codes called the outer code CSS$_1 = [[n_1,k_1,d_1]]$ and an inner code CSS$_2 = [[n_2,k_2,d_2]]$ by replacing the physical qubits of the inner code with the logical qubits of the outer code. The resulting concatenated code has parameters $[[n_1n_2,k_1k_2,d \ge d_1d_2]]$, although the resulting code is not symmetric under swapping the inner and outer code. By abuse of notation, we will denote the resulting concatenated code CSS$_1 \circ$ CSS$_2$. The symbol $\circ$ is only to denote the fact that concatenation is not symmetric.

 We begin by reiterating an example from the main text. Let CSS$_1$ be the Z-type repetition code and CSS$_2$ be the X-type repetition code. The concatenated code CSS$_1 \circ$ CSS$_2$ is a generalization of Shor's code, also known as the quantum parity code, which contain large weight $X$-stabilizers. Alternatively CSS$_2 \circ$ CSS$_1$ has large weight $Z$-stabilizers.

 There are two solutions to remove the need of measuring these large weight stabilizers. One is to gauge these stabilizers. This is performed by introducing gauge fields (qubits on the vertical edges in Fig~\ref{fig:Shorfull}), each initialized in the stabilizer $Z$. Next, we measure the Gauss law for these long stabilizers, these turn out to be exactly the plaquette terms of the surface code. the $ZZ$ checks of Shor's code gets minimally coupled to the gauge fields, resulting in the vertex terms of the surface code. The resulting code is nothing but the tensor product code, which is the surface code. 
 
 Alternatively, one can measure short versions of these large stabilizers. These corresponds to the Gauss law terms mentioned previously, but without introducing gauge fields. The short $XX$-measurements actually don't commute with the $ZZ$ stabilizers of Shor's code. Thus, the resulting code is actually a subsystem code, the Bacon-Shor code\cite{bacon2006operator,poulin2005stabilizer}. The two concatenated codes correspond to gauge-fixed versions of the Bacon-Shor code. We will denote the resulting subsystem code  CSS$_1 \otimes_\text{sub}$CSS$_2$, and call it the subsystem tensor product code~\cite{bacon2006quantum,ZengPryadko20,LiYoder20}. The relation between the two concatenated code, the tensor product code, and the subsystem tensor product code is shown in Fig.~\ref{fig:Shorfull}. 

 \begin{figure}
     \centering
     \includegraphics[width=\linewidth]{figures/Shor5.pdf}
     \caption{When CSS$_1$ and CSS$_2$ correspond to the $Z$-type and $X$-type repetition codes, the resulting concatenated code, tensor product code, and subsystem tensor product code, correspond to Shor's code, the surface code, and the Bacon-Shor code, respectively. All codes share a logical representative given by a row of $X$ (faint red) and a column of $Z$ (faint blue).}
     \label{fig:Shorfull}
 \end{figure}

We now present the coupled-layer construction of the concatenated code for general input CSS codes. We start with $n_2$ copies of the outer code CSS$_1 \otimes Q_2$. Denote the corresponding stabilizer group $\mathcal S_\text{outer,1} = \left \langle a_1^{(q_2)}, b_1^{(q_2)}\right\rangle $, where we recall that $a_1^{(q_2)} = \prod_{q_1\in a_1} X_{q_1, q_2}$ and $b_1^{(q_2)} = \prod_{q_1\in b_1} Z_{q_1, q_2}$. Then, we add $\mathcal S_\text{inner,2}$, which consists of combination of logicals of the outer code using the pattern of checks in CSS$_2$. To be precise, let $L_{X,1}$ and $L_{Z,1}$ be the set of logical operators in CSS$_1$ respectively, we have
\begin{align*}
    \mathcal{S}_\text{inner,2} = \left \langle 
    \prod_{q_2\in a_2} \mathcal{X}^{(q_2)},\,
    \prod_{q_2\in b_2} \mathcal{Z}^{(q_2)}
    \, | \,
    \mathcal{X}\in L_{X,1}, \,
    \mathcal{Z}\in L_{Z,1}, \,
    a_2 \in A_2, \,
    b_2\in B_2
   \right \rangle 
\end{align*}
Since the two stabilizers commute, the resulting code CSS$_1 \circ$ CSS$_2$ has stabilizer group $\mathcal S_\text{outer,1} \cup \mathcal S_\text{inner,2}$. Note that when CSS$_1$ is a $X$-type classical code, and CSS$_2$ is a $Z$-type classical code, the resulting concatenated code is also known as the generalized Shor code. Gauging the stabilizers in $\mathcal{S}_\text{inner,2}$ will result in the tensor product code.

If we swapped the roles of codes $1$ and $2$, we would instead have the concatenated code CSS$_2 \circ$ CSS$_1$ with stabilizers $\mathcal S_\text{outer,2} \cup \mathcal S_\text{inner,1}$ where
\begin{align}
    \mathcal S_\text{outer,2} &= \left \langle a_2^{(q_1)}, b_2^{(q_1)}\right\rangle\\
    \mathcal{S}_\text{inner,1} &= \left \langle 
    \prod_{q_1\in a_1} \mathcal{X}^{(q_1)},\,
    \prod_{q_1\in b_1} \mathcal{Z}^{(q_1)}
    \,|\,
    \mathcal{X}\in L_{X,2}, \,
    \mathcal{Z}\in L_{Z,2}, \,
    a_1 \in A_1, \,
    b_1\in B_1
   \right \rangle 
\end{align}
It is worth noting that this perspective is closely related to an example of code surgery, termed full block reading \cite{cowtan2025fast}. In full block reading, taking CSS$_1$ as the outer code, we start with layers of CSS$_1$ indexed by qubits of CSS$_2$. For each $X$-check $a_2$ ($Z$-check $b_2$) of CSS$_2$, we measure the product of $X$-logicals ($Z$-logicals) on all logical qubits in layers $q_2\in a_2$ ($q_2\in b_2$). That is, we measure $\prod_{q_2\in a_2} \bar X_i^{(q_2)}$ ($\prod_{q_2\in b_2} \bar Z_i^{(q_2)}$), for $i = 1, \cdots k_2$, where $\bar X_i^{(q_2)}$ is the logical $X$ acting on the $i$th logical qubit in layer $q_2$. At the level of stabilizer group, this coincides with code concatenation, and the surgery scheme in full block reading coincides with the gauging procedure we have introduced.

Now, let us define the subsystem tensor product code. Starting with the concatenated code with stabilizers $\mathcal S_\text{outer,1} \cup \mathcal S_\text{inner,2}$, we promote it to the check group $\mathcal G$ of a subsystem code, and add $S_\text{outer,2}$ to the check group. Note that $S_\text{outer,2}$ generates $\mathcal S_\text{inner,2}$, but nevertheless does not commute with $\mathcal S_\text{outer,1}$. Therefore, the resulting check group of the subsystem code is generated by the outer stabilizers of the two concatenated codes
\begin{align*}
    \mathcal{G} = \langle i,\,\mathcal S_\text{outer,1}, \,\mathcal S_\text{outer,2}
    \rangle
    =   \left\langle i,a_1^{(q_2)}, b_1^{(q_2)},  a_2^{(q_1)},  b_2^{(q_1)} \right\rangle,
\end{align*}
while resulting stabilizer group, which is the center of the check group (modulo $i$), is generated by the inner stabilizers of the two concatenated codes
\begin{align*}
    \mathcal Z(\mathcal{G}) = 
    \langle i, 
    \mathcal S_\text{inner,1}, \mathcal S_\text{inner,2}
    \rangle
\end{align*}

\subsection{Example: Toric code $\otimes [[4,2,2]]$}
To give an example of two quantum codes, we revisit the example where CSS$_1$ is the toric code and CSS$_2$ is the $[[4,2,2]]$ code. First, we consider the concatenated code CSS$_1 \circ$ CSS$_2$.  Start with four copies of the toric code
\begin{align*}
    \mathcal S_\text{outer,1} =  \left \langle   \begin{tikzpicture}
    [baseline=0ex]
        \node at (-0.9, 0) {$X_i$};
        \node at (0.9,0) {$X_i$};
        \node at (0, 0.9) {$X_i$};
        \node at (0,-0.9) {$X_i$};
        \draw[step = 2] (-1.3,-1.3) grid (1.3,1.3);
    \end{tikzpicture}
    \hspace{5pt}
    ,
    \hspace{5pt}
    \begin{tikzpicture}
    [baseline=5ex]
        \node at (0, 1) {$Z_i$};
        \node at (2,1) {$Z_i$};
        \node at (1, 0) {$Z_i$};
        \node at (1,2) {$Z_i$};
        \draw[   step = 2] (0,0) grid (2,2);
    \end{tikzpicture} 
    \hspace{5pt}
    \Bigg |
    \hspace{5pt}
    i = 1, \cdots , 4
    \right \rangle
\end{align*}
and add the following product of toric code logicals to the stabilizer group
\begin{align*}
    \mathcal S_\text{inner,2} =  \left \langle 
    \hspace{5pt}
    \begin{tikzpicture}[baseline = 1cm]
        \node at (1,2) {$\displaystyle\prod_{i=1}^4 X_i$};
        \node at (1,0) {$\displaystyle\prod_{i=1}^4 X_i$};
        \node at (1, 3.2) {$\vdots$};
        \node at (1, -1.0) {$\vdots$};
        % %
        % \node at (0.7, 2.9) {$l_v^*$};
        %
        \draw (0, -0.5) -- (0, 2.5)
        (2, -0.5) -- (2, 2.5)
        (0,0) -- (2,0)
        (0,2) -- (2,2);
        \draw [dotted] (1,-0.8) -- (1,2.8);
    \end{tikzpicture}
    \hspace{5pt}
    ,
    \hspace{5pt}    \begin{tikzpicture}[baseline = 1cm]
        \node at (-2,1) {$\displaystyle\prod_{i=1}^4 X_i$};
        \node at (0,1) {$\displaystyle\prod_{i=1}^4 X_i$};
        \node at (-3.2, 1) {$\cdots$};
        \node at (1.2, 1) {$\cdots$};
% %
%         \node at (2.7, 0.7) {$l_h^*$};
        %
        \draw (-2.5, 0) -- (0.5, 0)
        (-2.5, 2) -- (0.5, 2)
        (-2,0) -- (-2,2)
        (0,-0) -- (0,2);
        \draw [dotted] (-2.8,1) -- (0.8,1);
    \end{tikzpicture}
    \hspace{5pt}
    ,
    \hspace{5pt}    \begin{tikzpicture}[baseline = 0cm]
        \node at (0,1) {$\displaystyle\prod_{i=1}^4Z_i$};
        \node at (0,-1) {$\displaystyle\prod_{i=1}^4Z_i$};
        \node at (0, 2.85) {$\vdots$};
        \node at (0, -2.7) {$\vdots$};
% %
%         \node at (0, 2.8) {$l_v$};
        %
        \draw (0, -2.5) -- (0, 2.5)
        (-0.5,-2) -- (0.5,-2)
        (-0.5,0) -- (0.5,0)
        (-0.5,2) -- (0.5,2);
    \end{tikzpicture}
    \hspace{5pt}
    ,
    \hspace{5pt}
    \begin{tikzpicture}[baseline = 0cm]
        \node at (-1,0) {$\displaystyle\prod_{i=1}^4Z_i$};
        \node at (1,0) {$\displaystyle\prod_{i=1}^4Z_i$};
        \node at (-2.75, 0) {$\cdots$};
        \node at (2.75, 0) {$\cdots$};
% %
%         \node at (3.5, 0) {$l_h$};
        %
        \draw (-2.5, 0) -- (2.5, 0)
        (-2,-0.5) -- (-2,0.5)
        (0,-0.5) -- (0,0.5)
        (2,-0.5) -- (2,0.5);
    \end{tikzpicture}
    \right \rangle
\end{align*}
since they commute, the final stabilizer group is trivially $\mathcal S_\text{outer,1} \cup \mathcal S_\text{inner,2}$. Note that alternatively we can also start with the $[[4,2,2]]$ code on each edge, then add $\mathcal S_\text{outer}$ to the stabilizer group. The remaining stabilizers will be $\mathcal S_\text{inner}$. However, we will not use this construction in the following discussion.
%, which we can clearly see that is obtained by replacing each $X$ and $Z$ in the outer code \SZ{inner code?} with the stabilizers of the inner code \SZ{logicals of outer code?}. 

On the other hand, swapping the roles of the two codes, we obtain
\begin{align*}    \mathcal S_\text{outer,2}&=\left \langle
    \begin{tikzpicture}[baseline = -0.1cm, scale = 1]
        \node at (1,0) {$\displaystyle\prod_{i=1}^4 X$};
        \draw[step = 2] (-0.5,-0.5) grid (2.5,0.5);
    \end{tikzpicture}
    \hspace{5pt}
    ,
    \hspace{5pt}
    \begin{tikzpicture}[baseline = 0.8cm, scale = 1]
        \node at (0,1) {$\displaystyle\prod_{i=1}^4 X$};
        \draw[step = 2] (-0.5,-0.5) grid (0.5,2.5);
    \end{tikzpicture}
    \hspace{5pt}
    ,
    \hspace{5pt}
    \begin{tikzpicture}[baseline = -0.1cm, scale = 1]
        \node at (1,0) {$\displaystyle\prod_{i=1}^4 Z$};
        \draw[step = 2] (-0.5,-0.5) grid (2.5,0.5);
    \end{tikzpicture}
    \hspace{5pt}
    ,
    \hspace{5pt}
    \begin{tikzpicture}[baseline = 0.8cm, scale = 1]
        \node at (0,1) {$\displaystyle\prod_{i=1}^4 Z$};
        \draw[step = 2] (-0.5,-0.5) grid (0.5,2.5);
    \end{tikzpicture}
    \right \rangle\\
    \mathcal S_\text{inner,1} &=  \left \langle \begin{tikzpicture}
    [baseline=0ex]
        \node at (-0.9, 0) {$X_1X_2$};
        \node at (0.9,0) {$X_1X_2$};
        \node at (0, 0.9) {$X_1X_2$};
        \node at (0,-0.9) {$X_1X_2$};
        \draw[step = 2] (-1.3,-1.3) grid (1.3,1.3);
    \end{tikzpicture}
    \hspace{5pt}
    ,
    \hspace{5pt}\begin{tikzpicture}
    [baseline=0ex]
        \node at (-0.9, 0) {$X_1X_3$};
        \node at (0.9,0) {$X_1X_3$};
        \node at (0, 0.9) {$X_1X_3$};
        \node at (0,-0.9) {$X_1X_3$};
        \draw[step = 2] (-1.3,-1.3) grid (1.3,1.3);
    \end{tikzpicture}
    \hspace{5pt}
    ,
    \hspace{5pt}
    \begin{tikzpicture}
    [baseline=5ex]
        \node at (0, 1) {$Z_1Z_2$};
        \node at (2,1) {$Z_1Z_2$};
        \node at (1, 0) {$Z_1Z_2$};
        \node at (1,2) {$Z_1Z_2$};
        \draw[step = 2] (0,0) grid (2,2);
    \end{tikzpicture}\hspace{5pt}
    ,
    \hspace{5pt}
    \begin{tikzpicture}
    [baseline=5ex]
        \node at (0, 1) {$Z_1Z_3$};
        \node at (2,1) {$Z_1Z_3$};
        \node at (1, 0) {$Z_1Z_3$};
        \node at (1,2) {$Z_1Z_3$};
        \draw[step = 2] (0,0) grid (2,2);
    \end{tikzpicture}
    \right \rangle
\end{align*}
which has the same code parameters, and the same form of logical operators, but are realized by different stabilizer groups.

Now, we discuss the subsystem tensor product code. The check group is
\begin{align}
    \mathcal G = \mathcal S_\text{outer,1} \cup \mathcal S_\text{outer,2} = \left \langle
        \begin{tikzpicture}
    [baseline=0ex]
        %\node at (0,1.8) {\underline{$a_v^{(i)} = \prod_{e\in v}X_{e,i}$}};
        %
        \node at (-0.9, 0) {$X_i$};
        \node at (0.9,0) {$X_i$};
        \node at (0, 0.9) {$X_i$};
        \node at (0,-0.9) {$X_i$};
        \draw[step = 2] (-1.3,-1.3) grid (1.3,1.3);
    \end{tikzpicture} ,\begin{tikzpicture}
    [baseline=1cm]
        %\node at (1,2.8) {\underline{$b_p^{(i)} = \prod_{e\in p}Z_{e,i}$}};
        %
        \node at (0, 1) {$Z_i$};
        \node at (2,1) {$Z_i$};
        \node at (1, 0) {$Z_i$};
        \node at (1,2) {$Z_i$};
        \draw[   step = 2] (0,0) grid (2,2);
    \end{tikzpicture} ,
    \begin{tikzpicture}[baseline = -0.1cm, scale = 1]
       % \node at (1,1.65) {\underline{$a^{(e)} = \prod_{i=1}^4 X_{e,i}$}};
        %
        \node at (1,0) {$\displaystyle\prod_{i=1}^4 X_i$};
        \draw[step = 2] (-0.5,-0.5) grid (2.5,0.5);
    \end{tikzpicture},
    \begin{tikzpicture}[baseline = -0.1cm, scale = 1]
        %\node at (1,1.65) {\underline{$b^{(e)} = \prod_{i=1}^4 Z_{e,i}$}};
        %
        \node at (1,0) {$\displaystyle\prod_{i=1}^4 Z_i$};
        \draw[step = 2] (-0.5,-0.5) grid (2.5,0.5);
    \end{tikzpicture} \right \rangle
\end{align}
which we identify as a particular instance of a class of topological subsystem codes introduced in \cite{Ellison2023paulitopological}. Namely, this subsystem code corresponds to starting with four copies of the toric code and then ``gauging out" the anyons $e_1e_2e_3e_4$ and $m_1m_2m_3m_4$ (i.e. adding short strings of these anyons to the check group).

\subsection{Balanced concatenated code and balanced subsystem product code}
Let us now perform a balancing of the previous constructions. This will yield a balanced version of the subsystem product code, whose gauge fixings give a balanced concatenated code, and whose stabilizers can be gauged further to recover the balanced product code.
For simplicity, we assume the group action is free. In parallel to Section~\ref{sec:concat}, given two CSS codes
\begin{align*}
    \text{CSS}_i: \,
    A_i \xrightarrow{\delta_i} Q_i \xrightarrow{d_i^T} B_i
\end{align*}
first we consider the balanced concatenation by taking CSS$_1$ as the outer code, and CSS$_2$ as the inner code. We start with one copy of CSS$_1$ for each equivalence class $\tilde q_2 \in G\backslash Q_2$, giving CSS$_1\otimes_G Q_2$ with stabilizer group
\begin{align*}
    \mathcal S_\text{outer,1}
    =
    \langle\,
    a_1^{(\tilde q_2)}, b_1^{(\tilde q_2)}
    \,|\,
    a_1\in A_1,\,
    b_1\in B_1,\,
    \tilde q_2\in G\backslash Q_2
    \,\rangle 
\end{align*}
where the $a_1^{(\tilde q_2)}$ and $b_1^{\tilde q_2}$ denotes the stabilizers CSS$_1$ in layer $\tilde q_2$. Next, we promote the Pauli operators in CSS$_2$ to logical operators in  each layer of CSS$_1$. These are
\begin{align*}
    \mathcal S_\text{inner,2}
    = 
    \left\langle\,
    \prod_{q_2\in \tilde a_2}
    \mathcal{X}^{(\tilde q_2)} \cdot g_{q_2},\,
    \prod_{q_2\in \tilde b_2} \mathcal{Z}^{(\tilde q_2)} \cdot g_{q_2}
    \,\bigg|\,
    \mathcal{X} \in L_{X,1},\,
    \mathcal{Z} \in L_{Z,1},\,
    \tilde a_2 \in G\backslash A_2,\,
    \tilde b_2 \in G\backslash B_2
    \,\right\rangle
\end{align*}
These terms commute for the same reason that $\alpha(q_1, \tilde a_2)$ and $\beta(q_1, \tilde b_2)$ commute. Combining the two stabilizer groups $\mathcal{S}_\text{outer, 1} \cup \mathcal{S}_\text{inner, 2}$, we obtain the balanced concatenated code CSS$_1 \circ_G $CSS$_2$. Conversely, one can use CSS$_1$ as the inner code, and stack copies of CSS$_2$, this leads to the balanced concatenated code CSS$_2 \circ_G $CSS$_1$ with stabilizers group given by
\begin{align*}
    & \mathcal S_\text{outer,2}
    =
    \langle\,
    a_2^{(\tilde q_1)}, b_2^{(\tilde q_1)}
    \,|\,
    a_2\in A_2,\,
    b_2\in B_2,\,
    \tilde q_1\in  Q_1 / G
    \,\rangle \\
    & \mathcal S_\text{inner,1}
    = 
    \left\langle\,
    \prod_{q_1\in \tilde a_1} 
    g_{q_1}\cdot \mathcal{X}^{(\tilde q_1)},\,
    \prod_{q_1\in \tilde b_1} g_{q_1}\cdot \mathcal{Z}^{(\tilde q_1)}
    \,\bigg|\,
    \mathcal{X} \in L_{X,2},\,
    \mathcal{Z} \in L_{Z,2},\,
    \tilde a_1 \in  A_1 / G,\,
    \tilde b_1 \in  B_1 / G
    \,\right\rangle
\end{align*}
In either case, gauging the inner stabilizer group recovers the balanced product CSS$_1 \otimes_G$ CSS$_2$. This can be thought of as a balanced version of the full block reading in \cite{cowtan2025fast}. Assuming doing surgery on layers of CSS$_1$ (treating CSS$_1$ as the outer code) instead of measuring identical logicals $\prod_{q_1\in a_1}\bar X_i^{(q_1)}$ and $\prod_{q_1\in b_1}\bar Z_i^{(q_1)}$, for $i = 1, \cdots, k_1$, in layers of CSS$_1$, we measure logicals modulated by appropriate group actions in different layers.

Similar to the subsystem tensor product code, the two concatenations can be thought of as two gauge fixings of a subsystem code. Without making a choice of representatives, the Hilbert space of the subsystem code is $Q_1 \otimes_G Q_2$, and the gauge group is generated by
\begin{equation*}
    \mathcal{G} = \left\langle
    i, \,
    \alpha(a_1\otimes_G q_2),\,
    \alpha(q_1\otimes_G a_2),\,
    \beta(b_1\otimes_G q_2),\,
    \beta(q_1\otimes_G b_2)
    \right\rangle
\end{equation*}
where the first two terms are $X$-operators 
spanning the vector spaces $A_1\otimes_G Q_2$ and $Q_1\otimes_G A_2$ respectively. Concretely they are given by maps
\begin{align*}
    A_1\otimes_G Q_2 
    \xrightarrow{\delta_1\otimes_G id} Q_1\otimes_G Q_2
    \hspace{55pt}
    Q_1\otimes_G A_2 
    \xrightarrow{id \otimes_G \delta_2} Q_1\otimes_G Q_2
\end{align*}
Similarly, the last two terms are $Z$-operators spanning the vector spaces $B_1\otimes_G Q_2$ and $Q_1\otimes_G B_2$ given by
\begin{align*}
    B_1\otimes_G Q_2 
    \xrightarrow{d_1\otimes_G id} Q_1\otimes_G Q_2
    \hspace{55pt}
    Q_1\otimes_G B_2 
    \xrightarrow{id \otimes_G d_2} Q_1\otimes_G Q_2
\end{align*}
If we choose representatives by making the quotient on CSS$_1$, then $Q_1\otimes_G A_2$ and $Q_1\otimes_G B_2$ are given by $\mathcal{S}_\text{outer, 2}$, while
\begin{align*}
    \alpha(\tilde a_1\otimes_G q_2) = \prod_{q_1\in \tilde a_1} X_{\tilde q_1, g_{q_1} \cdot q_2}
    \hspace{35pt}
    \beta(\tilde b_1\otimes_G q_2) = \prod_{q_1\in \tilde b_1} X_{\tilde q_1, g_{q_1} \cdot q_2}    
\end{align*}
If we choose to quotient CSS$_2$, then $A_1\otimes_G Q_2$ and $B_1\otimes_G Q_2$ gives $\mathcal{S}_\text{outer, 1}$, while
\begin{align*}
    \alpha(q_1\otimes_G \tilde a_2 ) = \prod_{q_2\in \tilde a_2} X_{q_1 \cdot g_{q_1}, \tilde  q_2}
    \hspace{35pt}
    \beta(q_1\otimes_G \tilde b_2 ) = \prod_{q_2\in \tilde b_2} X_{q_1 \cdot g_{q_1}, \tilde  q_2}
\end{align*}
Schematically we may write
\begin{align*}
    \mathcal{G} = \langle i,\,
    \mathcal{S}_\text{outer, 1} ,\,
    \mathcal{S}_\text{outer, 2}
    \rangle
\end{align*}
However, we note $\mathcal{G}$ does not have a preferred basis, and only reduce to the two components upon making choices of appropriate bases. To switch between the above two bases, we can define a bijection on the basis sets
\begin{align*}
    \mu: Q_1/G \times Q_2
    \ra 
    Q_1 \times G\backslash Q_2
    \hspace{35pt}
    \tilde q_1\otimes_G q_2
    \mapsto
    \tilde q_1 \cdot g_{q_2} \otimes_G \tilde q_2
\end{align*}

In order to see the concatenations are gauge fixings of $\mathcal{G}$, we need to show that $\mathcal{S}_{\text{inner},1}$ and $\mathcal{S}_{\text{inner},2}$ are subgroups of $\mathcal{G}$. To see this, take $\prod_{q_1\in \tilde a_1} g_{q_1}\cdot\mathcal{X}^{(\tilde q_1)}$ in $\mathcal{S}_{\text{inner},1}$ with $\mathcal{X}\in L_{X,2}$, and rewrite it in the basis $Q_1/G \times Q_2$,
\begin{align*}
    \prod_{q_1\in \tilde a_1} g_{q_1} \cdot \mathcal{X}^{(\tilde q_1)}
    =
    \prod_{q_2\in \mathcal{X}}\prod_{q_1\in \tilde a_1} X_{\tilde q_1, g_{q_1} \cdot q_2}
    \,\xrightarrow{\,\mu\,} \,
    \prod_{q_2\in \mathcal{X}}\prod_{q_1\in \tilde a_1} X_{q_1 \cdot g_{q_2}, \tilde q_2}
    =
    \prod_{q_2\in \mathcal{X}}\prod_{q_1\in \tilde a_1\cdot g_{q_2}} X_{q_1, \tilde q_2}
\end{align*}
The right-hand-side is indeed a product of operators in $A_1\otimes_G Q_2$. In additive notation, it is $\sum_{q_2\in \mathcal{X}} (\tilde a_1\cdot g_{q_2}) \otimes \tilde q_2$. A similar calculation holds for $Z$-operators in $\mathcal{S}_{\text{inner},1}$, as well as the operators in $\mathcal{S}_{\text{inner},2}$. The center (modulo $i$) of the gauge group is the stabilizer group, given by $ Z(\mathcal{G}) = \left\langle i, \,\mathcal{S}_{\text{inner},1},\,
    \mathcal{S}_{\text{inner},2}\right\rangle $.

\subsection{Example: color code}

We again use color code as an example. Adapting the conventions in Section~\ref{sec:cc}, the two inputs are classical NM model, with the first one in $Z$-basis and the second in $X$-basis, and the group $G$ is $\mathbb{Z}\times\mathbb{Z}$ via translation. We choose the basis by quotient the second NM model. There is only one orbit for $G\backslash Q_2$ and $G\backslash A_2$ each, represented by a vertex $\tilde v$ and a triangle $\tilde \triangle$, given by
\begin{equation*}
    \begin{tikzpicture}[baseline = 0.2cm]
        \coordinate (A) at (90:1);
        \coordinate (B) at (-30:1);
        \coordinate (C) at (210:1);

        \node at (0:0) {$\tilde \triangle$};

        \node at (90:1.2) {$\tilde v$}; 
        
        \draw (A) -- (B) -- (C) -- cycle;
    \end{tikzpicture}
\end{equation*}
The gauge group in this basis is
\begin{equation*}
    \mathcal{G} = 
    \Bigg\langle i,
    \hspace{5pt}
    \alpha(v, \tilde \triangle) =  \begin{tikzpicture}[baseline = 0.2cm]
        \coordinate (A) at (90:1);
        \coordinate (B) at (-30:1);
        \coordinate (C) at (210:1);

        \foreach \X in {A,B,C}
        \node at (\X) {$X$};

        \node at (90:1.4) {$v$};
        
        \draw (A) -- (B) -- (C) -- cycle;
    \end{tikzpicture}
    \hspace{5pt},\hspace{5pt}
    \beta(\triangle, \tilde v) =  \begin{tikzpicture}[baseline = 0.2cm]
        \coordinate (A) at (90:1);
        \coordinate (B) at (-30:1);
        \coordinate (C) at (210:1);

        \foreach \X in {A,B,C}
        \node at (\X) {$Z$};
        
        \node at (0:0) {$ \triangle$};
        
        \draw (A) -- (B) -- (C) -- cycle;
    \end{tikzpicture}
    \Bigg\rangle
\end{equation*}
We may relabel $\alpha(v,\tilde \triangle)$ by $\alpha(v)$, and $\beta(\triangle,\tilde v)$ by $\beta(v)$, where $v$ is the top vertex of $\triangle$.

Let $\mathcal{F}$ be a fractal operator of the NM model. There are two ways of gauge fixing by taking either of the two NM models as the outer code. For example, taking the first component as the outer code, then we obtain the following balanced concatenated code
\begin{align*}
    \mathcal{S}_1 = \Big\langle\,
    \prod_{v\in \mathcal{F}} \alpha(v),\,
    \beta(v)
    \,\Big\rangle
\end{align*}
To gauge the fractal operator, we introduce an $Z$-ancilla for each triangle, and perform code switching by imposing
\begin{align*}
    \begin{tikzpicture}[scale = 1.5, baseline = 0cm]
        \coordinate (A) at (0:1);
        \coordinate (B) at (60:1);
        \coordinate (C) at (120:1);
        \coordinate (D) at (180:1);
        \coordinate (E) at (240:1);
        \coordinate (F) at (300:1);
        \node at (30:0.55) {$X$};
        \node at (150:0.55) {$X$};
        \node at (270:0.55) {$X$};
        \node at (0:0) {$X$};
        \node at (-60:1) {$X$};
        \node at (-120:1) {$X$};
        % Draw boundary
        \draw (A) -- (B) -- (C) -- (D) -- (E) -- (F) -- cycle;
        % Draw diagonal lines
        \foreach \X/\Y in {A/D,B/E,C/F}
        \draw (\X) -- (\Y);
    \end{tikzpicture}
\end{align*}
The deformed $Z$-stabilizer is thus
\begin{align*}
    \begin{tikzpicture}[scale = 1.5, baseline = 0cm]
        \coordinate (A) at (0:1);
        \coordinate (B) at (60:1);
        \coordinate (C) at (120:1);
        \coordinate (D) at (180:1);
        \coordinate (E) at (240:1);
        \coordinate (F) at (300:1);
        \node at (30:0.55) {$Z$};
        \node at (150:0.55) {$Z$};
        \node at (270:0.55) {$Z$};
        \node at (0:0) {$Z$};
        \node at (-60:1) {$Z$};
        \node at (-120:1) {$Z$};
        % Draw boundary
        \draw (A) -- (B) -- (C) -- (D) -- (E) -- (F) -- cycle;
        % Draw diagonal lines
        \foreach \X/\Y in {A/D,B/E,C/F}
        \draw (\X) -- (\Y);
    \end{tikzpicture}
\end{align*}

Conversely, we can take the second component as the outer code, in which case the stabilizer group is
\begin{align*}
     \mathcal{S}_1 = \Big\langle\,
     \alpha(v),\,
    \prod_{v\in \mathcal{F}}\beta(v)
    \,\Big\rangle
\end{align*}
To gauge the fractal operator, we introduce an $X$-ancilla for each triangle, and the code switching is given by imposing the six-body $Z$-operator above, and the deformed stabilizer is the $X$-operator. In both cases we have the balanced product of two NM models, which is the color code.

\hskip 1cm

\section{Metachecks}
\label{sec:metacheck}

In this section, we will demonstrate subtleties related to product construction that arises due to existence of low-weight metachecks, which can give rise to a tensor product code with bad distance. Then, we study the corresponding code obtained from our code switching procedure. We find that these low-weight logicals are automatically included in the stabilizer group, but it also includes large-weight stabilizers, which will not result in a qLDPC code. Next, we study the condensation from the Hamiltonian perspective using perturbation theory. We will show this viewpoint gives a qLDPC code. Lastly, we will show how to modify the product construction by including constant-weight metachecks into the chain complex, in order to get a product code that is also qLDPC. 

An example of 3D TC $\otimes$ Ising was already discussed in Sec.~\ref{sec:3Dx1D}. Here we give an even simpler example where this phenomenon occurs, which is a product of two classical codes. Let CSS$_1$ be the $Z$-type 2D Ising model. Qubits are placed on vertices and $Z$-stabilizers are on edges. We denote vertices, edges and plaquettes in 2D square lattice by $v$, $e$ and $p$ respectively. Let CSS$_2$ be the $X$-type 1D Ising model. For convenience, we put qubits on edges, and $X$-stabilizers on vertices. Vertices and edges in the 1D chain are denoted by $w$ and $\tilde w := [w,w+1]$ respectively. Note that 3D cubic lattice can be constructed by product of the 2D square lattice and 1D chain. Vertices on the 3D lattice are labelled by a pair $(v,w)$. Edges are labeled by $(v,\tilde w)$ and $(e,w)$, and so on.

First let us look at the tensor product at the chain complex level. This is constituted of the following terms
\begin{equation*}
    \begin{tikzpicture}[scale = 1]
        \node at (0,1.9,0) {\underline{$\alpha(v,w) = \prod_{\tilde w\in w}X_{v,\tilde w} \prod_{e\in v} X_{e,w}$}};
        
        % Define coordinates
        \coordinate (O) at (0,0,0);
        \coordinate (A) at (-1.5,0,0);
        \coordinate (B) at (1.5,0,0);
        \coordinate (C) at (0,-1.5,0);
        \coordinate (D) at (0,1.5,0);
        \coordinate (E) at (0,0,1.5);
        \coordinate (F) at (0,0,-1.5);
            
        % Draw edges
        \draw (A) -- (B);
        \draw (C) -- (D); 
        \draw (E) -- (F);
            
        \foreach \X/\Y in { O/E, O/F}
        \node at ($( \X )!0.5!( \Y )$) {$X$};
        \foreach \X/\Y in { O/C, O/D}
        \node[black] at ($( \X )!0.5!( \Y )$) {$X$};
            
        \foreach \X/\Y in {O/A, O/B}
        \node[black] at ($( \X )!0.5!( \Y )$) {$X$}; 
    \end{tikzpicture}
    \hspace{45pt}
    \begin{tikzpicture}[scale = 1, baseline = -3ex]
        \node at (2,2.95) {\underline{$\zeta(e,\tilde w) = \prod_{v\in e}Z_{v,\tilde w} \prod_{w\ni \tilde w} Z_{e,w}$}};

        % First plaquette
        % Define coordinates
        \coordinate (O) at (0,0,0); 
        \coordinate (A) at (2,0,0);
        \coordinate (B) at (2,2,0);
        \coordinate (C) at (0,2,0);
    
        % Draw edges
        \draw (O)-- (A) -- (B) -- (C) -- cycle;

        \foreach \X/\Y in { O/C, A/B} 
        \node[black] at ($( \X )!0.5!( \Y )$) {$Z$}; 

        \foreach \X/\Y in {O/A, C/B}
        \node[black] at ($( \X )!0.5!( \Y )$) {$Z$};

        % Second plaquette
        % Define coordinates
        \coordinate (O) at (4,0.5,0);
        \coordinate (A) at (4,2.5,0);
        \coordinate (B) at (4,2.5,2);
        \coordinate (C) at (4,0.5,2);
        % Draw edges
        \draw (O) -- (A) --(B) -- (C) -- cycle;
        
        \foreach \X/\Y in { A/B,C/O}
        \node at ($( \X )!0.5!( \Y )$) {$Z$}; 
        \foreach \X/\Y in {O/A,  B/C}
        \node[black] at ($( \X )!0.5!( \Y )$) {$Z$};
    \end{tikzpicture}
\end{equation*}
This is not the 3D TC as it misses the horizontal plaquette $Z$-stabilizers, which become $Z$-logical operators of this code. Hence, this code has a bad distance, $d=4$, regardless of the system size.

Instead of looking at the formal tensor product, let us perform the code switching explicitly to see what stabilizer group we obtain. We will actually find that the horizontal plaquette $Z$-stabilizers are generated by the code switching.

For each edge $\tilde w$ in the 1D chain, introduce a copy of CSS$_1$. Qubits in this layer can be placed on edges $(v,\tilde w)$ in 3D cubic lattice. Stabilizers in this layer are denoted $b_e^{(\tilde w)} = \prod_{v\in e}Z_{v,\tilde w }$. For each vertex $w$ in the 1D chain and each edge $e$ in the 2D lattice, introduce an ancilla, which can be put on the edge $(e,w)$ in 3D cubic lattice. The ancillas are on the edges, hence they are in the (1-form) $Z$-paramagnet state with stabilizers $Z_{e,w}$. The stacked system is shown below, and the stabilizer group is $\mathcal{S}_0 = \langle b_e^{(\tilde w)} , Z_{e,w}\rangle$
\begin{equation*}
    \begin{tikzpicture}[scale = 1, baseline = 0cm]

        % First plaquette
        % Define coordinates
        \coordinate (O) at (0,0,0); 
        \coordinate (A) at (2,0,0);
        \coordinate (B) at (2,2,0);
        \coordinate (C) at (0,2,0);
    
        % Draw edges
        \draw (O)-- (A) -- (B) -- (C) -- cycle;

        \foreach \X/\Y in { O/C, A/B} 
        \node[black] at ($( \X )!0.5!( \Y )$) {$Z$};

        % Second Plaquette
        % Define coordinates
        \coordinate (O) at (4,0.5,0);
        \coordinate (A) at (4,2.5,0);
        \coordinate (B) at (4,2.5,2);
        \coordinate (C) at (4,0.5,2);
        % Draw edges
        \draw (O) -- (A) --(B) -- (C) -- cycle;

        \foreach \X/\Y in {O/A,  B/C}
        \node[black] at ($( \X )!0.5!( \Y )$) {$Z$};
    \end{tikzpicture}
    \hspace{55pt}
    \begin{tikzpicture}[scale=1,baseline = -1cm]
        % First edge
        % Left star
        \coordinate (L) at (0,0,0);
        \foreach \x/\y/\z in {-1/0/0, 0/1/0, 0/-1/0, 0/0/1, 0/0/-1} {
        \draw (L) -- (0.5*\x,0.5*\y,0.5*\z);
        }
    
        % Right star
        \coordinate (R) at (2,0,0);
        \foreach \x/\y/\z in {1/0/0, 0/1/0, 0/-1/0, 0/0/1, 0/0/-1} {
        \draw (R) -- ($(R)+(0.5*\x,0.5*\y,0.5*\z)$);
        }
        
        % Connecting line
        \draw (L) -- (R);
    
        % Pauli Z's
        % \node at ([shift={(0.2,-0.2)}]L) {$Z$};
    
        \node[black] at ($(L)!0.5!(R)$) {$Z$};

        % Second edge
        % Front star
        \coordinate (F) at (4,0.5,0);
        \foreach \x/\y/\z in {1/0/0, -1/0/0, 0/1/0, 0/-1/0, 0/0/-1} {
        \draw (F) -- ($ (F)+(0.5*\x,0.5*\y,0.5*\z)$);
        }
    
        % Back star
        \coordinate (B) at (4,0.5,2);
        \foreach \x/\y/\z in {1/0/0, -1/0/0, 0/1/0, 0/-1/0, 0/0/1} {
        \draw (B) -- ($(B)+(0.5*\x,0.5*\y,0.5*\z)$);
        }
        
        % Connecting line
        \draw (F) -- (B);
    
        % Pauli Z's
        % \node at ([shift={(0.2,-0.2)}]L) {$Z$};
    
        \node[black] at ($(F)!0.5!(B)$) {$Z$};
    \end{tikzpicture}
\end{equation*}
\noindent We perform a code switching by measuring the stabilizers $\alpha(v,w)$. This eliminates any term in $\mathcal{S}_0$ that anti-commutes with the measurements, and leaving only the commuting terms. Besides the $Z$-stabilizers $\xi(e,\tilde w)$ above, we also have the following terms
\begin{equation*}
    \begin{tikzpicture}[scale = 1, baseline = 4ex]
        \node at (0.8,2.7,0) {\underline{$\nu(p,w) = \prod_{e\in p}Z_{e,w}$}};
        
        % Define coordinates
        \coordinate (B) at (2,2,0);%back-top-right
        \coordinate (C) at (0,2,0);%back-top-left
        \coordinate (F) at (2,2,2);
        \coordinate (G) at (0,2,2);

        % Draw edges    
        \draw  (B) -- (C); 
        \draw  (F) -- (G) ;
        \draw (B) -- (F);
        \draw (C) -- (G);

        \foreach \X/\Y in { B/F, C/G}
        \node[circle, inner sep=1.5pt] at ($( \X )!0.5!( \Y )$) {$Z$};
    
        \foreach \X/\Y in { B/C,  F/G}
        \node[circle, inner sep=1.5pt] at ($( \X )!0.5!( \Y )$) {$Z$}; 
    \end{tikzpicture}
    \hspace{35pt}
    \begin{tikzpicture}[scale=1, baseline = -3.5em]
        % Left star
        \coordinate (L) at (0,0,0);
        \foreach \x/\y/\z in {-1/0/0, 0/1/0, 0/-1/0, 0/0/1, 0/0/-1} {
        \draw (L) -- (0.5*\x,0.5*\y,0.5*\z);
        }
    
        % Middle star
        \coordinate (M) at (2,0,0);
        \foreach \x/\y/\z in {1/0/0, 0/1/0, 0/-1/0, 0/0/1, 0/0/-1} {
        \draw (M) -- ($(M)+(0.5*\x,0.5*\y,0.5*\z)$);
        }

        % Right star
        \coordinate (R) at (4,0,0);
        \foreach \x/\y/\z in {1/0/0, 0/1/0, 0/-1/0, 0/0/1, 0/0/-1} {
        \draw (R) -- ($(R)+(0.5*\x,0.5*\y,0.5*\z)$);
        }
        
        % Connecting line
        \draw (L) -- (M) -- (R);
    
        % Pauli Z's
        % \node at ([shift={(0.2,-0.2)}]L) {$Z$};
    
        \node at ($(L)!0.5!(M)$) {$Z$};
        \node at ($(R)!0.5!(M)$) {$Z$};

        \node at ([shift={(-0.8,0)}]L) {$\cdots$};
        \node at ([shift={(0.8,0)}]R) {$\cdots$};
    \end{tikzpicture}
\end{equation*}
The missing plaquette term $\nu(p,w)$ is generated from code switching. Together with $\alpha(v,w)$ and $\zeta(e,\tilde w)$, they form the stabilizer group of 3D TC. However, non-local terms as above are also generated. More specifically, given a non-contractible loop $l$ in the 2D lattice, the $Z$-operator
\begin{align*}
    \mathcal{Z}(l,w) := \prod_{e\in l}Z_{e,w}
\end{align*}
is also in $\mathcal{S}_0$ and clearly commutes with $\alpha(v,w)$. Since it cannot be generated from a product of the local $Z$-stabilizers, it is an independent term. Physically, $\mathcal{Z}(l,w)$ hops the anyon $\ee$ around the loop $l$. Thus, we find that the resulting stabilizer code contains two additional large weight logicals, corresponding to the $x$ and $y$ cycles of the three-torus. The corresponding code has only one logical qubit left: the $X$-logical is supported on a $xy$-plane in the dual lattice, and its conjugate $Z$-logical is supported on a line in $z$-direction on the direct lattice. Moreover, this code is not qLDPC.

As a final perspective, we instead consider the condensation in the Hamiltonian formulation. We start with the stabilizer Hamiltonian defined by $\mathcal{S}_0$, and we turn on the condensation terms $\alpha(v,w)$. The Hamiltonian of the system is
\begin{align*}
    H = -\sum_{e,\tilde w} b_e^{(\tilde w)} - \sum_{e,w} Z_{e,w} - \Lambda \sum _{v,w} \alpha(v,w),
\end{align*}
where $\Lambda=0$ corresponds to the initial stabilizer Hamiltonian $S_0$ and the condensation corresponds to tuning the parameter $\Lambda$ to infinity.
Working in degenerate perturbation theory in orders of $\frac{1}{\Lambda}$, we derive a low energy effective Hamiltonian
\begin{align*}
    H^\text{eff} = - \sum _{v,w} \alpha(v,w) - \frac{1}{\Lambda^3}\sum_{e,\tilde w}\zeta(e,\tilde w) - \frac{1}{\Lambda^4} \sum_{p,w} \nu(p,w) - \sum_{l,w}\frac{1}{\Lambda^{|l|}}\mathcal{Z}(l,w),
\end{align*}
where $|l|$ is the length of the non-contractible loop $l$, which scales linearly with respect to the system size $L$. If we consider this as a family of error correcting codes as a function of $L$, then as $L\rightarrow \infty$, the non-local operators are exponentially suppressed by $\Lambda$, hence they should be dropped from the stabilizer group of the family. The rest of the terms form the 3D TC, and define a family of qLDPC codes.

\begin{figure}[h!]
    \centering
    \resizebox{0.5\textwidth}{!}{\includegraphics[width=0.5\linewidth]{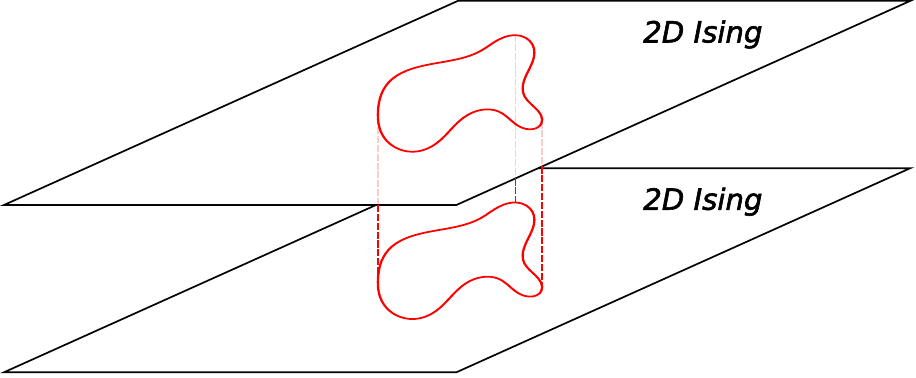}}
    \caption{Schematics of condensation. The red strings are 1D domain walls of the global $\mathbb{Z}_2$. Pairs of domain walls in neighboring layers are condensed.}
    \label{fig:2DIsing1DIsingPostCond}
\end{figure}

Let us look more closely at the code switching term $\alpha(v,w) = \prod_{\tilde w\in w}X_{v,\tilde w} \prod_{e\in v} X_{e,w}$. A single $X_{v,\tilde w}$ in the copy of 2D $Z$-Ising creates a 1D domain wall around the vertex $v$. On the other hand, the product $\prod_{e\in v} X_{e,w}$ creates a loop which is charged under the 1-form $\mathbb{Z}_2$ in the $Z$-paramagnet. Hence this term condenses pairs of 1D domain walls in each neighboring layers of 2D Ising, as shown pictorially in Figure~\ref{fig:2DIsing1DIsingPostCond}. Equivalently, this is the local symmetry action (Gauss law) used to gauge the diagonal global $\mathbb{Z}_2$ symmetry in each pair of neighboring layers of 2D Ising. Condensing pairs of domain walls gives rise to an equivalence class of a single domain wall across all layers. This is the $\mm$ loop of the 3D TC.

Note that if we start with 3D TC, and set $Z = 1$ for all the qubits on horizontal edges, then we recover $\mathcal{S}_0$. Since a single $Z$ creates a pair of charges $\ee$, we have proliferated $\ee$ in each horizontal plane. Said differently, condensing domain walls in stacks of 2D Ising can be thought of \textit{uncondensing} $\ee$ in each horizontal plane.

To summarize. We started with 2D $Z$-Ising as CSS$_1$, and 1D $X$-Ising as CSS$_2$. The CSS code of their product chain complex differs from the 3D TC by the horizontal plaquette terms. This code is LDPC, but the code distance is $4$. If we start with the stacked system $\mathcal{S}_0$, and performs a code switching by measuring $\alpha(v,w)$, then the stabilizer group strictly contains the 3D TC. The additional terms are the non-local logicals $\mathcal{Z}(l,w)$. The code distance is $L$, but it is not LDPC. Finally, we start with the stacked system, and condense pairs of domain walls. Under perturbation theory, the non-local operators are suppressed, and we correctly obtain the 3D TC, which is LDPC, and the code distance is $L$.

The term $\nu(p,w)$ is what makes tensor product differ from code switching and condensation different. Its appearance is due to low-weight metachecks in the $Z$-stabilzers of 2D Ising, given by $\prod_{e\in p} b_e = 1$ for each plaquette $p$. In terms of gauging, the metachecks become the dual $\mathbb{Z}_2$ 1-form symmetry.

It is not hard to recover the missing terms $\nu(p,w)$ on the chain complex level, such that we obtain the 3D TC from the tensor product perspective. We simply extend the chain complex of CSS$_1$ to include the metachecks, so the complex becomes
\begin{align*}
    \text{CSS$_1^{\text{ext}}$}: \,Q_1 \ra B_1 \ra M
\end{align*}
This is exactly the cellular complex of $\mathbb{R}^2$. The tensor product between the extended complex of CSS$_1^{\text{ext}}$ and CSS$_2$ is given in Figure~\ref{fig:R3complex}. By extending the complex by the metachecks $M$, we haven an additional terms $M\otimes A_2$ in the product complex, which is exactly the missing term $\nu(p,w)$. Note CSS$_1^{\text{ext}}\otimes$ CSS$_2$ is exactly the cellular complex of $\mathbb{R}^3$, with qubits on edges, which indeed gives the complex for 3D TC.

\begin{figure}
    \begin{tikzpicture}[scale=0.65, every node/.style={scale=0.8}]
  % column centers
    \def\LL{-6}
  \def\L{-3}
  \def\C{0}
  \def\R{3}
   \def\RR{6}
   \def\shadedwidth{60}
   \def\shadedheight{140}
    \def\H{1}
    \def\HH{2}
    \def\HHH{3}

  % Background stripes
  \begin{scope}
    \node[rounded corners=9pt, fill=red!20,  minimum width=\shadedwidth, minimum height=\shadedheight] at (\L,-1.5) {};
    \node[rounded corners=9pt, fill=black!10, minimum width=\shadedwidth, minimum height=\shadedheight] at (\C,-1.5) {};
    \node[rounded corners=9pt, fill=blue!20, minimum width=\shadedwidth, minimum height=\shadedheight] at (\R,-1.5) {};
  \end{scope}

  % Big labels and arrows along the bottom
  \node[font=\bfseries\Huge, text=red!80!black]  (A) at (\L, -3) {$\mathit{A}$};
  \node[font=\bfseries\Huge, text=black!75]      (Q) at (\C,-3) {$\mathit{Q}$};
  \node[font=\bfseries\Huge, text=blue!80!black] (B) at (\R, -3) {$\mathit{B}$};
  \draw[very thick,->] (A) -- (Q);
  \draw[very thick,->] (Q) -- (B);

  % Column subtitles
  \node[font=\Large, text=red!80!black,  anchor=north] at (\L, -3.5) {$X$-checks};
  \node[font=\Large, text=black!80,      anchor=north] at (\C, -3.5) {Qubits};
  \node[font=\Large, text=blue!80!black, anchor=north] at (\R, -3.5) {$Z$-checks};

  % Centered \oplus in each column
  \node[font=\Large, text=black]      at (\C, 0) {$\oplus$};
  \node[font=\Large, text=blue!80!black] at (\R, -1) {$\oplus$};

  % External boundary labels
  \node (MQ) at ( \RR, -1) {$M\!\otimes\! Q_{2}$};

  % Nodes inside stripes
  % Left stripe (A)
  \node[text=red!80!black]  (QA) at (\L, 0) {$Q_{1}\otimes A_{2}$};

  % Middle stripe (Q)
  \node[text=black!80]      (QQ) at (\C, 1) {$Q_{1} \otimes Q_{2}$};
  \node[text=black!80]      (BA) at (\C, -1) {$B_{1}\otimes A_{2}$};

  % Right stripe (B)
  \node[text=blue!80!black] (BQ) at (\R, 0) {$B_{1}\otimes Q_{2}$};
  \node[text=blue!80!black] (MA) at (\R, -2) {$M\otimes A_{2}$};

  % Arrows A stripe -> Q stripe
  \draw[->] (QA) -- (QQ);
  \draw[->] (QA) -- (BA);

  % Arrows Q stripe -> B stripe
  \draw[black, ->] (QQ) -- (BQ);
  \draw[black, ->] (BA) -- (BQ);
  \draw[black, ->] (BA) -- (MA);

  % Arrows from B stripe to far right
  \draw[black, ->] (BQ) -- (MQ);
  \draw[black, ->] (MA) -- (MQ);

\end{tikzpicture}
    \caption{The tensor product between the extended complex CSS$_1^\text{ext}$ and CSS$_2$ contans new $Z$-checks given by the map $B_1 \otimes A_2 \rightarrow M \otimes A_2$.}
    \label{fig:R3complex}
\end{figure}

It is interesting to note that such metachecks do not necessarily have to be constant weight in the code family. For example, in the 3D TC, the product of all $X$-stabilizers is a metacheck with weight $O(n)$. Therefore, even though CSS$_1$ and CSS$_2$, as well as their tensor product are LDPC codes, the resulting code after code switching  might not be LDPC. However, from the perspective of condensation, the terms associated to large-weight metachecks are suppressed via perturbation theory.

\subsection{General product construction}

We now generalize the above observation. Given the chain complex of a qLDPC code $A \ra Q \ra B$, assume we can extend it to the left and to the right into the following chain complex
\begin{align*}
      \text{CSS}^{\text{ext}}: 0 \ra A^{(-n)} \xrightarrow{\delta^{-n}} \cdots \xrightarrow{\delta^{-3}} A^{(-2)} \xrightarrow{\delta^{-2}} A^{(-1)} \xrightarrow{\delta^{-1}} Q \xrightarrow{\delta^{0}} B^{(1)} \xrightarrow{\delta^{1}} B^{(2)} \xrightarrow{\delta^{2}} \cdots \xrightarrow{\delta^{m-1}} B^{(m)} \ra 0
\end{align*}
with $A^{(1)} := A$ and $B^{(1)} := B$, which satisfies
\vspace{-0.5em}
\begin{itemize}
    \item Each $\delta^\bullet$ is $w$-sparse. This means that in addition to being an qLDPC code, the metachecks and relations between them are bounded in weight by $w$.
    \item There does not exists $x$ in $\Ker \delta^\bullet - \Imaa \delta^{\bullet-1}$ such that $|x|\leq w$.
\end{itemize}
\vspace{-0.5em}
The space $A^{(-2)}$ describes all the $w$-local $X$-metachecks. $A^{(-3)}$ describes all the $w$-local metachecks of $X$-metachecks, and so forth. Similarly, $B^{(*)}$ describes a complex of $Z$-metachecks. For example, if CSS is the Toric code in $n$ dimensions with qubits live on $k$-cells $C^{k-1}\ra C^k \ra C^{k+1}$, then we should extend this to the full chain complex of $\mathbb{R}^n$.

Given two CSS codes which have been extended to such chain complexes CSS$_1^{\text{ext}}$ and CSS$_2^{\text{ext}}$, define the product code CSS$_1^{\text{ext}}$ $\otimes$ CSS$_2^{\text{ext}}$ to be the CSS code described by the three terms centered on degree $0$.

Now, we show how to realize this product through the coupled layer construction. We will find that we need to introduce more ancilla qubits, and perform additional code switching. For each qubit $q_2\in Q_2$, introduce a layer of CSS$_1$ (which from the perspective of the $X$ and $Z$ stabilizers, is the same data as CSS$_1^{\text{ext}}$). Next we introduce ancillas:

\begin{itemize}
    \item $Z$-ancilla for each $B^{(i)}_1 \otimes A^{(-i)}_2$, $i = 1, \cdots, \text{min}(m_1,n_2)$
    \item $X$-ancilla for each $A^{(-i)}_1\otimes B^{(i)}_2$, $i = 1, \cdots, \text{min}(n_1,m_2)$
\end{itemize}

Next we perform code switching, which is carried out by adding

\begin{itemize}
    \item $X$-stabilizers: $Q_1\otimes A_2^{(-1)}$ and $B^{(i)}_1 \otimes A^{(-i-1)}_2$, $i = 1, \cdots, \text{min}(m_1, n_2-1)$
    \item $Z$-stabilizers: $Q_1 \otimes B^{(1)}_2$ and $A^{(-i)}_1 \otimes B^{(i+1)}_2$, $i = 1, \cdots,  \text{min}(n_1, m_2-1)$ 
\end{itemize}

The schematics of stacking and code switching is shown by Figure~\ref{fig:productextcomplex} below.

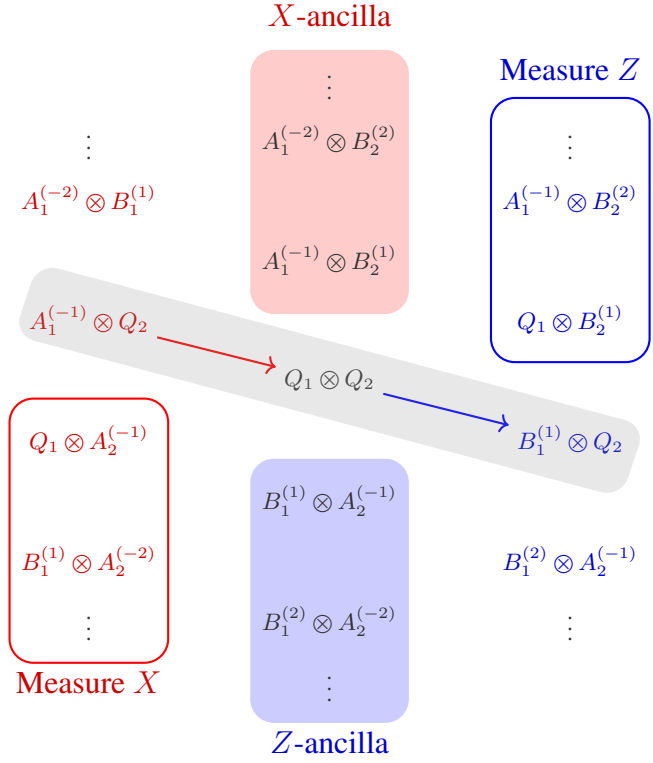
\begin{figure}[h!]
    \begin{tikzpicture}[scale=0.8]
  % column centers
    \def\LL{-8}
    \def\L{-4}
    \def\C{0}
    \def\R{4}
    \def\RR{8}
    \def\shadedwidth{60}
    \def\shadedheight{100}
    \def\H{1}
    \def\HH{2}
    \def\HHH{3}
    \def\HHHH{4}
    \def\HHHHH{5}

  % Background stripes
    \begin{scope}
    
    % Middle colors
    \node[rounded corners=9pt, fill=blue!20,  minimum width=\shadedwidth, minimum height=\shadedheight] at (\C,-3.5) {};
    
    \node[rounded corners=9pt, fill=red!20, minimum width=\shadedwidth, minimum height=\shadedheight] at (\C, 3.3) {};

    % Left colors
    \node[rounded corners=9pt, draw=red, thick,  minimum width=\shadedwidth, minimum height=\shadedheight] at (\L,-2.5) {};
    
    % Right colors
    \node[rounded corners=9pt, draw=blue, thick , minimum width=\shadedwidth, minimum height=\shadedheight] at (\R, 2.5) {};
    \end{scope}

    % % Big labels and arrows along the botton
  % \node[font=\bfseries\Huge, text=red!80!black]  (A) at (\L, -3) {$\mathit{A}$};
  % \node[font=\bfseries\Huge, text=black!75]      (Q) at (\C,-3) {$\mathit{Q}$};
  % \node[font=\bfseries\Huge, text=blue!80!black] (B) at (\R, -3) {$\mathit{B}$};
  % \draw[very thick,->] (A) -- (Q);
  % \draw[very thick,->] (Q) -- (B);

  % Column subtitles
  \node[font=\Large, text=red!80!black,  anchor=north] at (\L, -4.7) {\large Measure $X$};
  % \node[font=\Large, text=black!80,      anchor=north] at (\C, -3.5) {Qubits};
  \node[font=\Large, text=blue!80!black, anchor=north] at (\R, 5.5) {\large Measure $Z$};
  \node[font=\Large, text=blue!80!black,  anchor=north] at (\C, -5.7) {\large $Z$-ancilla};
  \node[font=\Large, text=red!80!black,  anchor=north] at (\C, 6.4) {\large $X$-ancilla};

  % % Centered \oplus in each column
  % \node[font=\Large, text=red!80!black]  at (\L, 0) {$\oplus$};
  % \node[font=\Large, text=black]      at (\C, \H) {$\oplus$};
  % \node[font=\Large, text=black]      at (\C, -\H) {$\oplus$};
  % \node[font=\Large, text=blue!80!black] at (\R, 0) {$\oplus$};

  % % External boundary labels
  % \node (AA) at (\LL, 0) {$A_{1}\!\otimes\! A_{2}$};
  % \node (BB) at ( \RR, 0) {$B_{1}\!\otimes\! B_{2}$};

  % Nodes inside stripes
  % Left stripe (A)
  \node[text=red!80!black]  (R1B) at (\L,  \HHH) {$A^{(-2)}_{1}\otimes B^{(1)}_{1}$};
  \node[text=red!80!black]  (AQ) at (\L,  \H) {$A^{(-1)}_{1}\otimes Q_{2}$};
  \node[text=red!80!black]  (QA) at (\L, -\H) {$Q_{1}\otimes A^{(-1)}_{2}$};
  \node[text=red!80!black]  (BR1) at (\L, -\HHH) {$B^{(1)}_{1}\otimes A^{(-2)}_{2}$};
  \node[text=black!80]       at (\L, -\HHHH) {$\vdots$};
  \node[text=black!80]       at (\L, \HHHH) {$\vdots$};

  % Middle stripe (Q)
  \node[text=black!80]      (R1M1) at (\C,  \HHHH) {$A^{(-2)}_{1}\otimes B^{(2)}_{2}$};
  \node[text=black!80]      (AB) at (\C,  \HH) {$A^{(-1)}_{1}\otimes B^{(1)}_{2}$};
  \node[text=black!80]      (QQ) at (\C, 0.0) {$Q_{1} \otimes Q_{2}$};
  \node[text=black!80]      (BA) at (\C, -\HH) {$B^{(1)}_{1}\otimes A^{(-1)}_{2}$};
  \node[text=black!80]      (M1R1) at (\C, -\HHHH) {$B^{(2)}_{1}\otimes A^{(-2)}_{2}$};
  \node[text=black!80]       at (\C, -\HHHHH) {$\vdots$};
  \node[text=black!80]       at (\C, \HHHHH) {$\vdots$};

  % Right stripe (B)
  \node[text=blue!80!black] (AM1) at (\R,  \HHH) {$A^{(-1)}_1\otimes B^{(2)}_{2}$};
  \node[text=blue!80!black] (QB) at (\R,  \H) {$Q_{1}\otimes B^{(1)}_{2}$};
  \node[text=blue!80!black] (BQ) at (\R, -\H) {$B^{(1)}_{1}\otimes Q_{2}$};
  \node[text=blue!80!black] (M1A) at (\R,  -\HHH) {$B^{(2)}_{1}\otimes A^{(-1)}_2$};
  \node[text=black!80]       at (\R, -\HHHH) {$\vdots$};
  \node[text=black!80]       at (\R, \HHHH) {$\vdots$};

    % Arrows A stripe -> Q stripe
    \draw[ ->, thick,red] (AQ) -- (QQ);

    % Arrows Q stripe -> B stripe
    \draw[ ->,  thick, blue] (QQ) -- (BQ);

    \def\pad{0.35}   % how much to extend *past* A1⊗Q2 and B1⊗Q2 along the line
    \def\th{0.2}    % half-thickness; try 0.45–0.70
    \fill[black!40, fill opacity=0.22, draw opacity=0, rounded corners=8pt]
    % upper edge from left to right
    ($ (AQ) + ({-3*\pad},{\pad}) + ({\th},{3*\th}) $) --
    ($ (BQ) + ({ 3*\pad},{-\pad}) + ({\th},{3*\th}) $) --
    % lower edge back from right to left
    ($ (BQ) + ({ 3*\pad},{-\pad}) + ({-\th},{-3*\th}) $) --
    ($ (AQ) + ({-3*\pad}, {\pad}) + ({-\th},    {-3*\th}) $) -- cycle;
  
    \end{tikzpicture}
    \begin{minipage}{0.95\textwidth}
        \centering\caption{The stacked system $\mathcal{S}_0$ before code switching. The gray shade covers stacks of CSS$_1$. The blue/ red shade are $Z$/ $X$-ancillas, and blue/ red arrows are $Z$/ $X$ stabilizers in each layer. The blue/ red box are terms to be measured in $Z$/ $X$ basis. Only the arrows correspond to the terms in $\mathcal{S}_0$ are shown.}
        \label{fig:productextcomplex}
    \end{minipage}
\end{figure}

\begin{figure}[h!]
    \begin{tikzpicture}[scale=0.8]
  % column centers
    \def\L{-4}
    \def\C{0}
    \def\R{4}
    \def\shadedwidth{60}
    \def\shadedheight{100}
    \def\H{1}
    \def\HH{2}
    \def\HHH{3}
    \def\HHHH{4}
    \def\HHHHH{5}

  % Background stripes
    \begin{scope}
    % Middle colors
    \node[rounded corners=9pt, fill=blue!20,  minimum width=\shadedwidth, minimum height=50] at (\C,-4.5) {};
    
    % Left box
    \node[rounded corners=9pt, draw=red, thick,  minimum width=\shadedwidth, minimum height=25] at (\L,-\H) {};

    \node[rounded corners=9pt, draw=red, thick,  minimum width=\shadedwidth, minimum height=50] at (\L,-3.5) {};

    \end{scope}

    % Column subtitles
    \node[font=\Large, text=red!80!black,  anchor=north]  at (\L, 0.3) {\large Measured};

    \node[font=\Large, text=red!80!black,  anchor=north]  at (\L, -4.6) {\large To be measured};
  
    \node[font=\Large, text=blue!80!black,  anchor=north] at (4.3,-1.6) {\large Generated};
  
    \node[font=\Large, text=blue!80!black,      anchor=north] at (\C, -5.5) {\large $Z$-ancilla};

    % Left column
  \node[text=red!80!black]  (QA) at (\L, -\H) {$Q_{1}\otimes A^{(-1)}_{2}$};
  \node[text=red!80!black]  (BR1) at (\L, -\HHH) {$B^{(1)}_{1}\otimes A^{(-2)}_{2}$};
  \node[text=black!80]       at (\L, -\HHHH) {$\vdots$};

  % Middle column
  \node[text=black!80]      (QQ) at (\C, 0.0) {$Q_{1} \otimes Q_{2}$};
  \node[text=black!80]      (BA) at (\C, -\HH) {$B^{(1)}_{1}\otimes A^{(-1)}_{2}$};
  \node[text=black!80]      (M1R1) at (\C, -\HHHH) {$B^{(2)}_{1}\otimes A^{(-2)}_{2}$};
  \node[text=black!80]       at (\C, -\HHHHH) {$\vdots$};

  % Right column
  
  \node[text=blue!80!black] (BQ) at (\R, -\H) {$B^{(1)}_{1}\otimes Q_{2}$};
  
  \node[text=blue!80!black] (M1A) at (4.15,  -\HHH) {$B^{(2)}_{1}\otimes A^{(-1)}_2$};
  \node[text=black!80]       at (\R, -\HHHH) {$\vdots$};

    % Arrows A stripe -> Q stripe
    \draw[->, thick, red] (-2.6,-0.7) -- (QQ);
    \draw[->, thick, red] (-2.6,-1.3) -- (BA);

    % Arrows Q stripe -> B stripe
    \draw[ ->] (QQ) -- (BQ);
    \draw[->, thick, blue] (BA) -- (BQ);
    \draw[blue, thick,->] (BA) -- (M1A);

    \def\pad{0.29}   % how much to extend *past* A1⊗Q2 and B1⊗Q2 along the line
    \def\th{0.1}    % half-thickness; try 0.45–0.70
    % \fill[black!40, fill opacity=0.22, draw opacity=0, rounded corners=8pt]
    % upper edge from left to right
    
    \draw [blue, rounded corners, thick]
    ($ (BA) + ({3*\pad}, {\pad}) + ({-\th},    {3*\th}) + (0.25,0.0) - (0,1.1) + (0.1,0)$)
    --
    ($ (BQ) + ({ -3*\pad},{-\pad}) + ({-\th},{3*\th}) - (0,2)$) 
    --
    % lower edge back from right to left
    ($ (BQ) + ({ -3*\pad},{-\pad}) + ({-\th},{3*\th}) $) 
    --
    ($ (BA) + ({3*\pad}, {\pad}) + ({-\th},    {3*\th}) + (0.25,0.0)+(0.1,0)$) -- cycle;

    \def\pad{0.35}   % how much to extend *past* A1⊗Q2 and B1⊗Q2 along the line
    \def\th{0.2}    % half-thickness; try 0.45–0.70
    \fill[black!40, fill opacity=0.22, draw opacity=0, rounded corners=8pt]
    % upper edge from left to right
    ($ (BA) + ({-3*\pad},{\pad}) + ({\th},{3*\th}) $) --
    ($ (M1A) + ({ 3*\pad},{-\pad}) + ({\th},{3*\th}) $) --
    % lower edge back from right to left
    ($ (M1A) + ({ 3*\pad},{-\pad}) + ({-\th},{-3*\th}) $) --
    ($ (BA) + ({-3*\pad}, {\pad}) + ({-\th},    {-3*\th}) $) -- cycle;
    
    \end{tikzpicture}
    \begin{minipage}{0.95\textwidth}
        \centering\caption{Stabilizers of the system after the first code switching. After the top-left $X$-stabilizers are measured, the two blue arrows are generated. The gray shade is to show that analogy with Figure~\ref{fig:productextcomplex}.}
        \label{fig:condextcomplex1}
    \end{minipage}
\end{figure}
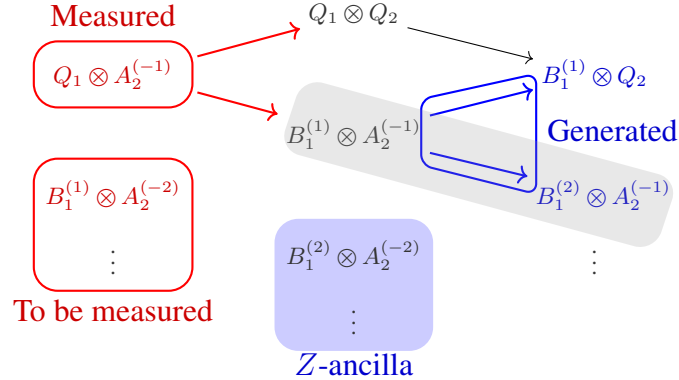

To see why this produces the desired chain complex, we perform the code switching sequentially. To illustrate the workflow, look at the half of the system on which $Z$-ancillas are supported below $Q_1\otimes Q_2$ in Figure~\ref{fig:productextcomplex}. First we only include the $Z$-ancillas on $B_1^{(1)}\otimes A_2^{(-1)}$, and add all the terms in $Q_1\otimes A^{(-1)}_2$ to the stabilizer group. The remaining  $Z$-stabilizers the commute are of the form
\vspace{-0.5em}
\begin{itemize}
    \item $B^{(1)}_1 \otimes Q_2$. Each pair $(b^{(1)}_1, q_2)$ gives $b^{(1)}_1(q_2) \prod_{a^{(-1)}_2 \ni q_2}Z_{b^{(1)}_1, a^{(-1)}_2}$.
    \item $B^{(2)}_1 \otimes A^{(-1)}_2$. Each pair $(b^{(2)}_1, a^{(-1)}_2)$ gives $\prod_{b^{(1)}_1 \in  b^{(2)}_1} Z_{b^{(1)}_1, a^{(-1)}_2}$.
\end{itemize}
\vspace{-0.5em}
Figure~\ref{fig:condextcomplex1} shows the stabilizers of the system after the first code switching. Note that in comparison to Figure~\ref{fig:productextcomplex}, $B^{(1)}_1 \otimes A^{(-1)}_2$ and $B^{(2)}_1 \otimes A^{(-1)}_2$ now plays the role of $Q_1\otimes Q_2$ and $B^{(1)}_1 \otimes Q_2$ respectively. This analogy is shown by the gray shade in Figure~\ref{fig:condextcomplex1}.

\begin{figure}[h!]
    \begin{tikzpicture}[scale=0.45, every node/.style={scale=0.8}]
  % column centers
    \def\LL{-6}
    \def\L{-3}
    \def\C{0}
    \def\R{3}
    \def\RR{6}
    \def\shadedwidth{60}
    \def\shadedheight{100}
    \def\H{2}
    \def\HH{4}
    \def\HHH{6}

  % Left column
  \node[text=red!80!black]  (QA) at (\L, -\H) {$B^{(m_1-1)}_{1}\otimes A^{(-m_1)}_{2}$};
  \node[text=red!80!black]  (BR1) at (\L, -\HHH) {$B^{(m_1)}_{1}\otimes A^{(-m_1-1)}_{2}$};
  
  \node   at (\L, 0) {$\vdots$};

  % Middle column
  \node[text=black!80] (QQ) at (\C, 0.5) {$\vdots$};

  \node[text=black!80]      (BA) at (\C, -\HH) {$B^{(m_1)}_{1}\otimes A^{(-m_1)}_{2}$};

  % Right column
  
  \node[text=blue!80!black] (BQ) at (\R, -\H) {$B^{(m_1)}_{1}\otimes A^{(-m_1+1)}_{2}$};

  \node   at (\R, 0) {$\vdots$};

  % Arrows A stripe -> Q stripe
  \draw[->] (QA) -- (-0.5,-0.5);
  \draw[->] (QA) -- (BA);
  \draw[->] (BR1) -- (BA);
  % \draw[->] (BR1) -- (M1R1);

    % Arrows Q stripe -> B stripe
    \draw[ ->] (0.5, -0.5) -- (BQ);
    \draw[->] (BA) -- (BQ);
    % \draw[blue, ultra thick,->] (BA) -- (M1A);

    % % Arrows from B stripe to far right
    % \draw[black, ->] (Q1B2) -- (BB);
    % \draw[black, ->] (B1Q2) -- (BB);
    \end{tikzpicture}
    \hspace{15pt}
    \begin{tikzpicture}[scale=0.45, every node/.style={scale=0.8}]
  % column centers
    \def\LL{-6}
    \def\L{-3}
    \def\C{0}
    \def\R{3}
    \def\RR{6}
    \def\shadedwidth{60}
    \def\shadedheight{100}
    \def\H{2}
    \def\HH{4}
    \def\HHH{6}

  % Left column
  \node[text=red!80!black]  (QA) at (\L, -\H) {$B^{(n_2-1)}_{1}\otimes A^{(-n_2)}_{2}$};

  \node   at (\L, 0) {$\vdots$};

  % Middle column
  \node[text=black!80] (QQ) at (\C, 0.5) {$\vdots$};

  \node[text=black!80]      (BA) at (\C, -\HH) {$B^{(n_2)}_{1}\otimes A^{(-n_2)}_{2}$};
  % \node[text=black!80]      (M1R1) at (\C, -\HHHH) {$B^{(1)}_{1}\otimes A^{(1)}_{2}$};

  % Right column
  
  \node[text=blue!80!black] (BQ) at (\R, -\H) {$B^{(n_2)}_{1}\otimes A^{(-n_2+1)}_{2}$};

  \node   at (\R, 0) {$\vdots$};
  
  \node[text=blue!80!black] (M1A) at (\R,  -\HHH) {$B^{(n_2+1)}_{1}\otimes A^{(-n_2)}_2$};

  % % Arrows from the far left into A stripe
  % \draw[->] (AA) -- (A1Q2);
  % \draw[ ->] (AA) -- (Q1A2);

  % Arrows A stripe -> Q stripe
  \draw[->] (QA) -- (-0.5,-0.5);
  \draw[->] (QA) -- (BA);
  % \draw[->] (BR1) -- (M1R1);

    % Arrows Q stripe -> B stripe
    \draw[ ->] (0.5, -0.5) -- (BQ);
    \draw[->] (BA) -- (BQ);
    \draw[->] (BA) -- (M1A);

    % % Arrows from B stripe to far right
    % \draw[black, ->] (Q1B2) -- (BB);
    % \draw[black, ->] (B1Q2) -- (BB);
    \end{tikzpicture}
    \hspace{15pt}
    \begin{tikzpicture}[scale=0.45, every node/.style={scale=0.8},baseline = -9.3em]
  % column centers
    \def\LL{-6}
    \def\L{-3}
    \def\C{0}
    \def\R{3}
    \def\RR{6}
    \def\shadedwidth{60}
    \def\shadedheight{100}
    \def\H{2}
    \def\HH{4}

  % Left column
  \node[text=red!80!black]  (QA) at (\L, -\H) {$B^{(n-1)}_{1}\otimes A^{(-n)}_{2}$};

  \node   at (\L, 0) {$\vdots$};

  % Middle column
  \node[text=black!80] (QQ) at (\C, 0.5) {$\vdots$};

  \node[text=black!80]      (BA) at (\C, -\HH) {$B^{(n)}_{1}\otimes A^{(-n)}_{2}$};
  % \node[text=black!80]      (M1R1) at (\C, -\HHHH) {$B^{(1)}_{1}\otimes A^{(1)}_{2}$};

  % Right column
  
  \node[text=blue!80!black] (BQ) at (\R, -\H) {$B^{(n)}_{1}\otimes A^{(-n+1)}_{2}$};

  \node   at (\R, 0) {$\vdots$};

  % % Arrows from the far left into A stripe
  % \draw[->] (AA) -- (A1Q2);
  % \draw[ ->] (AA) -- (Q1A2);

  % Arrows A stripe -> Q stripe
  \draw[->] (QA) -- (-0.5,-0.5);
  \draw[->] (QA) -- (BA);
  % \draw[->] (BR1) -- (M1R1);

    % Arrows Q stripe -> B stripe
    \draw[ ->] (0.5, -0.5) -- (BQ);
    \draw[->] (BA) -- (BQ);

    % % Arrows from B stripe to far right
    % \draw[black, ->] (Q1B2) -- (BB);
    % \draw[black, ->] (B1Q2) -- (BB);
    \end{tikzpicture}
    \begin{minipage}{0.95\textwidth}
        \centering\caption{From left to right: $m_1 < n_2$, $n_2 < m_1$ and $n:= m_1 = n_2$.}
        \label{fig:condextcomplexbot}
    \end{minipage}
\end{figure}
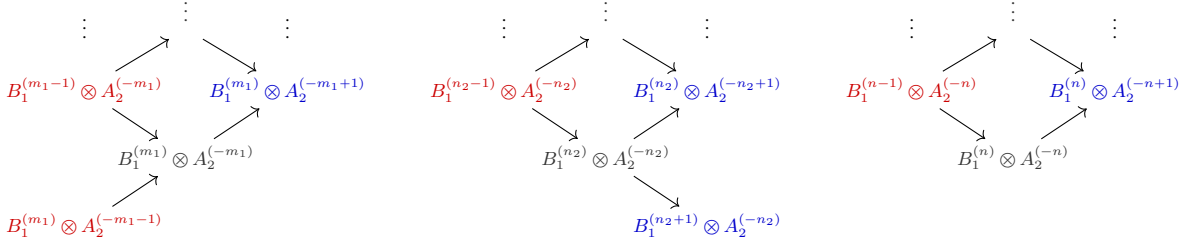

The procedure is now clear, next we add $Z$-ancillas $B^{(2)}_1 \otimes A^{(-2)}_2$, and add $X$-stabilizers in $B^{(1)}_1 \otimes A^{(-2)}_2$ to the stabilizer group, and so on. As we add more and more $X$-stabilizers, $Z$-stabilizers in the right column will be generated, and arrows in the diagram cascades from top to bottom. Depending on the length of the chain complexes, the bottom of the cascade will terminate in one of the three cases depending on the values of $m_1$ and $n_2$ in Figure~\ref{fig:condextcomplexbot}.

Exactly the same procedure applies to the upper half of Figure~\ref{fig:productextcomplex}, where $X$-ancillas are added step-by-step, and we perform code switching by adding $Z$-stabilizers. By the construction of the extended complex and applying the K\"unneth formula, there are no logicals of weight less than $w$ in the resulting product code.

This procedure can also be understood as gauging sequentially. Again, focus on the lower half of Figure~\ref{fig:productextcomplex}, starting with the chain complex $A_1^{(-1)} \otimes Q_2 \ra Q_1 \otimes Q_2 \ra B_{1}^{(1)}\otimes Q_2$, which is stacks of disjoint copies of CSS$_1$ labeled by $q_2\in Q_2$. We gauge all $X$-logicals and $X$-stabilizers that is diagonal in layers $q_2\in a_2^{(-1)}$ for all $a_2^{(-1)}\in A_2^{(-1)}$. This introduces the gauge qubits labeled by $B_1^{(1)}\otimes A_2^{(-1)}$, and the Gauss's law are $X$-stabilizers in the space $Q_1\otimes A_2^{(-1)}$. The deformed $Z$-stabilizers in space $B_1^{(1)}\otimes Q_2$ are the minimal coupling terms; the generated $Z$-stabilizers given by $B_1^{(1)}\otimes A_2^{(-1)} \ra B_1^{(2)}\otimes A_2^{(-1)}$ in Figure~\ref{fig:condextcomplex1} are the dual symmetries after gauging. Combining these terms, we have arrived at Figure~\ref{fig:condextcomplex1}.
Adding the dual symmetry operators to the Hamiltonian, and focus on the part of the system given by the complex $Q_1\otimes A_2^{(-1)} \ra B_1^{(1)}\otimes A_2^{(-1)} \ra B_1^{(2)} \otimes A_2^{(-1)}$. We gauge all the $X$-logicals and $X$-stabilizers in the subsystem $B_1^{(1)}\otimes A_2^{(-1)}$ diagonal in layers $a_2^{(-1)} \in a_2^{(-2)}$ for all $a_2^{(-2)}\in A_2^{(-2)}$. The layers in which the $X$-logicals are diagonal must be chosen so that it commutes with the arrow $B_1^{(1)} \otimes A_2^{(-1)} \ra B_1^{(1)} \otimes Q_2$ in Figure~\ref{fig:condextcomplex1}, which is guaranteed by the three terms $A_2^{(-2)}\ra A_2^{(-1)} \ra  Q_2$ in CSS$_2^{\text{ext}}$. The diagram is generated downward following this procedure.

A similar discussion holds for balanced product. If a CSS code has symmetry $G$, then one can extend the chain complex as above while preserving the $G$ symmetry. For instance, if there is an $X$-metacheck $\delta^{-1}(\sum a) = 0$, then $\delta^{-1}(\sum g\cdot a) = g\cdot \delta^{-1}(\sum a) = 0$, so group action preserves metachecks. After the complexes have been extended, one can take the balanced product of the extended complexes, forming
\begin{align*}
    \left(\text{CSS}_1^{\text{ext}} \otimes_G \text{CSS}_2^{\text{ext}}\right)^n = \bigoplus_{i+j = n} \left(\text{CSS}_1^{\text{ext}} \right)^i \otimes_G \left(\text{CSS}_2^{\text{ext}}\right)^j
\end{align*}
The CSS code is taken to be the three terms centered at degree $0$. This code can be analogously constructed by the cascade of code switching described. For instance, after the first round of code switching, besides $\xi([a_1]^{(\tilde q_2)}, \tilde q_2)$ and $ \zeta([b_1]^{(\tilde q_2)}, \tilde q_2)$, we also generate 
\begin{align*}
    \mu([r_X]^{(\tilde b_2)}, \tilde b_2) = \prod_{[a_1]^{(\tilde b_2)} \in [r_X]^{(\tilde b_2)}}X_{[a_1]^{(\tilde b_2)}, \tilde b_2} 
    \hspace{35pt}
    r_X \in A_1^{(1)},\\
    \nu([r_Z]^{(\tilde a_2)}, \tilde a_2) = \prod_{[b_1]^{(\tilde a_2)} \in [r_Z]^{(\tilde a_2)}}Z_{[b_1]^{(\tilde a_2)}, \tilde a_2}
    \hspace{35pt}
    r_Z \in B_1^{(1)},
\end{align*}
where $[a_1]^{(\tilde b_2)} \in [r_X]^{(\tilde b_2)}$ is the boundary map of the complex CSS$_1^{\text{ext}}/K_{\tilde b_2}$, and  $[b_1]^{(\tilde a_2)} \in [r_Z]^{(\tilde a_2)}$ is the transpose boundary map of the complex CSS$_1^{\text{ext}}/K_{\tilde a_2}$. These terms are logical operators in the code CSS$_1\otimes_G$CSS$_2$ of weight at most $w$, and they are obviously in $\mathcal{S}_0$: recall $\mathcal{S}_0 = \langle [a_1]^{({\Tilde{q}_2})}, [b_1]^{({\Tilde{q}_2})}, X_{[a_1]^{(\Tilde{b}_2)},\Tilde{b}_2}, Z_{[b_1]^{(\Tilde{a}_2)},\Tilde{a}_2} \rangle$. Keep going down the cascade we generate the CSS code associate to $\text{CSS}_1^{\text{ext}} \otimes_G \text{CSS}_2^{\text{ext}}$.

%\onecolumngrid
\clearpage
\twocolumngrid

\bibliography{references}
\end{document}